\documentclass[a4paper, traditabstract, longauth]{aa} 
\usepackage{amsmath}
\usepackage{graphicx}
\usepackage{color}
\usepackage[breaklinks, colorlinks, citecolor=blue]{hyperref}
\usepackage{natbib}
\usepackage{longtable,lscape}
\usepackage{txfonts}
\usepackage{amsfonts}
\usepackage{amssymb}
\usepackage{newlfont}
\usepackage{textcomp}
\usepackage{times}
\usepackage{units}
\usepackage{mathbbol}
\usepackage[switch]{lineno}
\usepackage{epstopdf}
\usepackage{ifthen}

\def\setsymbol#1#2{\expandafter\def\csname #1\endcsname{#2}}
\def\getsymbol#1{\csname #1\endcsname}

\def\Planck{\textit{Planck}}

\newbox\tablebox    \newdimen\tablewidth
\def\leaderfil{\leaders\hbox to 5pt{\hss.\hss}\hfil}
\def\endPlancktable{\tablewidth=\columnwidth 
    $$\hss\copy\tablebox\hss$$
    \vskip-\lastskip\vskip -2pt}

\def\tablenote#1 #2\par{\begingroup \parindent=0.8em
    \abovedisplayshortskip=0pt\belowdisplayshortskip=0pt
    \noindent
    $$\hss\vbox{\hsize\tablewidth \hangindent=\parindent \hangafter=1 \noindent
    \hbox to \parindent{$^#1$\hss}\strut#2\strut\par}\hss$$
    \endgroup}
\def\doubleline{\vskip 3pt\hrule \vskip 1.5pt \hrule \vskip 5pt}

\def\L2{\ifmmode L_2\else $L_2$\fi}

\def\DeltaT{\ifmmode \Delta T\else $\Delta T$\fi}
\def\deltat{\ifmmode \Delta t\else $\Delta t$\fi}
\def\fknee{\ifmmode f_{\rm knee}\else $f_{\rm knee}$\fi}
\def\Fmax{\ifmmode F_{\rm max}\else $F_{\rm max}$\fi}
\def\solar{\ifmmode{\rm M}_{\mathord\odot}\else${\rm M}_{\mathord\odot}$\fi}
\def\Msolar{\ifmmode{\rm M}_{\mathord\odot}\else${\rm M}_{\mathord\odot}$\fi}
\def\Lsolar{\ifmmode{\rm L}_{\mathord\odot}\else${\rm L}_{\mathord\odot}$\fi}
\def\inv{\ifmmode^{-1}\else$^{-1}$\fi}
\def\mo{\ifmmode^{-1}\else$^{-1}$\fi}
\def\sup#1{\ifmmode ^{\rm #1}\else $^{\rm #1}$\fi}
\def\expo#1{\ifmmode \times 10^{#1}\else $\times 10^{#1}$\fi}
\def\,{\thinspace}
\def\lsim{\mathrel{\raise .4ex\hbox{\rlap{$<$}\lower 1.2ex\hbox{$\sim$}}}}
\def\gsim{\mathrel{\raise .4ex\hbox{\rlap{$>$}\lower 1.2ex\hbox{$\sim$}}}}

\def\simprop{\mathrel{\raise .4ex\hbox{\rlap{$\propto$}\lower 1.2ex\hbox{$\sim$}}}}
\def\deg{\ifmmode^\circ\else$^\circ$\fi}
\def\pdeg{\ifmmode $\setbox0=\hbox{$^{\circ}$}\rlap{\hskip.11\wd0 .}$^{\circ}
          \else \setbox0=\hbox{$^{\circ}$}\rlap{\hskip.11\wd0 .}$^{\circ}$\fi}
\def\arcs{\ifmmode {^{\scriptstyle\prime\prime}}
          \else $^{\scriptstyle\prime\prime}$\fi}
\def\arcm{\ifmmode {^{\scriptstyle\prime}}
          \else $^{\scriptstyle\prime}$\fi}
\newdimen\sa  \newdimen\sb
\def\parcs{\sa=.07em \sb=.03em
     \ifmmode \hbox{\rlap{.}}^{\scriptstyle\prime\kern -\sb\prime}\hbox{\kern -\sa}
     \else \rlap{.}$^{\scriptstyle\prime\kern -\sb\prime}$\kern -\sa\fi}
\def\parcm{\sa=.08em \sb=.03em
     \ifmmode \hbox{\rlap{.}\kern\sa}^{\scriptstyle\prime}\hbox{\kern-\sb}
     \else \rlap{.}\kern\sa$^{\scriptstyle\prime}$\kern-\sb\fi}
\def\ra[#1 #2 #3.#4]{#1\sup{h}#2\sup{m}#3\sup{s}\llap.#4}
\def\dec[#1 #2 #3.#4]{#1\deg#2\arcm#3\arcs\llap.#4}
\def\deco[#1 #2 #3]{#1\deg#2\arcm#3\arcs}
\def\rra[#1 #2]{#1\sup{h}#2\sup{m}}

\def\dots{\relax\ifmmode \ldots\else $\ldots$\fi}
\def\WHzsr{\ifmmode $W\,Hz\mo\,sr\mo$\else W\,Hz\mo\,sr\mo\fi}
\def\mHz{\ifmmode $\,mHz$\else \,mHz\fi}
\def\GHz{\ifmmode $\,GHz$\else \,GHz\fi}
\def\mKs{\ifmmode $\,mK\,s$^{1/2}\else \,mK\,s$^{1/2}$\fi}
\def\muKs{\ifmmode \,\mu$K\,s$^{1/2}\else \,$\mu$K\,s$^{1/2}$\fi}
\def\muKRJs{\ifmmode \,\mu$K$_{\rm RJ}$\,s$^{1/2}\else \,$\mu$K$_{\rm RJ}$\,s$^{1/2}$\fi}
\def\muKHz{\ifmmode \,\mu$K\,Hz$^{-1/2}\else \,$\mu$K\,Hz$^{-1/2}$\fi}
\def\MJysr{\ifmmode \,$MJy\,sr\mo$\else \,MJy\,sr\mo\fi}
\def\MJysrmK{\ifmmode \,$MJy\,sr\mo$\,mK$_{\rm CMB}\mo\else \,MJy\,sr\mo\,mK$_{\rm CMB}\mo$\fi}
\def\microns{\ifmmode \,\mu$m$\else \,$\mu$m\fi}

\def\muK{\ifmmode \,\mu$K$\else \,$\mu$\hbox{K}\fi}
\def\microK{\ifmmode \,\mu$K$\else \,$\mu$\hbox{K}\fi}
\def\muW{\ifmmode \,\mu$W$\else \,$\mu$\hbox{W}\fi}
\def\kms{\ifmmode $\,km\,s$^{-1}\else \,km\,s$^{-1}$\fi}
\def\kmsMpc{\ifmmode $\,\kms\,Mpc\mo$\else \,\kms\,Mpc\mo\fi}
\providecommand{\sorthelp}[1]{}

\def\reff@jnl#1{{\rm#1\/}}
\def\apj{\reff@jnl{ApJ}}       % Astrophysical Journal
\def\apjs{\reff@jnl{ApJS}}     % Astrophysical Journal, Supplement
\def\aaps{\reff@jnl{A\&AS}}    % Astronomy and Astrophysics, Supplement
\def\mnras{\reff@jnl{MNRAS}}   % Monthly Notices of the RAS
\def\prd{\reff@jnl{Phys.\ Rev.\ D}}    % Physical Review D

\newcommand{\beq}{\begin{equation}}
\newcommand{\eeq}{\end{equation}}

\newcommand{\be}{\begin{equation}}
\newcommand{\ee}{\end{equation}}
\newcommand{\bea}{\begin{eqnarray}}
\newcommand{\eea}{\end{eqnarray}}

\def\L{\mathcal{L}}

{\begin{enumerate}\setlength{\itemsep}{0mm}}%
{\end{enumerate}}
{\begin{enumerate}\setlength{\itemsep}{0mm}}%
{\end{enumerate}}

\newcommand{\bc}{\begin{center}}
\newcommand{\ec}{\end{center}}
\newcommand{\bi}{\begin{itemize}}
\newcommand{\ei}{\end{itemize}}
\newcommand{\ben}{\begin{enumerate}}
\newcommand{\een}{\end{enumerate}}

\newfont{\gwpfont}{cmssq8 scaled 1000}

\def\gtorder{\mathrel{\raise.3ex\hbox{$>$}\mkern-14mu
             \lower0.6ex\hbox{$\sim$}}}
\def\ltorder{\mathrel{\raise.3ex\hbox{$<$}\mkern-14mu
             \lower0.6ex\hbox{$\sim$}}}

\def\ba{\begin{eqnarray}}
\def\ea{\end{eqnarray}}

\newcommand{\comm}[1]{}
\renewcommand{\comm}[1]{ \textcolor{magenta}{#1}}

\newcommand{\dthree}[1]{d^3 \! #1}

\begin{document}
%\alltwentyfifteenresultspapers
%\alltwentythirteenresultspapers
%
%\linenumbers

   \title{{\Planck} 2015 results. XIX. Constraints on primordial magnetic fields}

%This author list corresponds to \title{Author list for A22\_Primordial\_magnetic\_fields}
%Prepared by M. Lopez-Caniego (Marcos.Lopez.Caniego@sciops.esa.int), ESAC/ESA
%This version is from Tue Feb  9 10:24:47 2016 CET
%\subtitle{There are 232 co-authors in this list}
\author{\small
Planck Collaboration: P.~A.~R.~Ade\inst{96}
\and
N.~Aghanim\inst{63}
\and
M.~Arnaud\inst{79}
\and
F.~Arroja\inst{71, 85}
\and
M.~Ashdown\inst{75, 6}
\and
J.~Aumont\inst{63}
\and
C.~Baccigalupi\inst{94}
\and
M.~Ballardini\inst{51, 53, 34}
\and
A.~J.~Banday\inst{107, 10}
\and
R.~B.~Barreiro\inst{70}
% ...
}
\institute{\small
APC, AstroParticule et Cosmologie, Universit\'{e} Paris Diderot, CNRS/IN2P3, CEA/lrfu, Observatoire de Paris, Sorbonne Paris Cit\'{e}, 10, rue Alice Domon et L\'{e}onie Duquet, 75205 Paris Cedex 13, France\goodbreak
\and
Aalto University Mets\"{a}hovi Radio Observatory and Dept of Radio Science and Engineering, P.O. Box 13000, FI-00076 AALTO, Finland\goodbreak
\and
African Institute for Mathematical Sciences, 6-8 Melrose Road, Muizenberg, Cape Town, South Africa\goodbreak
\and
Agenzia Spaziale Italiana Science Data Center, Via del Politecnico snc, 00133, Roma, Italy\goodbreak
\and
Aix Marseille Universit\'{e}, CNRS, LAM (Laboratoire d'Astrophysique de Marseille) UMR 7326, 13388, Marseille, France\goodbreak
\and
Astrophysics Group, Cavendish Laboratory, University of Cambridge, J J Thomson Avenue, Cambridge CB3 0HE, U.K.\goodbreak
\and
Astrophysics \& Cosmology Research Unit, School of Mathematics, Statistics \& Computer Science, University of KwaZulu-Natal, Westville Campus, Private Bag X54001, Durban 4000, South Africa\goodbreak
\and
CGEE, SCS Qd 9, Lote C, Torre C, 4$^{\circ}$ andar, Ed. Parque Cidade Corporate, CEP 70308-200, Bras\'{i}lia, DF, Brazil\goodbreak
\and
CITA, University of Toronto, 60 St. George St., Toronto, ON M5S 3H8, Canada\goodbreak
\and
CNRS, IRAP, 9 Av. colonel Roche, BP 44346, F-31028 Toulouse cedex 4, France\goodbreak
\and
CRANN, Trinity College, Dublin, Ireland\goodbreak
\and
California Institute of Technology, Pasadena, California, U.S.A.\goodbreak
\and
Centre for Theoretical Cosmology, DAMTP, University of Cambridge, Wilberforce Road, Cambridge CB3 0WA, U.K.\goodbreak
\and
Centro de Estudios de F\'{i}sica del Cosmos de Arag\'{o}n (CEFCA), Plaza San Juan, 1, planta 2, E-44001, Teruel, Spain\goodbreak
\and
Computational Cosmology Center, Lawrence Berkeley National Laboratory, Berkeley, California, U.S.A.\goodbreak
\and
Consejo Superior de Investigaciones Cient\'{\i}ficas (CSIC), Madrid, Spain\goodbreak
\and
DSM/Irfu/SPP, CEA-Saclay, F-91191 Gif-sur-Yvette Cedex, France\goodbreak
\and
DTU Space, National Space Institute, Technical University of Denmark, Elektrovej 327, DK-2800 Kgs. Lyngby, Denmark\goodbreak
\and
D\'{e}partement de Physique Th\'{e}orique, Universit\'{e} de Gen\`{e}ve, 24, Quai E. Ansermet,1211 Gen\`{e}ve 4, Switzerland\goodbreak
\and
Departamento de Astrof\'{i}sica, Universidad de La Laguna (ULL), E-38206 La Laguna, Tenerife, Spain\goodbreak
\and
Departamento de F\'{\i}sica, Universidad de Oviedo, Avda. Calvo Sotelo s/n, Oviedo, Spain\goodbreak
\and
Department of Astronomy and Astrophysics, University of Toronto, 50 Saint George Street, Toronto, Ontario, Canada\goodbreak
\and
Department of Astrophysics/IMAPP, Radboud University Nijmegen, P.O. Box 9010, 6500 GL Nijmegen, The Netherlands\goodbreak
\and
Department of Physics \& Astronomy, University of British Columbia, 6224 Agricultural Road, Vancouver, British Columbia, Canada\goodbreak
\and
Department of Physics and Astronomy, Dana and David Dornsife College of Letter, Arts and Sciences, University of Southern California, Los Angeles, CA 90089, U.S.A.\goodbreak
\and
Department of Physics and Astronomy, Johns Hopkins University, Bloomberg Center 435, 3400 N. Charles St., Baltimore, MD 21218, U.S.A.\goodbreak
\and
Department of Physics and Astronomy, University College London, London WC1E 6BT, U.K.\goodbreak
\and
Department of Physics, Florida State University, Keen Physics Building, 77 Chieftan Way, Tallahassee, Florida, U.S.A.\goodbreak
\and
Department of Physics, Gustaf H\"{a}llstr\"{o}min katu 2a, University of Helsinki, Helsinki, Finland\goodbreak
\and
Department of Physics, Princeton University, Princeton, New Jersey, U.S.A.\goodbreak
\and
Department of Physics, University of California, Santa Barbara, California, U.S.A.\goodbreak
\and
Department of Physics, University of Illinois at Urbana-Champaign, 1110 West Green Street, Urbana, Illinois, U.S.A.\goodbreak
\and
Dipartimento di Fisica e Astronomia G. Galilei, Universit\`{a} degli Studi di Padova, via Marzolo 8, 35131 Padova, Italy\goodbreak
\and
Dipartimento di Fisica e Astronomia, ALMA MATER STUDIORUM, Universit\`{a} degli Studi di Bologna, Viale Berti Pichat 6/2, I-40127, Bologna, Italy\goodbreak
\and
Dipartimento di Fisica e Scienze della Terra, Universit\`{a} di Ferrara, Via Saragat 1, 44122 Ferrara, Italy\goodbreak
\and
Dipartimento di Fisica, Universit\`{a} La Sapienza, P. le A. Moro 2, Roma, Italy\goodbreak
\and
Dipartimento di Fisica, Universit\`{a} degli Studi di Milano, Via Celoria, 16, Milano, Italy\goodbreak
\and
Dipartimento di Fisica, Universit\`{a} degli Studi di Trieste, via A. Valerio 2, Trieste, Italy\goodbreak
\and
Dipartimento di Matematica, Universit\`{a} di Roma Tor Vergata, Via della Ricerca Scientifica, 1, Roma, Italy\goodbreak
\and
Discovery Center, Niels Bohr Institute, Blegdamsvej 17, Copenhagen, Denmark\goodbreak
\and
Discovery Center, Niels Bohr Institute, Copenhagen University, Blegdamsvej 17, Copenhagen, Denmark\goodbreak
\and
European Space Agency, ESAC, Planck Science Office, Camino bajo del Castillo, s/n, Urbanizaci\'{o}n Villafranca del Castillo, Villanueva de la Ca\~{n}ada, Madrid, Spain\goodbreak
\and
European Space Agency, ESTEC, Keplerlaan 1, 2201 AZ Noordwijk, The Netherlands\goodbreak
\and
Gran Sasso Science Institute, INFN, viale F. Crispi 7, 67100 L'Aquila, Italy\goodbreak
\and
HGSFP and University of Heidelberg, Theoretical Physics Department, Philosophenweg 16, 69120, Heidelberg, Germany\goodbreak
\and
Helsinki Institute of Physics, Gustaf H\"{a}llstr\"{o}min katu 2, University of Helsinki, Helsinki, Finland\goodbreak
\and
INAF - Osservatorio Astrofisico di Catania, Via S. Sofia 78, Catania, Italy\goodbreak
\and
INAF - Osservatorio Astronomico di Padova, Vicolo dell'Osservatorio 5, Padova, Italy\goodbreak
\and
INAF - Osservatorio Astronomico di Roma, via di Frascati 33, Monte Porzio Catone, Italy\goodbreak
\and
INAF - Osservatorio Astronomico di Trieste, Via G.B. Tiepolo 11, Trieste, Italy\goodbreak
\and
INAF/IASF Bologna, Via Gobetti 101, Bologna, Italy\goodbreak
\and
INAF/IASF Milano, Via E. Bassini 15, Milano, Italy\goodbreak
\and
INFN, Sezione di Bologna, Via Irnerio 46, I-40126, Bologna, Italy\goodbreak
\and
INFN, Sezione di Roma 1, Universit\`{a} di Roma Sapienza, Piazzale Aldo Moro 2, 00185, Roma, Italy\goodbreak
\and
INFN, Sezione di Roma 2, Universit\`{a} di Roma Tor Vergata, Via della Ricerca Scientifica, 1, Roma, Italy\goodbreak
\and
INFN/National Institute for Nuclear Physics, Via Valerio 2, I-34127 Trieste, Italy\goodbreak
\and
IPAG: Institut de Plan\'{e}tologie et d'Astrophysique de Grenoble, Universit\'{e} Grenoble Alpes, IPAG, F-38000 Grenoble, France, CNRS, IPAG, F-38000 Grenoble, France\goodbreak
\and
IUCAA, Post Bag 4, Ganeshkhind, Pune University Campus, Pune 411 007, India\goodbreak
\and
Imperial College London, Astrophysics group, Blackett Laboratory, Prince Consort Road, London, SW7 2AZ, U.K.\goodbreak
\and
Infrared Processing and Analysis Center, California Institute of Technology, Pasadena, CA 91125, U.S.A.\goodbreak
\and
Institut N\'{e}el, CNRS, Universit\'{e} Joseph Fourier Grenoble I, 25 rue des Martyrs, Grenoble, France\goodbreak
\and
Institut Universitaire de France, 103, bd Saint-Michel, 75005, Paris, France\goodbreak
\and
Institut d'Astrophysique Spatiale, CNRS, Univ. Paris-Sud, Universit\'{e} Paris-Saclay, B\^{a}t. 121, 91405 Orsay cedex, France\goodbreak
\and
Institut d'Astrophysique de Paris, CNRS (UMR7095), 98 bis Boulevard Arago, F-75014, Paris, France\goodbreak
\and
Institut f\"ur Theoretische Teilchenphysik und Kosmologie, RWTH Aachen University, D-52056 Aachen, Germany\goodbreak
\and
Institute for Space Sciences, Bucharest-Magurale, Romania\goodbreak
\and
Institute of Astronomy, University of Cambridge, Madingley Road, Cambridge CB3 0HA, U.K.\goodbreak
\and
Institute of Theoretical Astrophysics, University of Oslo, Blindern, Oslo, Norway\goodbreak
\and
Instituto de Astrof\'{\i}sica de Canarias, C/V\'{\i}a L\'{a}ctea s/n, La Laguna, Tenerife, Spain\goodbreak
\and
Instituto de F\'{\i}sica de Cantabria (CSIC-Universidad de Cantabria), Avda. de los Castros s/n, Santander, Spain\goodbreak
\and
Istituto Nazionale di Fisica Nucleare, Sezione di Padova, via Marzolo 8, I-35131 Padova, Italy\goodbreak
\and
Jet Propulsion Laboratory, California Institute of Technology, 4800 Oak Grove Drive, Pasadena, California, U.S.A.\goodbreak
\and
Jodrell Bank Centre for Astrophysics, Alan Turing Building, School of Physics and Astronomy, The University of Manchester, Oxford Road, Manchester, M13 9PL, U.K.\goodbreak
\and
Kavli Institute for Cosmological Physics, University of Chicago, Chicago, IL 60637, USA\goodbreak
\and
Kavli Institute for Cosmology Cambridge, Madingley Road, Cambridge, CB3 0HA, U.K.\goodbreak
\and
Kazan Federal University, 18 Kremlyovskaya St., Kazan, 420008, Russia\goodbreak
\and
LAL, Universit\'{e} Paris-Sud, CNRS/IN2P3, Orsay, France\goodbreak
\and
LERMA, CNRS, Observatoire de Paris, 61 Avenue de l'Observatoire, Paris, France\goodbreak
\and
Laboratoire AIM, IRFU/Service d'Astrophysique - CEA/DSM - CNRS - Universit\'{e} Paris Diderot, B\^{a}t. 709, CEA-Saclay, F-91191 Gif-sur-Yvette Cedex, France\goodbreak
\and
Laboratoire Traitement et Communication de l'Information, CNRS (UMR 5141) and T\'{e}l\'{e}com ParisTech, 46 rue Barrault F-75634 Paris Cedex 13, France\goodbreak
\and
Laboratoire de Physique Subatomique et Cosmologie, Universit\'{e} Grenoble-Alpes, CNRS/IN2P3, 53, rue des Martyrs, 38026 Grenoble Cedex, France\goodbreak
\and
Laboratoire de Physique Th\'{e}orique, Universit\'{e} Paris-Sud 11 \& CNRS, B\^{a}timent 210, 91405 Orsay, France\goodbreak
\and
Lawrence Berkeley National Laboratory, Berkeley, California, U.S.A.\goodbreak
\and
Lebedev Physical Institute of the Russian Academy of Sciences, Astro Space Centre, 84/32 Profsoyuznaya st., Moscow, GSP-7, 117997, Russia\goodbreak
\and
Leung Center for Cosmology and Particle Astrophysics, National Taiwan University, Taipei 10617, Taiwan\goodbreak
\and
Max-Planck-Institut f\"{u}r Astrophysik, Karl-Schwarzschild-Str. 1, 85741 Garching, Germany\goodbreak
\and
McGill Physics, Ernest Rutherford Physics Building, McGill University, 3600 rue University, Montr\'{e}al, QC, H3A 2T8, Canada\goodbreak
\and
National University of Ireland, Department of Experimental Physics, Maynooth, Co. Kildare, Ireland\goodbreak
\and
Nicolaus Copernicus Astronomical Center, Bartycka 18, 00-716 Warsaw, Poland\goodbreak
\and
Niels Bohr Institute, Blegdamsvej 17, Copenhagen, Denmark\goodbreak
\and
Niels Bohr Institute, Copenhagen University, Blegdamsvej 17, Copenhagen, Denmark\goodbreak
\and
Nordita (Nordic Institute for Theoretical Physics), Roslagstullsbacken 23, SE-106 91 Stockholm, Sweden\goodbreak
\and
Optical Science Laboratory, University College London, Gower Street, London, U.K.\goodbreak
\and
SISSA, Astrophysics Sector, via Bonomea 265, 34136, Trieste, Italy\goodbreak
\and
SMARTEST Research Centre, Universit\`{a} degli Studi e-Campus, Via Isimbardi 10, Novedrate (CO), 22060, Italy\goodbreak
\and
School of Physics and Astronomy, Cardiff University, Queens Buildings, The Parade, Cardiff, CF24 3AA, U.K.\goodbreak
\and
School of Physics and Astronomy, University of Nottingham, Nottingham NG7 2RD, U.K.\goodbreak
\and
Sorbonne Universit\'{e}-UPMC, UMR7095, Institut d'Astrophysique de Paris, 98 bis Boulevard Arago, F-75014, Paris, France\goodbreak
\and
Space Research Institute (IKI), Russian Academy of Sciences, Profsoyuznaya Str, 84/32, Moscow, 117997, Russia\goodbreak
\and
Space Sciences Laboratory, University of California, Berkeley, California, U.S.A.\goodbreak
\and
Special Astrophysical Observatory, Russian Academy of Sciences, Nizhnij Arkhyz, Zelenchukskiy region, Karachai-Cherkessian Republic, 369167, Russia\goodbreak
\and
Stanford University, Dept of Physics, Varian Physics Bldg, 382 Via Pueblo Mall, Stanford, California, U.S.A.\goodbreak
\and
Sub-Department of Astrophysics, University of Oxford, Keble Road, Oxford OX1 3RH, U.K.\goodbreak
\and
The Oskar Klein Centre for Cosmoparticle Physics, Department of Physics,Stockholm University, AlbaNova, SE-106 91 Stockholm, Sweden\goodbreak
\and
Theory Division, PH-TH, CERN, CH-1211, Geneva 23, Switzerland\goodbreak
\and
UPMC Univ Paris 06, UMR7095, 98 bis Boulevard Arago, F-75014, Paris, France\goodbreak
\and
Universit\'{e} de Toulouse, UPS-OMP, IRAP, F-31028 Toulouse cedex 4, France\goodbreak
% ...
}

%   \date{Received January 5, 2010}

 \abstract { 
We compute and investigate four types of imprint of a stochastic background of primordial magnetic fields (PMFs) on the cosmic microwave background (CMB) anisotropies: 
the impact of PMFs on the CMB temperature and polarization spectra, related to their contribution to cosmological perturbations; the effect on CMB polarization induced by Faraday rotation; the impact of PMFs on the ionization history; magnetically-induced non-Gaussianities and related non-zero bispectra; and the magnetically-induced breaking of statistical isotropy.
We present constraints on the amplitude of PMFs derived from different \Planck\ data products, depending on the specific effect that is analysed. Overall, \Planck\ data constrain the amplitude of PMFs to less than a few nanogauss, with different bounds depending on the considered model. In particular, individual limits coming from the analysis of the CMB angular power spectra, using the {\Planck} likelihood, are $B_{1\,\mathrm{Mpc}}< 4.4$ nG (where $B_{1\,\mathrm{Mpc}}$ is the comoving field amplitude at a scale of 1\,Mpc) at 95\,\% confidence level, assuming zero helicity. By considering the \Planck\ likelihood, based only on parity-even angular power spectra, we obtain $B_{1\,\mathrm{Mpc}}< 5.6$ nG for a maximally helical field.
For nearly scale-invariant PMFs we obtain $B_{1\,\mathrm{Mpc}}<2.0$\,nG and $B_{1\,\mathrm{Mpc}}<0.9$\,nG if the impact of PMFs on the ionization history of the Universe is included in the analysis.
From the analysis of magnetically-induced non-Gaussianity
we obtain three different values, corresponding to three applied methods, all below 5\,nG. The constraint from the magnetically-induced 
passive-tensor bispectrum is $B_{1\,\mathrm{Mpc}}< 2.8$\,nG. A search for preferred directions in the magnetically-induced passive bispectrum yields $B_{1\,\mathrm{Mpc}}< 4.5$\,nG, whereas the 
the compensated-scalar bispectrum gives $B_{1\,\mathrm{Mpc}}< 3$\,nG.
The analysis of the Faraday rotation of CMB polarization by PMFs uses the \Planck\ power spectra in $EE$ and $BB$ at 70\,GHz and gives $B_{1\,\mathrm{Mpc}}< 1380$\,nG.
In our final analysis, we consider the harmonic-space correlations produced by Alfv\'en waves, finding no significant evidence for the presence of these waves.
Together, these results comprise a comprehensive set of constraints on possible PMFs with {\Planck} data.
}


\authorrunning{Planck Collaboration}
\titlerunning{Constraints on primordial magnetic fields}

   \maketitle
%\allearlypapers

\section{Introduction}
\label{sec:introduction}

\subsection{Cosmic magnetism}
Magnetic fields are one of the fundamental and ubiquitous components of our Universe. They are a common feature of many astrophysical objects, starting from the smallest up to the largest observed scales (for reviews see \citealt{2012SSRv..166....1R} and \citealt{2012SSRv..166...37W}).
In particular, large-scale magnetic fields are observed in almost every galaxy, starting from the Milky Way, 
with possible hints of their presence also in high-redshift galaxies \citep{2000IAUJD..14E...4B,2008Natur.454..302B,2008Natur.455..638W}, suggesting an early origin for the galactic fields.

Large-scale magnetic fields are also probed in galaxy clusters, both through the measurement of the Faraday rotation effect on the light of background galaxies and through radio emission
from the halos and relics of the clusters \citep{2004IJMPD..13.1549G,2008SSRv..134...93F,2012A&ARv..20...54F}. 
These large-scale magnetic fields have measured amplitudes that range from a few to several microgauss. 
Recent Faraday rotation measurements from low-density intercluster regions suggest the presence of large-scale magnetic fields also in cosmic structure filaments \citep{2013arXiv1305.1450N}.
A recent addition to the large-scale magnetic field constraints comes from the interpretation of \textit{Fermi}-LAT data \citep{2010Sci...328...73N}. High-energy $\gamma$-rays (in the TeV band) emitted by blazars generate 
electron-positron pairs when interacting with the optical and IR background light. These pairs re-emit at lower energies (GeV), but \textit{Fermi}-LAT data do not show this re-emitted 
flux in the GeV band. 
One possible explanation for this observation is a deflection of the electron-positron pairs due to the presence of a diffuse magnetic field  
(see \citealt{2010Sci...328...73N}, \citealt{2011A&A...529A.144T}, \citealt{2011MNRAS.414.3566T}, \citealt{2012ApJ...747L..14V}, and \citealt{2013A&A...554A..31N} for details on this explanation and see \citealt{2012ApJ...752...22B} for alternative scenarios). 
Constraints on the GeV emission provide lower limits on the amplitude
of intergalactic fields of the order of $10^{-18}$--$10^{-15}$\,G \citep{2010MNRAS.406L..70T,2011A&A...529A.144T,2011ApJ...733L..21D,2012ApJ...747L..14V}, if this scenario is correct. 

The origin of large-scale magnetic fields is strongly debated. Several mechanisms have been proposed and one popular hypothesis is that the observed large-scale fields are remnants 
of fields that existed from the earliest times, i.e., primordial fields.
During structure formation the adiabatic compression and turbulent shock flows would naturally lead to an amplification of initial seeds 
(which may act in addition to astrophysical mechanisms of large-scale magnetic field generation, like AGN ejection and galactic dynamos; for reviews see \citealt{2002RvMP...74..775W} and \citealt{2004IJMPD..13..391G}). A Kolmogorov-like magnetic power spectrum has been observed in the
central region of the Hydra cluster \citep{2011A&A...529A..13K}, supporting the
idea that the observed extragalactic magnetic fields are largely shaped
and amplified by hydrodynamical processes. Primordial magnetic fields (PMFs) can naturally provide the initial seeds to be amplified into the observed large-scale fields. 
Several early-Universe scenarios predict the generation of cosmological magnetic fields, either during inflation \citep{1992ApJ...391L...1R}, with a suitable breaking of conformal 
invariance of electromagnetism \citep{1988PhRvD..37.2743T}, during phase transitions motivated by particle physics \citep{1991PhLB..265..258V, 1998PhLB..418..258G}, or 
via other physical processes \citep{2003JCAP...11..010D,2006Sci...311..827I}. 
The importance of PMF studies lies not only in the possibility of PMFs being the progenitors of the observed cosmic magnetic fields, but also 
in them providing a new potential observational window to 
the early Universe (for reviews see \citealt{2005NewAR..49...79K}, \citealt{2008LNP...737..863G}, \citealt{2013PPCF...55l4026K}, and \citealt{2013A&ARv..21...62D}).

\subsection{Imprints of primordial magnetism}
PMFs leave imprints on several cosmological observables and can be constrained with different cosmological data sets. In particular, interesting constraints come from their 
influence on Big Bang nucleosynthesis, which provides upper limits of the order of 0.1\,$\mu$G \citep{1995APh.....3...95G,2010PhRvD..82h3005K}, and from their impact on large-scale structure formation (see, e.g., \citealt{2012PhRvD..86d3510S} and \citealt{2012JCAP...11..055F}). 

Another limit on PMFs based on Big Bang Nucleosynthesis (BBN) is derived by \cite{2002PhRvD..65b3517C}. Gravitational waves can be produced by PMFs, in particular before neutrino free streaming.
The upper limit on the amount of gravitational waves allowed at nucleosynthesis to not spoil the BBN predictions poses a constraint on the amplitude of PMFs, which is especially strong for causal magnetogenesis mechanisms. 

It has been suggested that PMFs could have an influence on the formation of the filamentary large-scale structure,
in the pioneering work of \cite{1978ApJ224337W} and by \cite{1996ApJ46828K}. In particular, \cite{1997A&A32613B} conclude that magnetic fields with comoving strengths lower than 1\,nG have negligible effect and that for strengths higher
than 10\,nG the influence is too high to be compatible with observations. 

Other constraints on PMFs come from their impact on the thermal spectrum of the CMB radiation.
PMFs can induce spectral distortions of both early and late type via injection of dissipated magnetic energy
into the plasma through damping processes \citep{2000PhRvL..85..700J, Sethi2005, 2014JCAP...01..009K}.
PMFs imply also photon emission and absorption through the cyclotron process, which, depending on the amplitude of PMFs,
could in principle play a role in the generation and evolution of spectral distortions and in particular
in the thermalization process (for a general overview see \citealt{1991AA...246...49B}, \citealt{1993PhRvD..48..485H}, and \citealt{2012MNRAS.419.1294C}). On the other hand, the cyclotron process is relevant only at very long wavelengths. Thus,
for realistic shapes of distorted spectra and PMFs with amplitudes compatible with current constraints,
the cyclotron contribution is found to be much less important than radiative Compton and bremsstrahlung contributions
in re-establishing a blackbody spectrum \citep{2006AN....327..424B} and also extremely small in generating
a polarized signal \citep{2005NewA...11....1Z}. Thus, the current constraints derived from COBE-FIRAS present good
prospects for possible future observations \citep{2000PhRvL..85..700J,2014JCAP...01..009K,Chluba2014PMF}
while polarization anisotropies directly induced by PMFs are not significantly affected by the cyclotron process associated with PMFs.

Many of the stronger and more robust constraints come from the impact of PMFs on the CMB anisotropies. 
CMB data are a crucial source of information for investigating 
and constraining PMF characteristics, understanding their origin, and exploring the possibility of them being the seeds that generated the observed large-scale magnetic fields.

The first release of {\Planck}\footnote{\Planck\ (\url{http://www.esa.int/Planck}) is a project of the European Space Agency  (ESA) with instruments provided by two scientific consortia funded by ESA member states and led by Principal Investigators from France and Italy, telescope reflectors provided through a collaboration between ESA and a scientific consortium led and funded by Denmark, and additional contributions from NASA (USA).} data in 2013 has led to some of most stringent constraints on PMFs \citep[see][]{planck2013-p11}. The scope of this paper is to provide the ``{\Planck} constraints on PMFs'' through combined analyses of the temperature and polarization data.
The results of this paper are derived from \Planck\ products, which are based on the work done in \cite{planck2014-a01}, \cite{planck2014-a03}, \cite{planck2014-a04}, \cite{planck2014-a05}, \cite{planck2014-a06}, \cite{planck2014-a07}, \cite{planck2014-a08}, and \cite{planck2014-a09}.
Before going into the description of our analysis, we briefly discuss the most important PMF models. 

The simplest PMF model is that of a homogeneous field. This model cannot be included in a homogeneously and isotropically expanding cosmological model.
It needs to be analysed in the context of an anisotropic cosmological model with associated isotropy-breaking predictions \citep{2008PhRvD..78f3012K}, and has already been  
strongly constrained by COBE data \citep{1997PhRvL..78.3610B}. More recently, \cite{2011JCAP...06..017A} have reconsidered the impact of a homogeneous large-scale magnetic field on the CMB anisotropies, with the addition of the contribution from free-streaming particles like neutrinos. The presence of the anisotropic neutrino stress induces a compensation of the magnetic field effect, allowing for a larger homogeneous-field amplitude than concluded from previous analyses.

The most widely used model of PMFs is a stochastic background modelled as a fully inhomogeneous component of the cosmological plasma (we neglect PMF energy density and anisotropic stress contributions at the homogeneous level), with the energy momentum tensor components (quadratic in the fields) 
on the same footing as cosmological perturbations. Within this model, PMFs leave several signatures in the CMB temperature and polarization anisotropy patterns, inducing also non-Gaussianities.

In this paper, we predict and analyse four different types of PMF signatures: the impact on the CMB power spectra in temperature and polarization; 
the impact on polarization power spectra induced by Faraday rotation; the impact on CMB non-Gaussianities and the related non-zero magnetically-induced bispectra; and finally 
the impact on the statistics of CMB anisotropies and in particular the breaking of statistical isotropy, given by induced correlations in harmonic space.

The energy momentum tensor of PMFs sources all types of cosmological perturbations, i.e., scalar, vector, and tensor perturbations. Magnetically-induced pertubations have some crucial 
differences with respect to the primary perturbations. Firstly, PMFs generate vector perturbations that are sourced by the Lorentz force and, unlike the primary ones, are not decaying. Secondly, magnetically-induced perturbations are not suppressed by Silk damping \citep{1997ApJ...479..568H,1998PhRvD..58h3502S}. Their impact on the CMB temperature power spectrum is dominant on small 
angular scales, where the primary CMB is suppressed, and therefore they can be strongly constrained by high-resolution CMB data. 
Of additional interest are helical PMFs, which may be generated during inflation
through mechanisms like pseudoscalar coupling and may be crucial for connecting these PMFs to the magnetic fields observed on large scales.  
A helical PMF generates parity-violating correlations, such as $TB$ and $EB$ \citep{2004PhRvD..69f3006C,2005PhRvD..71j3006K,2014PhRvD..90h3004K}, which can be used to constrain PMFs with CMB polarization data.
After recombination, PMFs are dissipated via two additional effects that take place in the magnetized, not fully ionized, plasma: ambipolar diffusion and magnetohydrodynamical (MHD) turbulence.
The dissipation injects magnetic energy into the plasma, heating it and thereby modifying the optical depth of recombination, with an impact on the primary CMB power spectra
\citep{Sethi2005,2014JCAP...01..009K,2015arXiv150100142K,Chluba2014PMF}.

PMFs have another effect on the primary polarization anisotropies. They induce Faraday rotation, which rotates $E$-modes into $B$-modes, thus generating a new $B$-mode signal, and vice versa. This signal grows with decreasing observational frequency and therefore is 
a good target for {\Planck}'s low-frequency channels.

PMFs modelled as a stochastic background have a non-Gaussian contribution to CMB anisotropies, even if the magnetic fields themselves are Gaussian distributed, since the components of the energy momentum tensor are quadratic in the fields and therefore approximately follow $\chi^2$ statistics \citep{2005PhRvD..72f3002B}. In particular, PMFs
generate non-zero higher-order statistical moments. The third-order moment, the CMB bispectrum, can be used as a probe to derive constraints on PMFs that are complementary to the previously mentioned ones. 
{\Planck} polarization data thus provide a new way of probing PMFs, namely through the magnetically-induced polarization bispectrum.

The presence of PMFs induces and sustains the propagation of Alfv\'en waves. These waves have an impact on the statistics of the CMB anisotropies and 
in particular induce specific correlations between harmonic modes \citep{2008PhRvD..78f3012K}. It is possible to use this effect to constrain the amplitude of Alfv\'en waves and thus indirectly constrain the PMF amplitude \citep{DKY,kim09,planck2013-p09a}.

In the MHD limit, assuming that the fields are only modified by cosmic expansion, 
the magnetic field strength decreases as $a^{-2}$, where $a$ is the cosmological 
scale factor. Throughout, we will use a ``comoving'' magnetic field, defined as 
$B = a^2 \, B^{\mathrm{(phys)}}$, where $B^{\mathrm{(phys)}}$ is the physical strength of the magnetic field.

\subsection{Structure of the paper}
The paper is structured as follows.
In Sect.~\ref{sec:APS} we describe the analysis of the impact of helical and non-helical PMFs on CMB power spectra in temperature and polarization 
and we derive the constraints on the PMF amplitude and spectral index coming from {\Planck} data.
In Sect.~\ref{sec:non-gaussianities} we present three different analyses of the magnetically-induced bispectrum, specifically two analyses of the magnetically-induced passive bispectrum and an analytical 
treatment of the magnetically-induced scalar bispectrum. In all cases we derive the constraints on the amplitude of PMFs with a scale-invariant spectrum using {\Planck} non-Gaussianity measurements.
In Sect.~\ref{sec:Faraday} we present our analysis of the Faraday rotation signal induced by PMFs and we derive constraints from {\Planck} low frequency polarization data.
In Sect.~\ref{alfven} we present the analysis of the impact of Alv\'en waves on statistical correlations in harmonic space and the associated constraints on Alfv\'en waves derived from {\Planck} data. We summarize our conclusions in Sect.~\ref{sec:conclusions}

\section{Impact of primordial magnetic fields on the CMB power spectra}
\label{sec:APS}

PMFs affect cosmological perturbations and may leave significant imprints on the 
CMB power spectra in temperature and polarization.
Accurate prediction of these signatures allows us to derive constraints on PMF characteristics from CMB anisotropy data from {\Planck} using the {\Planck} likelihood.
In this section we derive the predictions for the magnetically-induced power spectra in temperature and polarization, 
considering helical and non-helical PMFs, and present the resulting constraints on PMFs.

\subsection{Magnetic modes}

When considering a stochastic background of PMFs, we can neglect the contribution of energy density and anisotropic stress at the homogeneous level. The magnetic energy momentum tensor can be seen as describing perturbations 
carrying energy density and anisotropic stress and inducing a Lorentz force on the charged particles of the plasma. 
PMFs source all types of perturbations; scalar, vector, and tensor. 
In the past years, several different analyses of magnetically-induced perturbations have been performed. Some examples from the wide literature of the field 
concern magnetically-induced scalar perturbations \citep{2004PhRvD..70l3507G,2007PhRvD..75b3002K,2007AIPC..957..449Y,2008PhRvD..77d3005Y,2008PhRvD..78b3510F,2008PhRvD..77f1301G,
2008PhRvD..77l3001G,2008PhRvD..77f3003G,2010JCAP...05..022B,2010arXiv1005.3332B,2011PhRvD..83b3006K}, while other treatments also include magnetically-induced vector and tensor perturbations \citep{1998PhRvL..81.3575S,2000PhRvD..61d3001D,2001AIPC..555..451K,2002PhRvD..65l3004M,2002PhRvD..65b3517C,2002MNRAS.335L..57S,2003MNRAS.344L..31S,2004PhRvD..70d3011L,2006AN....327..422C,2009MNRAS.396..523P,2010PhRvD..81d3517S}.
We can identify three different classes of intial conditions for magnetically-induced perturbations; compensated \citep{2004PhRvD..70l3507G,2008PhRvD..78b3510F}, passive
 \citep{2004PhRvD..70d3011L,2010PhRvD..81d3517S}, and inflationary \citep{2013PhRvD..88h3515B}.

In this paper we focus on the two magnetically-induced modes 
that are present for all types of PMFs produced prior to decoupling, independent of their generation mechanism, i.e., the compensated and passive modes. We do not consider specific inflationary initial conditions \citep{2013PhRvD..88h3515B} to maintain the generality of the PMFs we constrain. For the same reason we neither consider a possible cross-correlation between the magnetically-induced and the adiabatic mode  motivated by inflation \citep{2012PhRvD..86l3528J}.

\subsubsection{Compensated modes}
The compensated modes are the {\em regular} magnetically-induced modes. These are the regular (finite at $\tau\rightarrow 0$) solutions of the perturbed Einstein-Boltzmann equations, including the magnetic contributions after neutrino decoupling. 
These modes are called ``compensated'' because the magnetic contributions to the metric perturbations in
the initial conditions are compensated by fluid modes to leading order.
The initial conditions are the solutions of the Einstein-Boltzmann equation system for large wavelengths at early times, with the perturbed quantities expanded in power series of 
$k\tau$ (where $k$ is the perturbation wavenumber and $\tau$ is the conformal time). When performing this calculation for the magnetically-induced modes, the growing regular mode  
requires the source terms in the equations for the metric perturbations to vanish at the lowest order. This can only be realized by a compensation between the
magnetic terms and the perturbed quantities of the fluid.

\subsubsection{Passive modes}
The second class, the passive modes, is generated by the presence of a PMF \textit{before} neutrino decoupling. 
Without neutrinos free-streaming, there is no counterpart in the fluid to balance the anisotropic stress of the PMF.
This generates a logarithmically growing  mode (in conformal time, which diverges for early times). 
After neutrino decoupling, the anisotropic neutrino stress compensates the anisotropic stress due to the PMF, leading back
to the compensated case described before. But an imprint of this logarithmically growing mode survives neutrino decoupling in form of a constant offset on the amplitude of the inflationary non-magnetic mode 
(the primary cosmological pertubations of the standard model without PMFs).
This amplitude offset is due to the continuity condition for the matching of the initial conditions before and after neutrino decoupling. 
Passive modes have a logarithmic dependence on the ratio between the neutrino decoupling time and the generation time of the PMF, i.e., their amplitudes grow as $h(k)\propto \ln(\tau_\nu/\tau_B)$ (where $\tau_\nu$ is the neutrino decoupling time and $\tau_B$ is the PMF generation time). The passive modes, 
unlike the compensated ones, evolve following the standard non-magnetic equations and influence only scalar and tensor perturbations.

\subsection{Impact of non-helical PMFs on the CMB angular power spectra}

Our analysis is based on previous treatments of magnetically-induced compensated and passive scalar, vector, and tensor modes presented by 
\cite{2004PhRvD..70d3011L}, \cite{2008PhRvD..78b3510F}, \cite{2009MNRAS.396..523P}, and \cite{2010PhRvD..81d3517S}.

At linear order, PMFs evolve like a stiff source and we can therefore discard the back-reaction of gravity 
onto the stochastic background of PMFs. Prior to the recombination epoch the electric conductivity of the 
primordial plasma is very large. We therefore consider the limit of
infinite conductivity, in which the induced electric field is zero. 
In this limit, the temporal evolution of the PMF reduces to ${\vec B}^{(\mathrm{phys})}{({\vec x},\tau)}={\vec B}({\vec x})/a(\tau)^2$, where  ${\vec B}({\vec x})$ is the comoving field.\footnote{We choose the standard convention in which the scale factor is $a(\tau_0) = 1$ at the present time $\tau_0$.}

We model a non-helical stochastic PMF with a power-law power spectrum, where the two-point correlation function is described by\footnote{For the Fourier transform and its inverse, we use
\ba 
Y ({\vec k}{,}\tau)& = & \int \dthree{x} \, \mathrm{e}^{i {\vec k} \cdot {\vec x}} Y ({\vec x}{,}\tau)\,, \nonumber \\
%\, ,\,\,\,\,\, 
Y ({\vec x}{,}\tau) &=& \int \frac{\dthree{k}}{(2 \pi)^3} \, \mathrm{e}^{-i {\vec k} \cdot {\vec x}} Y ({\vec k}{,}\tau) \,, \nonumber 
\label{Fourier}
\ea
where $Y$ is a generic function.}
\be
\Big\langle B_i({\vec k}) \, B_j^*({\vec k}')\Big\rangle=\frac{(2\pi)^3}{2} \, \delta^{(3)}({\vec k}-{\vec k}') \, \left(\delta_{ij}-\hat k_i\hat k_j\right) \, P_B(k)\,,
\label{PSpectrum}
\ee
where $P_B(k)=A_B \, k^{n_B}$ and $\hat{k}_i$ denotes a cartesian component of a normalized wave vector.
In this model, the PMF is characterized by two quantities, the amplitude of the power spectrum, $A_B$, and the spectral index $n_B$. The latter  
is one of the major discriminating factors between generation mechanisms, since different mechanisms generate fields with 
different spectral indices (for example, causal mechanisms generate fields with $n_B \geq 2$; \citealt{2003JCAP...11..010D}).
While magnetically-induced compensated perturbations do not suffer from Silk damping, PMFs are nevertheless suppressed on small scales by radiation viscosity \citep{1998PhRvD..57.3264J,2001AIPC..555..451K}. 
To account for this damping we introduce a sharp cut-off in the PMF power spectrum at the damping scale $k_\mathrm{D}$.

For the amplitude we use the convention to smooth over a comoving scale of $\lambda = 1$\,Mpc, 
\ba
B^2_\lambda &=& \int_0^{\infty} \frac{d{k \, k^2}}{2 \pi^2} \, \mathrm{e}^{-k^2 \lambda^2} P_B (k) = \frac{A_B}{4 \pi^2 \lambda^{n_B+3}} \, \Gamma \left( \frac{n_B+3}{2} \right)\,.
\label{gaussian}
\ea
For the damping scale we use \citep{1998PhRvD..58h3502S, 2002PhRvD..65l3004M}\footnote{Note that, unlike in previous analyses \citep{2008PhRvD..78b3510F,2009MNRAS.396..523P,2011PhRvD..83l3533P}, we explicitly include the dependence of the damping scale on the baryon density to account for the updated cosmology with respect to the previous treatments.}
\ba
k_\mathrm{D}&=&(5.5 \times 10^4)^{\frac{1}{n_B+5}} \left(\frac{B_{\lambda}}{\mathrm{nG}}\right)^{-\frac{2}{n_B+5}} 
\left(\frac{2\pi}{\lambda/{\mathrm{Mpc}}}\right)^{\frac{n_B+3}{n_B+5}}\times\nonumber\\
&& h^{\frac{1}{n_B+5}} \left(\frac{\Omega_{\mathrm b} h^2}{0.022}\right)^{\frac{1}{n_B+5}}\Big|_{\lambda=1\,{\mathrm Mpc}} \mathrm{Mpc}^{-1}\,,
\label{kd_def}
\ea
where $h$ is the reduced Hubble constant, $H_0 = 100\,h \, \mathrm{km}\,\mathrm{s^{-1}}\,\mathrm{Mpc^{-1}}$, and $\Omega_{\mathrm b}$ is the baryon density parameter.\footnote{
For a quasi-scale-invariant PMF power spectrum ($n_B\approx -3$), the estimated value for $k_{\mathrm D}$ varies slightly, by about 30--40\,\%, for different approaches 
\citep{1998PhRvD..58h3502S,2000PhRvL..85..700J,2015arXiv150100142K}. However, this does not affect the main results of this paper significantly.}
Magnetically-induced scalar, vector, and tensor perturbations are sourced by the energy momentum tensor components due to PMFs, together with the Lorentz force contribution.
The energy momentum tensor of the PMFs is
\ba
\kappa^0_0&=&-\rho_B=-\frac{B^2({\vec x})}{8\pi a^4(\tau)}\,,\\
\kappa_i^0&=&0\,,\\
\kappa_j^i&=&\frac{1}{4\pi a^4(\tau)}
\left(\frac{B^2({\vec x})}{2}\,\delta_j^i-B_j({\vec x})\,B^i({\vec x})\right)\,,
\ea
where the components are all quadratic in the magnetic field. The power spectra of the perturbations are therefore fourth-order in the magnetic field and given by convolutions of the magnetic power spectrum.
The two-point correlation function of the spatial part of the 
energy momentum tensor is\footnote{We use the convention that Latin indices run 
from 1 to 3, while Greek indices run from 0 to 3.}
\ba
\Big\langle\kappa^*_{ab}({\vec k}) \, \kappa_{cd}({\vec k'})\Big\rangle&=&
\int \frac{ \dthree{q} \, \dthree{p}}{64\pi^5} \, \delta_{ab} \, \delta_{cd} \,\nonumber\\ 
&&\Big\langle B_l({\vec q}) \, B_l({\vec k}-{\vec q}) \, B_m({\vec p}) \, B_m({\vec k}'-{\vec p})\Big\rangle\nonumber\\
&& - \int \frac{ \dthree{q} \, \dthree{p}}{32\pi^5}\nonumber\\
&& \Big\langle B_a({\vec q}) \, B_b({\vec k}-{\vec q}) \, B_c({\vec p}) \,
B_d({\vec k}'-{\vec p})\Big\rangle\,.\nonumber
\ea
We can then obtain scalar, vector, and tensor correlation functions,
\ba
&&\Big\langle\Pi^{*(\mathrm{S})}({\vec k}) \, \Pi^{(\mathrm{S})}({\vec k'})\Big\rangle 
= \delta_{ab} \, \delta_{cd} \, \Big\langle\kappa^*_{ab}({\vec k}) \, \kappa_{cd}({\vec k'})\Big\rangle\,,\nonumber\\
&&\Big\langle\Pi_{i}^{*(\mathrm{V})}({\vec k}) \, \Pi_j^{(\mathrm{V})}({\vec k'})\Big\rangle 
= k_a \, P_{ib}({\vec k}) \, k'_c \, P_{jd}({\vec k'}) \, \Big\langle\kappa^*_{ab}({\vec k}) \,
\kappa_{cd}({\vec k'})\Big\rangle \,,\nonumber\\
&&\Big\langle\Pi_{ij}^{*(\mathrm{T})}({\vec k}) \, \Pi_{tl}^{(\mathrm{T})}({\vec k'})\Big\rangle 
= \left[P_{ia}({\vec k}) \, P_{jb}({\vec k})-\frac{1}{2} P_{ij}({\vec k}) \, P_{ab}({\vec k})\right]
\times\nonumber\\
&&~~~~~\left[P_{tc}({\vec k'}) \, P_{ld}({\vec k'})-\frac{1}{2} P_{tl}({\vec k'}) \,
P_{cd}({\vec k'})\right] \, \Big\langle\kappa^*_{ab}({\vec k}) \, \kappa_{cd}({\vec k'})\Big\rangle \,,
\ea
where the $\Pi^{(\mathrm{X})}$ are the scalar, vector, and tensor components of the energy momentum tensor, $P_{ij}=\delta_{ij}-\hat k_i\hat k_j$, and we sum over repeated indices.
Such convolutions can be written in terms of spectra as
\ba
\Big\langle\Pi^{*(\mathrm{S})}({\vec k}) \, \Pi^{(\mathrm{S})}({\vec k'})\Big\rangle &=& 
\left|\Pi^{(\mathrm{S})}(k)\right|^2 \, \delta({\vec k}-{\vec k'})\,,\nonumber\\
\Big\langle\Pi_i^{*(\mathrm{V})}({\vec k}) \, \Pi_j^{(\mathrm{V})}({\vec k'})\Big\rangle &=& 
\frac{1}{2} \left|\Pi^{(\mathrm{V})}(k)\right|^2 \, P_{ij}({\vec k}) \, \delta({\vec k}-{\vec k'})\,,\nonumber\\
\Big\langle\Pi_{ij}^{*(\mathrm{T})}({\vec k}) \, \Pi_{tl}^{(\mathrm{T})}({\vec k'})\Big\rangle &=& 
\frac{1}{4}\left|\Pi^{(\mathrm{T})}(k)\right|^2 \, \mathcal{M}_{ijtl} ({\vec k}) \, 
\delta({\vec k}-{\vec k'})\,,\nonumber 
\ea
where $\mathcal{M}_{ijtl}=P_{it}P_{jl}+P_{il}P_{jt}-P_{ij}P_{tl}$.
With this convention, the relevant components of the energy momentum tensor become
\ba
\left|\rho_B(k)\right|^2&=&\frac{1}{1024\,\pi^5}\int_\Omega \dthree{p} \, P_B(p) \, P_B(|{\vec k}-{\vec p}|) \, (1+\mu^2)\,,
\label{density}\\
\left|L_{B}^{(\mathrm{S})} (k)\right|^2 &=&\frac{1}{128\,\pi^2\,a^8}\int_\Omega \dthree{p} \, P_B(p) \nonumber\\
&&\times P_B(|{\vec{k}}-{\vec{p}}|) \, \left[1 + \mu^2 + 4 \gamma \beta(\gamma \beta - \mu)\right] \,,
\label{spectrum_LF}\\
\left|\Pi^{(\mathrm{V})}(k)\right|^2&=&\frac{1}{512\,\pi^5}\int_\Omega \dthree{p} \, P_B(p)\, P_B(|{\vec k}-{\vec p}|)\nonumber\\
&&\times\left[(1+\beta^2)(1-\gamma^2) + \gamma\beta(\mu-\gamma\beta)\right] \,,
\label{vector}\\
\left|\Pi^{(\mathrm{T})}(k)\right|^2&=&\frac{1}{512\,\pi^5} \int_\Omega \dthree{p} \, P_B(p) \, P_B(|{\vec k}-{\vec p}|)\nonumber\\
 &&\times(1+2\gamma^2+\gamma^2\beta^2) \,,
\label{tensor}
\ea
where $\mu = \hat {\vec p} \cdot ({\vec k} -{\vec p})/|{\vec k} -{\vec p}|$,
$\gamma= \hat {\vec k} \cdot \hat {\vec p}$,
$\beta= \hat {\vec k} \cdot ({\vec k} -{\vec p})/|{\vec k} -{\vec p}|$, and $\Omega$ denotes the volume with $p<k_\mathrm{D}$.

The conservation equations for the fields give a relation between the scalar projection of the anisotropic stress, the energy density, and the Lorentz force, $\sigma_B=\frac{\rho_B}{3}+L_B$ \footnote{Note that we use the notation of \cite{1995ApJ...455....7M} for the scalar anisotropic stress $\sigma_B$.}
(which reduces to a simple relation between Lorentz force and anisotropic stress for vector modes, $\Pi_B^{(\mathrm{V})}=k^i L_i^{(\mathrm{V})}$). This relation simplifies
the treatment, reducing the number of correlators to be computed by a factor of 2.
We use the analytic solutions to the convolutions derived by \cite{2008PhRvD..78b3510F} and \cite{2009MNRAS.396..523P} for fixed spectral indices. For general $n_B$, we 
use the fits to the generic analytic solutions provided by \cite{2011PhRvD..83l3533P}. These fits simplify the computation due to the presence of hypergeometric functions in the analytic solutions. 
The infrared behaviour of the spectra, 
which is relevant for CMB anisotropies, depends on the spectral index. In particular, the spectra describe white noise for indices greater than $n_B= -\frac{3}{2}$,
whereas they are infrared-dominated, as $k^{2 n_B+3}$, for smaller indices.
We use the initial conditions derived by \cite{2004PhRvD..70d3011L}, \cite{2009MNRAS.396..523P}, \cite{2011PhRvD..83l3533P}, and \cite{2014Finelli} for scalar and compensated tensor modes, vector modes, and 
passive modes, respectively.
We use an extended version of the \texttt{CAMB} code \citep{2011ascl.soft02026L} that includes all magnetic contributions to calculate predictions for 
the CMB power spectra in temperature and polarization.

\begin{figure*}
\includegraphics[width=88mm]{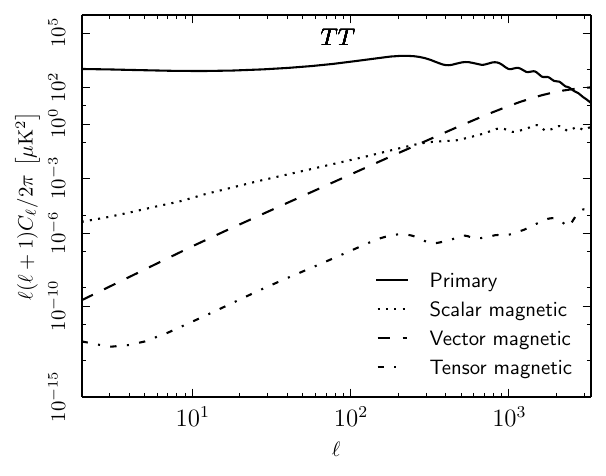}\includegraphics[width=88mm]{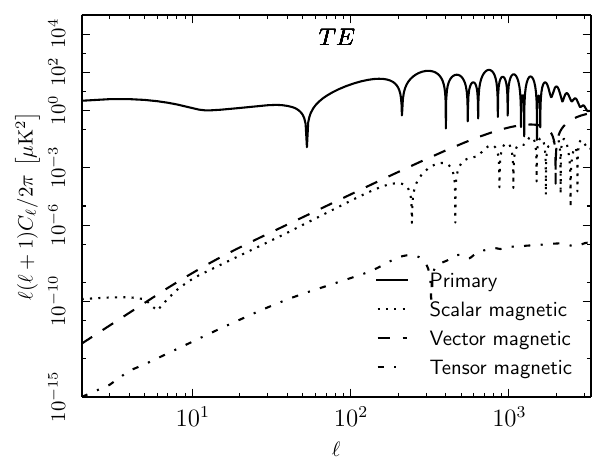}\\
\includegraphics[width=88mm]{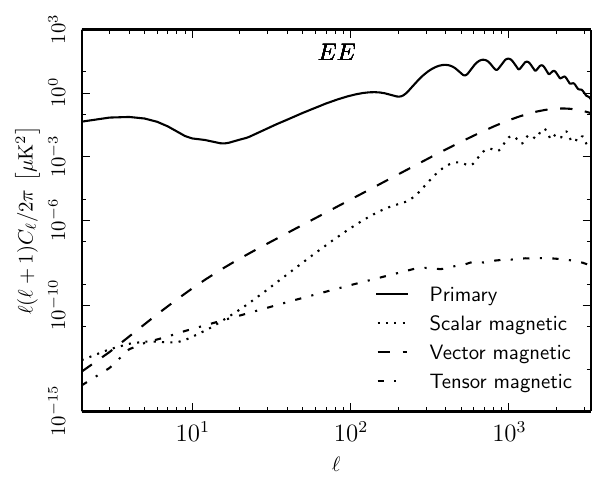}\includegraphics[width=88mm]{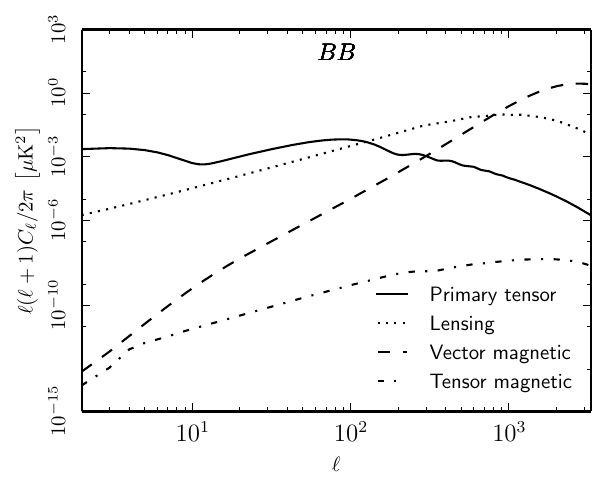}
\caption{Magnetically-induced CMB $TT$ (\emph{top left}), $TE$ (\emph{top right}), $EE$ (\emph{bottom left}), and $BB$ (\emph{bottom right}) power spectra. The solid lines represent primary 
CMB anisotropies, the dotted lines represent magnetically-induced compensated scalar modes (except for the $BB$ panel, where it represents the lensing contributions and the solid line represents primary tensor modes with a tensor-to-scalar ratio of $r=0.1$), the dashed lines represent vector modes, whereas the dot-dashed 
\label{fig:CMBAPS}
lines represent magnetically-induced compensated tensor modes. We consider PMFs with $B_{1\,\mathrm{Mpc}}=4.5$\,nG  and $n_B=-1$.}
\end{figure*}
\begin{figure*}
\includegraphics[width=88mm]{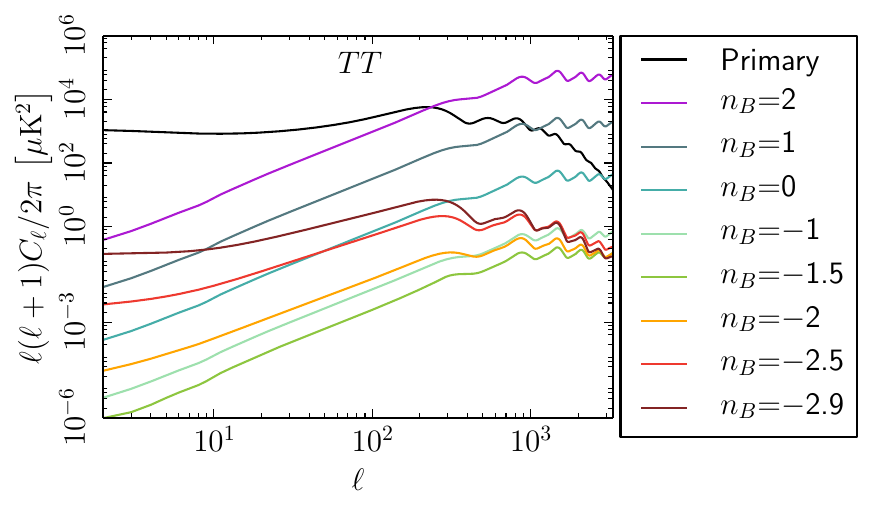}\includegraphics[width=88mm]{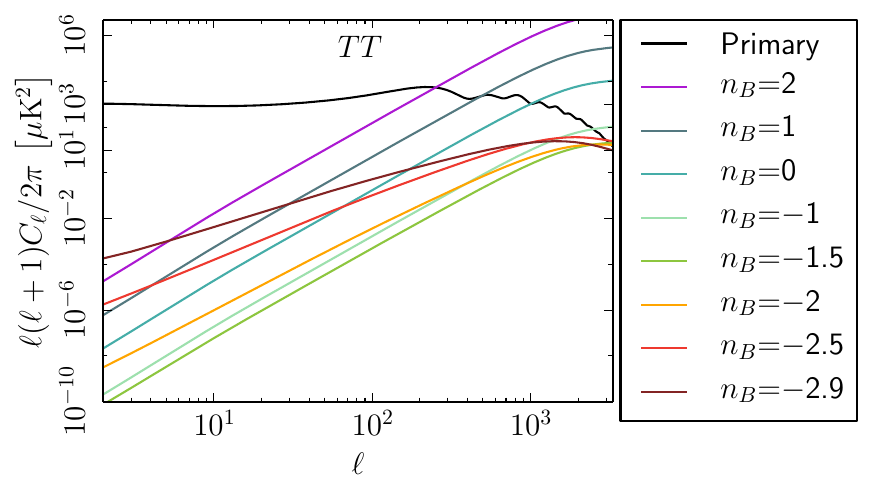}
\caption{Dependence of the magnetically-induced CMB power spectrum on the spectral index. For all plotted cases, the amplitude is $B_{1\,\mathrm{Mpc}}=4.5$\,nG. The black lines show primary CMB anisotropies; for the other colours we refer to the legend. \emph{Left}: scalar contributions, \emph{right}: vector contributions.}
\label{fig:CMBAPSVar}
\end{figure*}
\subsubsection{Compensated modes}
In Fig.~\ref{fig:CMBAPS}, we show the predictions for magnetically-induced compensated modes. This shows that the dominant compensated 
contributions to the angular power spectra are given by the scalar and vector modes.
In particular, because magnetically-induced perturbations are not suppressed by Silk damping, a significant contribution of magnetically-induced modes arises on small angular scales,
where the primary CMB fluctuations are suppressed. As we will show, the impact of PMFs on the CMB power spectrum at high multipoles is particularly relevant for the high-precision {\Planck} data, allowing us to derive strong constraints on the PMF amplitude.
Since magnetically-induced perturbations are solely sourced by energy momentum tensor components due to PMFs, the shape of the magnetically-induced spectra strongly depends on the PMF
spectral index. In Fig.~\ref{fig:CMBAPSVar}, we show this dependence for the temperature power spectrum for scalar and vector perturbations. We note
the qualitatively different dependence for two regimes. For $n_B>-3/2$ the angular power spectrum remains flat (i.e., $\ell(\ell+1)C_\ell \propto \ell^2$) with a rescaling of the amplitude due to the amplitude
of the Fourier spectra, whereas for $n_B<-3/2$ the shape varies according to the infrared domination of the energy momentum tensor of the PMFs.

\subsubsection{Passive modes}
\begin{figure*}
\includegraphics[width=88mm]{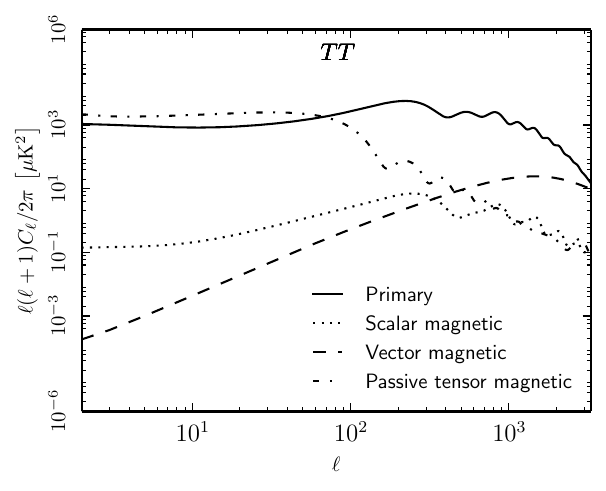}\includegraphics[width=88mm]{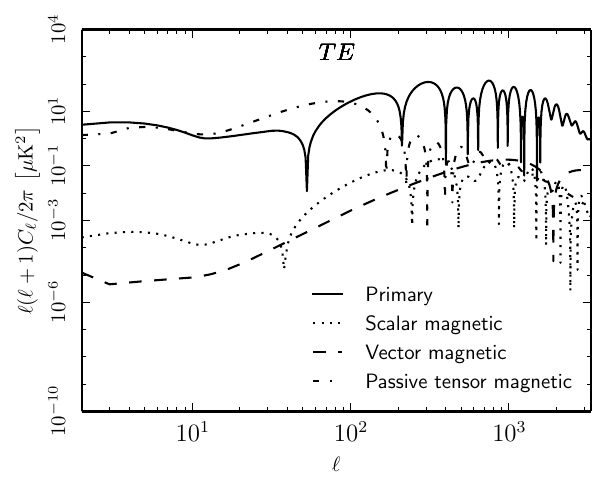}\\
\includegraphics[width=88mm]{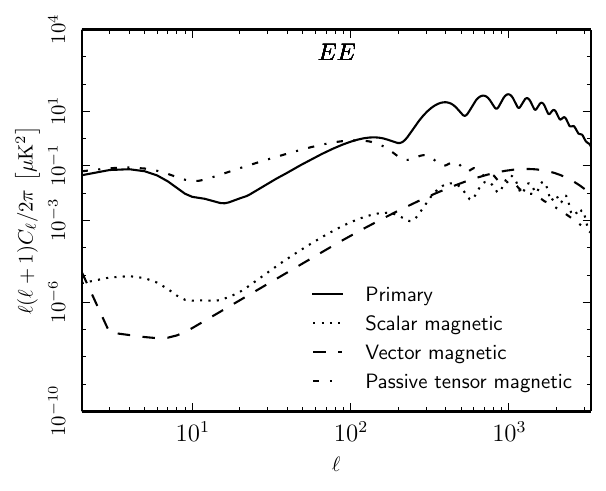}\includegraphics[width=88mm]{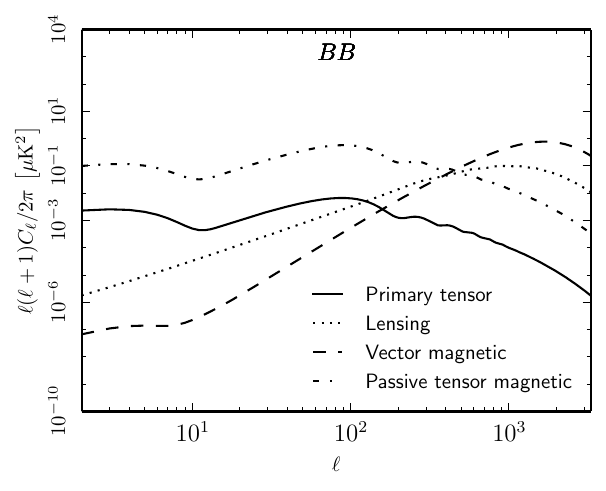}
\caption{Magnetically-induced CMB $TT$ (\emph{top left}), $TE$ (\emph{top right}), $EE$ (\emph{bottom left}), and $BB$ (\emph{bottom right}) power spectra due to passive tensor modes, compared with the ones due to compensated modes.
The solid lines represent primary 
CMB anisotropies, the dotted lines represent magnetically-induced compensated scalar modes (except for the $BB$ panel, where it represents the lensing contribution), the dashed lines represent vector modes, whereas dot-dashed 
lines represent magnetically-induced passive tensor modes. We consider PMFs with $B_{1\,\mathrm{Mpc}}=4.5$\,nG  and $n_B=-2.9$.}
\label{fig:CMBAPSPASS}
\end{figure*}

\begin{figure*}
\begin{tabular} cc
\includegraphics[width=88mm]{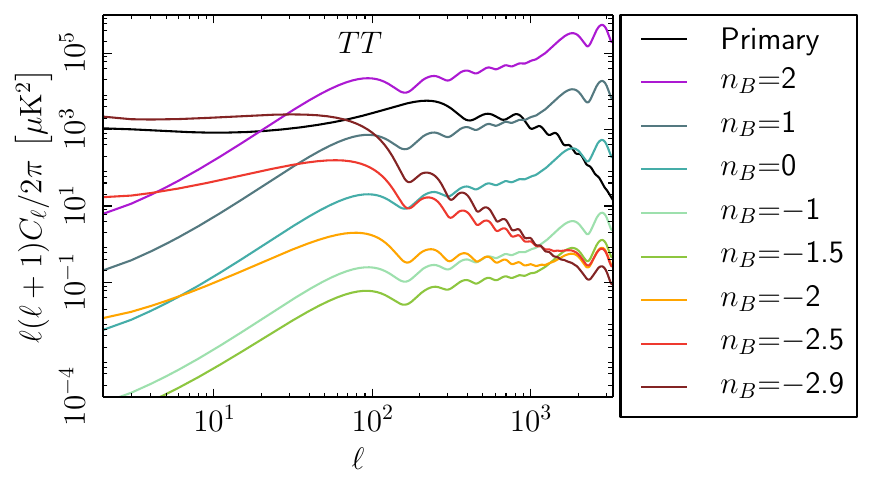}\includegraphics[width=88mm]{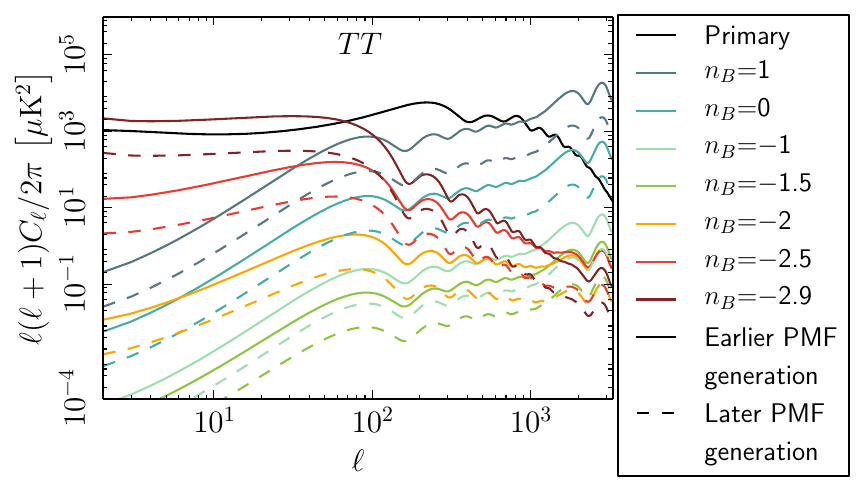}
\end{tabular}
\caption{Dependence of the magnetically-induced CMB power spectrum due to passive tensor modes on the spectral index for a GUT-scale PMF (\emph{left}) and comparison between the two extremes for the time ratio $\tau_\nu/\tau_B$ (\emph{right}). The black lines show the primary CMB anisotropies; for the other colours we refer to the legend. Solid lines represent PMFs generated at the GUT scale, $\tau_\nu/\tau_B=10^{17}$, whereas dashed lines represent PMFs generated at late times, $\tau_\nu/\tau_B=10^6$.}
\label{fig:CMBAPSPVar}
\end{figure*}

In addition to compensated initial conditions, we also consider passive tensor modes. The magnetically-induced passive modes are not completely determined by the amplitude and spectral index of the PMF, but
depend also on the ratio $\tau_\nu/\tau_B$. For tensors we specifically have $h(k)\propto \Pi^{(\mathrm{T})}(k) \ln(\tau_\nu/\tau_B)$, where $h(k)$ is the tensor metric perturbation.
This ratio may vary between $10^{17}$ and $10^{6}$ for fields originating at the grand unification energy scale (GUT) and at later phase transitions, \citep{2010PhRvD..81d3517S}. We consider the passive tensor modes, since, for red spectra, they give the dominant contribution on large angular scales, where the compensated 
modes are subdominant. In Fig.~\ref{fig:CMBAPSPASS}, we compare the magnetically-induced passive tensor modes for a nearly scale-invariant spectrum, $n_B=-2.9$,
for PMFs generated at the GUT scale 
(for which $\tau_\nu/\tau_B= 10^{17}$) with the corresponding dominant compensated modes. 
In Fig.~\ref{fig:CMBAPSPVar}, we show the dependence of the CMB power spectrum due to passive tensor modes on the spectral index, for GUT scale PMFs, and its dependence on the time ratio, showing
the two extreme values of the possible range. We note that the compensated vector modes dominate at small angular scales,
but the passive tensor modes give a contribution at low and intermediate multipoles for red spectra. For bluer spectra, the passive spectrum becomes steeper and therefore subdominant with respect to primary CMB fluctuations on large angular scales and with respect to vector modes on small angular scales. 
A similar behaviour can be observed in the scalar passive mode, which has the same origin as the tensor one but in the scalar sector. \cite{2010PhRvD..81d3517S} have shown that, just like the tensor mode, the scalar pasive mode becomes relevant on large angular scales for nearly scale-invariant power spectra. We will show below that the dominant passive contribution to the constraints on the PMF amplitude is given by tensor modes.

\subsection{Impact of helical PMFs on CMB anisotropies}

In addition to the ubiquitous presence of magnetic fields in the Universe, astrophysical observations show that some galaxies
 might have a helical magnetic field structure \citep{2002RvMP...74..775W,2004NewAR..48..763V}.
Following the hypothesis that the amplification of PMFs may have played a role in the generation of large-scale magnetic fields, 
the observed magnetic helicity may be related to helicity of the PMFs.
A helical intergalactic magnetic field could also be related to a possible CP violation recently hypothesized by \cite{Tashiro:2013ita} in an indirect study of cosmological large-scale magnetic fields in voids using secondary $\gamma$-ray data.
Helicity of the PMFs influences MHD processes in the early plasma, as well as cosmological perturbation 
dynamics, allowing different processes of energy transport, for example the inverse cascade mechanism \citep{2003matu.book.....B}.
These processes of energy transport play a role in the early time evolution of the PMFs and may have an impact on our understanding of their 
generation mechanisms, especially if PMFs with a non-zero helicity are generated.
Moreover, the presence of helicity would test possible modifications of Maxwell's theory by constraining parameters describing the gauge invariance 
(i.e., mass of the photon) and Lorentz invariance (i.e., existence of a preferred frame of reference) as discussed by \cite{1990PhRvD..41.1231C}. Thus it would carry information about particle physics at very high temperatures (above 1\,TeV).

A possible way to detect magnetic helicity directly from CMB data is to study the polarized CMB (cross-) power spectra.
A non-zero helicity in the PMFs changes the amplitudes of the parity-even power spectra and induces parity-odd cross-correlations between the temperature and $B$-polarization anisotropies and $E$- and $B$-polarization anisotropies \citep{2002PhRvD..65h3502P,2004PhRvD..69f3006C,2005PhRvD..71j3006K,2014Ballard}.
Such parity-odd cross-correlators are also generated by Faraday rotation, but only due to homogeneous PMFs, not due to the stochastic model for PMFs considered in this paper \citep{kosowsky05}.
Parity-odd signals may therefore give a more direct possibility for studying helical PMFs.

Our present study for a stochastic background of helical PMFs is an extension of the model described in the previous section. 
We consider the impact of such PMFs on CMB anisotropies in temperature and polarization. 
Using the exact expressions for the energy-momentum tensor components including the helical contribution discussed by \cite{2014Ballard}, we derive the predictions for the impact of a helical PMF on the CMB power spectra in temperature and polarization. 

The most general ansatz for the two-point correlation function, built on Eq.~\eqref{PSpectrum}, but taking into account an antisymmetric part \citep{2002PhRvD..65h3502P}, is
\ba
\Big\langle B_i(\vec{k}) \, B_j^*(\vec{k'}) \Big\rangle &=& \frac{(2\pi)^3 }{2} \, \delta^{(3)}(\vec{k}-\vec{k'}) \,\nonumber\\
&& \left[ (\delta_{ij}-\hat{k}_i\hat{k}_j) \, P_B(k) + i \, \epsilon_{ijl} \, \hat{k}_l \, P_H(k) \right]\,,
\label{HPSpectrum}
\ea
with $P_H(k) = A_H\,k^{n_H}$.
From a geometrical point of view $P_B(k)$ denotes the symmetric part and $P_H(k)$ the antisymmetric part of the correlator. 
The totally antisymmetric tensor $\epsilon_{ijl}$  is related to the parity violation under a transformation $\vec{k} \to -\vec{k}$.
The realizability condition gives $\left|P_H(k)\right|< P_B(k)$.

In this case the model is described by four parameters, two amplitudes and two spectral indices, where $A_B$ is the amplitude of the power spectrum of the fields, the same as defined in Eq.~\eqref{gaussian}, and $A_H$ is the amplitude of the power spectrum of the helical part of the PMFs.
These amplitudes can be expressed in terms of mean-square values of the magnetic field and of the helical component, respectively. As done for the amplitude 
of the magnetic field in Eq.~\eqref{gaussian}, we can express the amplitude of the helical component on a comoving scale of $\lambda$ as \citep{2014Ballard}
\be
\mathcal{B}^2_{\lambda} = \lambda \int_0^{\infty} \frac{d{k \, k^3}}{2 \pi^2} \mathrm{e}^{-k^2 \lambda^2}
\left| P_H (k) \right| = \frac{\left|A_H\right|}{4 \pi^2 \lambda^{n_H+3}} \, \Gamma \left( \frac{n_H+4}{2} \right)\,.
\label{Hgaussian}
\ee
The helical spectral index needs to satisfy $n_H>-4$ for convergence.

\begin{figure*}
\includegraphics[width=88mm]{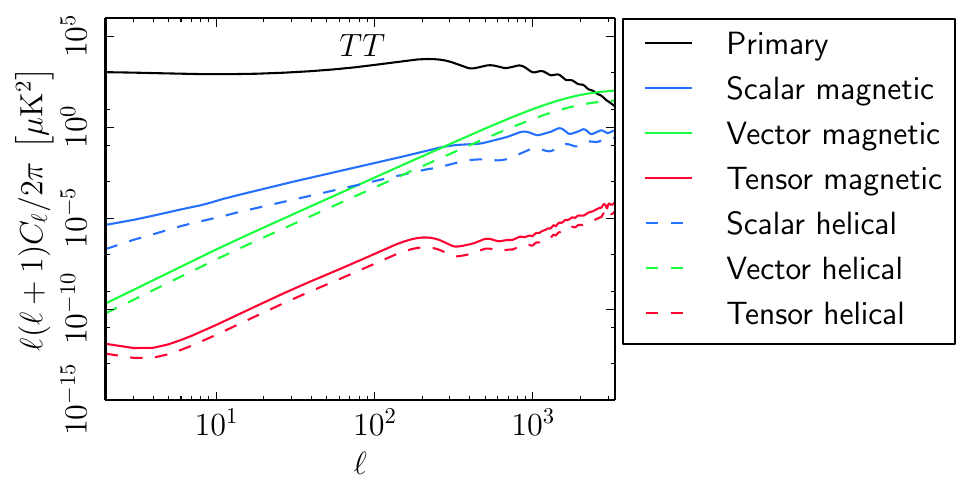}\includegraphics[width=88mm]{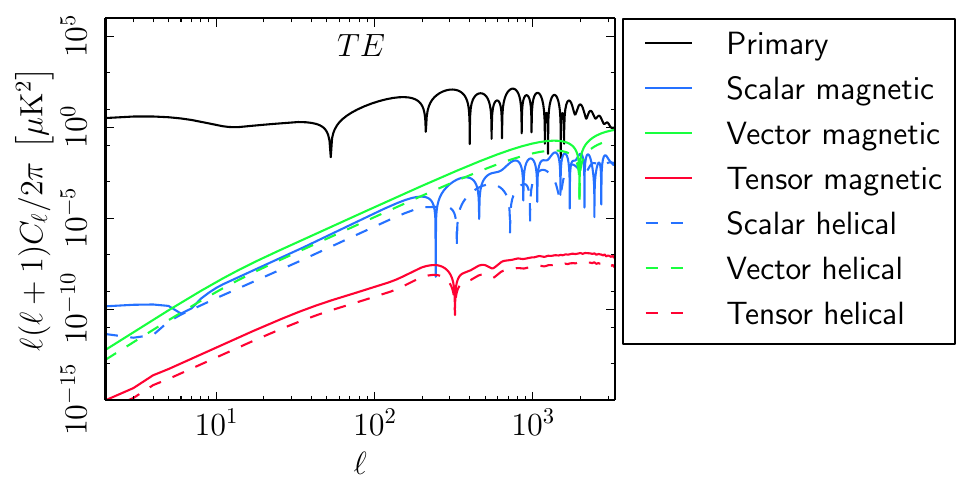}\\
\includegraphics[width=88mm]{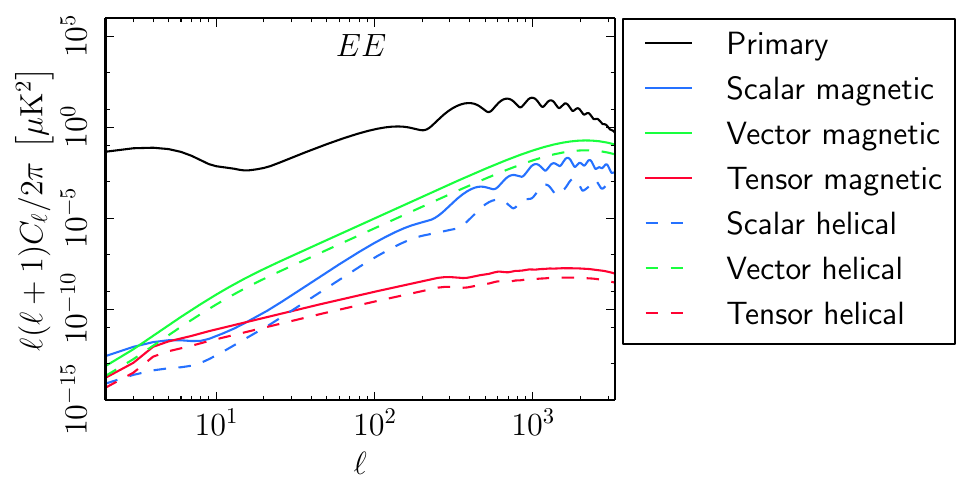}\includegraphics[width=88mm]{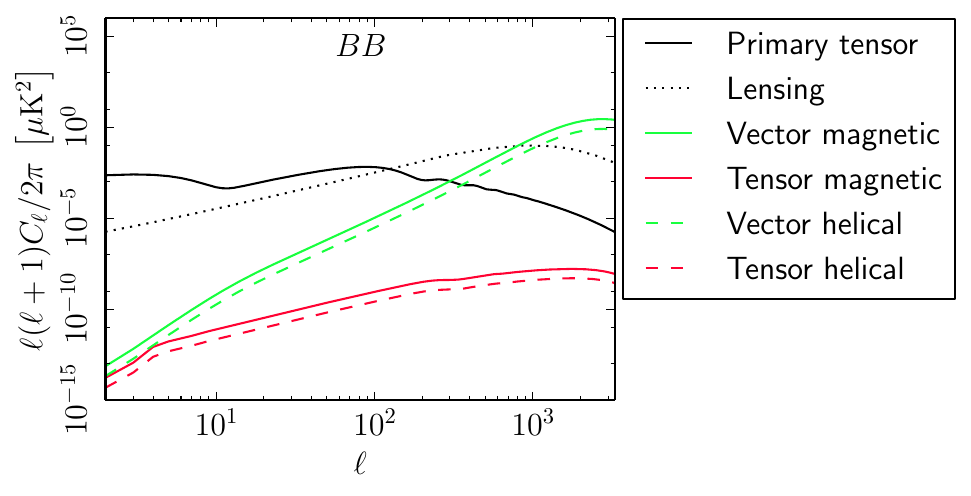}
\caption{CMB $TT$ (\emph{top left}), $TE$ (\emph{top right}), $EE$ (\emph{bottom left}), and $BB$ (\emph{bottom right}) power spectra due to helical PMFs compared to the ones due to non-helical PMFs.
Solid lines are non-helical predictions, while dashed lines are helical predictions. Blue are the scalar modes, green the vector, and red the compensated tensor modes. We consider PMFs with $B_{1\,\mathrm{Mpc}}=4.5$\,nG  and $n_B=-1$.}
\label{fig:CMBAPSH}
\end{figure*}

The helical term in Eq.~\eqref{HPSpectrum} generates new sources that contribute to the energy momentum tensor. All components of the energy momentum tensor are quadratic in the fields, so the Fourier-space two-point correlation function generates symmetric and antisymmetric sources: symmetric ones due to products of $P_B$ with $P_B$ (the components found in the non-helical case) and products of $P_H$ with $P_H$; antisymmetric ones, which generate odd-parity angular power spectra, from products of $P_B$ with $P_H$. The Fourier components of the energy momentum tensor are
\ba
|\rho_B(k)|^2&=&|\rho_B(k)|^2_{\mathrm{non-helical}} \notag\\
&&-\frac{1}{512\,\pi^5}\int_\Omega \dthree{p}\, P_H(p) \, P_H(|{\vec k}-{\vec p}|)\,\mu\,,
\label{Hdensity}\\
\left|L_{B}^{(\mathrm{S})} (k)\right|^2 &=&\left|L_{B}^{(\mathrm{S})} (k)\right|^2_{\mathrm{non-helical}} + \frac{1}{64\,\pi^2\,a^8} \notag\\
&&\times\int_\Omega \dthree{p}\, p \, P_H(p) \, P_H(|{\vec{k}}-{\vec{p}}|) \, \left(\mu - 2 \gamma \beta\right) \,,
\label{Hspectrum_LF}\\
\left|\Pi^{(\mathrm{V})}(k)\right|^2&=&\left|\Pi^{(\mathrm{V})}(k)\right|^2_{\mathrm{non-helical}} \notag\\
&&+\frac{1}{512\,\pi^5}\int_\Omega \dthree{p}\, P_H(p)\, P_H(|{\vec k}-{\vec p}|) \, \left(\mu - \gamma\beta\right) \,,
\label{Hvector}\\
\left|\Pi^{(\mathrm{T})}(k)\right|^2&=&\left|\Pi^{(\mathrm{T})}(k)\right|^2_{\mathrm{non-helical}} \notag\\
&&+\frac{1}{128\,\pi^5} \int_\Omega \dthree{p}\, P_H(p) \, P_H(|{\vec k}-{\vec p}|)\,\gamma\,\beta \,,
\label{Htensor}
\ea
where the non-helical parts are given by Eqs.~\eqref{density}--\eqref{tensor}. The system of equations for the background and the perturbations, as well as the initial conditions, are unmodified.

In Fig.~\ref{fig:CMBAPSH} we show the predictions for the magnetically-induced compensated modes, considering the additional contributions to the sources 
due to helicity. We consider the maximally helical case, $A_H=A_B$, with equal spectral indices, $n_B=n_H$, using the solutions
to the helical energy momentum tensor components  derived by \cite{2014Ballard}. 
The predictions show no difference in the shape and in the slope of the angular power spectra, with a small shift 
in the amplitude, which is always smaller for the helical case, at least for scales relevant for the CMB.
Magnetic helicity induces parity-odd cross-correlations between the $E$- and $B$-polarization anisotropies, as well as between temperature 
and $B$-polarization anisotropies. The parity-odd cross-correlations are sourced by the mixed terms in the correlation function 
of the energy momentum tensor, proportional to $\int\dthree{p} \, P_B(\vec p) \, P_H{\left(\left|\vec{k}-\vec{p}\right|\right)}$. 
These terms, after decomposition, contribute to the vector and tensor sources as
\ba
\left|A^{(\mathrm{V})}(k)\right|^2&=&\frac{1}{1024\,\pi^5}\int_\Omega \dthree{p}\,\Biggl\{ P_B(p)\, P_H(|{\vec k}-{\vec p}|)\notag\\
&&\left[\beta\left(1-\gamma^2\right)-\left(\gamma\beta-\mu\right)\gamma\right] \notag\\
&&+ P_H(p)\, P_B(|{\vec k}-{\vec p}|)\left[\gamma\left(1-\beta^2\right)-\left(\gamma\beta-\mu\right)\beta\right]\Biggr\}\,,
\label{Hvector}\\
\left|A^{(\mathrm{T})}(k)\right|^2&=&\frac{1}{256\,\pi^5}\int_\Omega \dthree{p}\,\Biggl\{ P_B(p)\, P_H(|{\vec k}-{\vec p}|)\left[\beta\left(1+\gamma^2\right)\right] \notag\\
&&+ P_H(p)\, P_B(|{\vec k}-{\vec p}|)\left[\gamma\left(1+\beta^2\right)\right]\Biggr\} \,.
\ea

In the limit of small momenta, i.e., $k \ll k_\mathrm{D}$, the spectra never show a white-noise behaviour, $\left|A^{(\mathrm{X})}(k)\right| = \mathrm{const}$, contrary to what happens for the symmetric and non-helical parts. 
For $n_B+n_H > -2$ the behaviour is proportional to $k\,k_\mathrm{D}^{n_B+n_H+2}$ and for $n_B+n_H < -2$ it does not depend on the damping scale and is 
proportional to $k^{n_B+n_H+2}$. 

\subsection{Constraints from the CMB temperature and polarization power spectra}

Here we present the constraints from {\Planck} on helical and non-helical PMFs. 
In the literature there are several previous studies that already derived constraints on PMFs using different combinations of observed 
CMB power spectra \citep{2010tsra.confE.222C,2010PhRvD..81b3008Y,2011PhRvD..83l3533P,2012PhRvD..86d3510S,2013PhLB..726...45P,2015arXiv150902461P}.
We use an extended version of the \texttt{CosmoMC} code \citep{2011ascl.soft06025L},
modified to include the magnetic contributions to the CMB power spectra, as described in the previous subsections, 
and to include the parameters characterizing the PMFs in the Markov chain Monte Carlo analysis.

We assume a flat Universe and a CMB temperature $T_0=2.7255$\,K, and we use the BBN consistency condition \citep{2006PhRvD..73f3528I,2008JCAP...03..004H}. 
We restrict our analysis to three massless neutrinos. A non-vanishing 
neutrino mass would not modify the results since it would only enhance the power on large scales in the presence of PMFs for the compensated modes, 
where the PMF contribution is less relevant \citep{2010PhRvD..81d3517S}.
The pivot scale of the primordial scalar is set to
$k_*=0.05$\,Mpc$^{-1}$. We consider the lensing effect for the primary CMB power spectrum  and follow the method implemented in
the {\Planck} likelihood to marginalize over astrophysical residuals and secondary anisotropy contamination
of the small-angular-scale data \citep{planck2014-a13}. 
This contamination is particularly relevant for the PMF scenario, 
since PMFs impact mainly small angular scales. If this contamination is not properly considered it may lead to biased constraints on 
PMFs \citep{2013PhLB..726...45P}.  We sample the posterior using the
Metropolis-Hastings algorithm \citep{Hastings70}, generating between four and sixteen 
parallel chains and imposing a conservative Gelman-Rubin convergence
criterion \citep{Gelman92} of $R-1 < 0.01$.\footnote{This convergence criterion is based on the analysis of the variance within a chain and between chains. 
The posterior marginal variance is a weighted average of the different variances. If all the chains have reached convergence,
this value will be very close to the variance within each single chain. The parameter $R$ is defined as the square root of the ratio of the posterior marginal variance and
the variance within a chain. With this definition, the closer $R$ is to unity, the closer to convergence the chains are.}  
We vary the baryon density $\omega_\mathrm{b}=\Omega_\mathrm{b} h^2$, the cold dark matter density $\omega_\mathrm{c}= \Omega_\mathrm{c}h^2$ , the reionisation optical depth $\tau_\mathrm{reion}$, the ratio of the sound horizon to the 
angular diameter distance at decoupling $\theta$, the scalar amplitude $\ln(A_{\mathrm{s}} 10^{10})$, and the scalar slope $n_\mathrm{s}$. 
In the MCMC analysis, we include the magnetic parameters $B_{1\,\mathrm{Mpc}}$ and $n_B$ for the compensated modes,
and add the parameter $\tau_\mathrm{rat}=\tau_{\mathrm{\nu}}/\tau_B$ whenever we also consider the passive tensor mode. 
We use flat priors for the magnetic parameters in the ranges
 $\left[0{,}10\right]$ for $B_{1\,\mathrm{Mpc}}/\mathrm{nG}$,  $\left[-2.9{,}3\right]$ for $n_B$ 
($n_B > -3$ to avoid infrared divergence in the PMF energy momentum tensor correlations). We sample  
$\tau_\mathrm{rat}$ logarithmically, with a flat prior on $\log_{10}{\tau_\mathrm{rat}}$ in the range $\left[4{,}17\right]$.

\subsubsection{Likelihood}

We derive the constraints on PMFs using the {\Planck} likelihood, which is described 
in detail in \cite{planck2014-a13}. Here we give a brief summary of the main points.
The {\Planck} likelihood is based on the \Planck\ 2015 data and considers both temperature and polarization.  
As in 2013, we use a hybrid approach with the combination of two likelihoods, one dedicated to low $\ell$ and 
the other to high $\ell$.

The \Planck\ low-$\ell$ likelihood is a fully pixel-based likelihood with temperature and polarization 
treated jointly and at the same resolution, $N_\mathrm{side}=16$. 
The $\ell$-range is $2<\ell<29$ in $TT$, $TE$, $EE$, and $BB$. 
The likelihood is based on the foreground-cleaned LFI maps at 70\,GHz and the temperature map derived 
by the component separation method {\tt Commander} using 94\,\% of the sky at frequencies 
from $30$ to $353$\,GHz \citep{planck2014-a11}. The polarization map covers 54\,\% of the sky and 
is derived from the $70$\,GHz $Q$ and $U$ maps cleaned with the $30$\,GHz map as a synchrotron template and the 
$353$\,GHz map as a dust template \citep[see][]{planck2014-a13}. This likelihood is denoted as ``lowP'' throughout the paper. 
Contrary to the 2013 analysis, where a combination of {\Planck} temperature and WMAP9 polarization data was used,
the 2015 low-$\ell$ likelihood is based entirely on {\Planck} data for both temperature and polarization. 

The \Planck\ high-$\ell$ likelihood is based on a Gaussian approximation (\cite{planck2013-p08} and \mbox{\cite{planck2014-a13}} for polarization) and covers the $\ell$-range $30<\ell<2500$. 
It uses the half-mission cross-power spectra of the 100\,GHz, 143\,GHz, and 217\,GHz channels,  measured in the cleanest 
region of the sky far from the Galactic plane and bright point sources. The sky fractions considered 
are 66\,\% of the sky for 100\,GHz, 57\,\% for 143\,GHz, and 47\,\% for 217\,GHz in temperature, whereas in polarization 
they are 70\,\%, 50\,\%, and 41\,\%, respectively. The likelihood takes foregrounds and secondary anisotropies into account. In particular, for the temperature spectra it considers the contributions of dust, clustered Cosmic Infrared Background (CIB), thermal and 
kinetic Sunyaev Zeldovich effect (tSZ and kSZ), the cross-correlation between tSZ and CIB, and a Poissonian term for unresolved 
point sources for the temperature spectra. In polarization, only the dust contribution is considered. 
Each model is parameterized as a template contribution to the $C_\ell$ with a free amplitude. 
The dominant contribution for $\ell<500$ is dust, whereas high $\ell$-modes are dominated by point sources and in 
particular the CIB for the $217$\,GHz auto-correlation. For the details of the foreground modelling see \citet{planck2014-a13}. 
This high-$\ell$ likelihood is denoted as ``{\Planck} $TT$'', for temperature only,  
or ``{\Planck} $TT$, $TE$, $EE$'', for temperature plus polarization, throughout the paper.

\subsubsection{Constraints with compensated scalar and vector contributions}

\begin{figure}
\includegraphics[width=88mm]{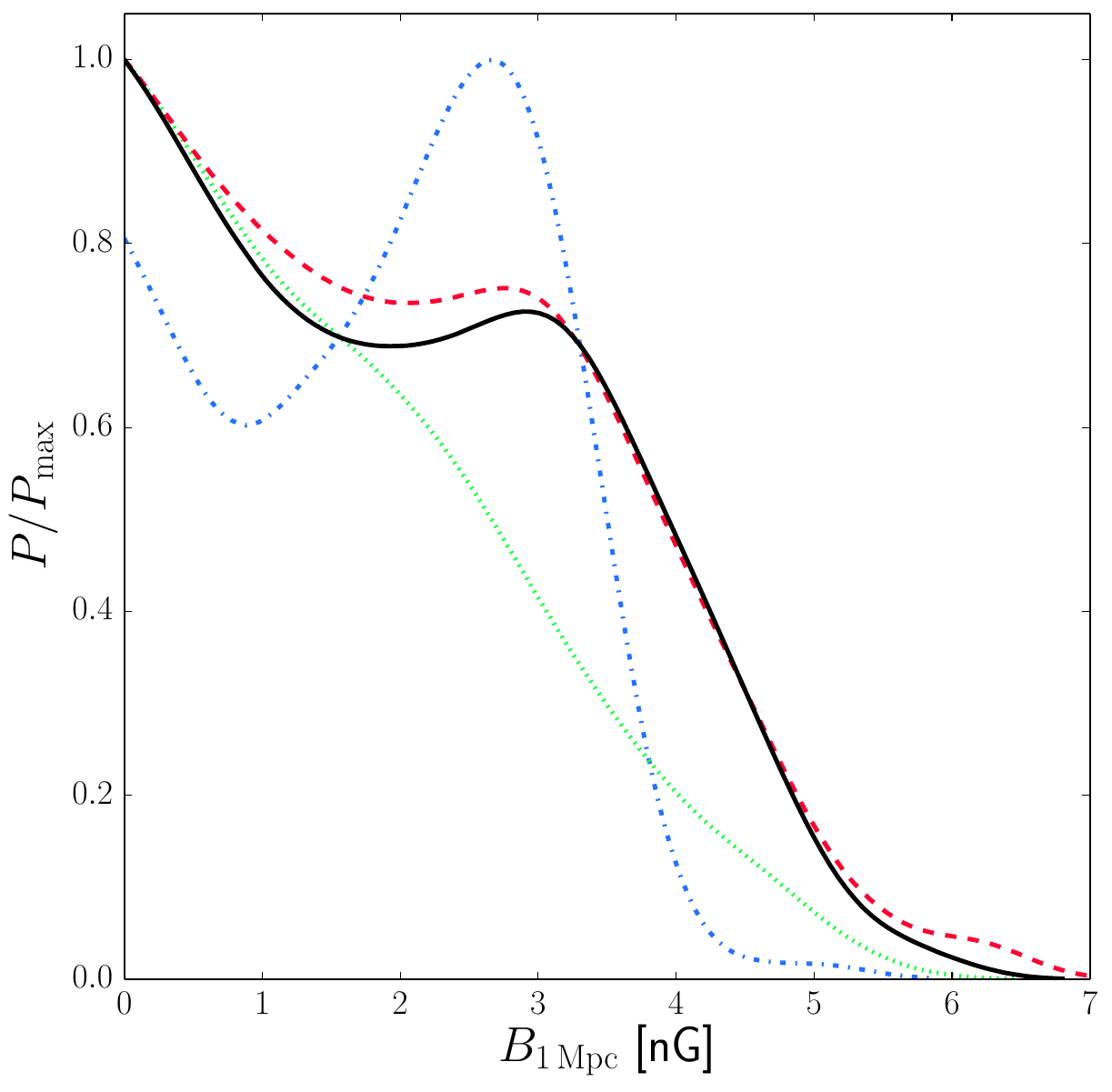}\\
\includegraphics[width=88mm]{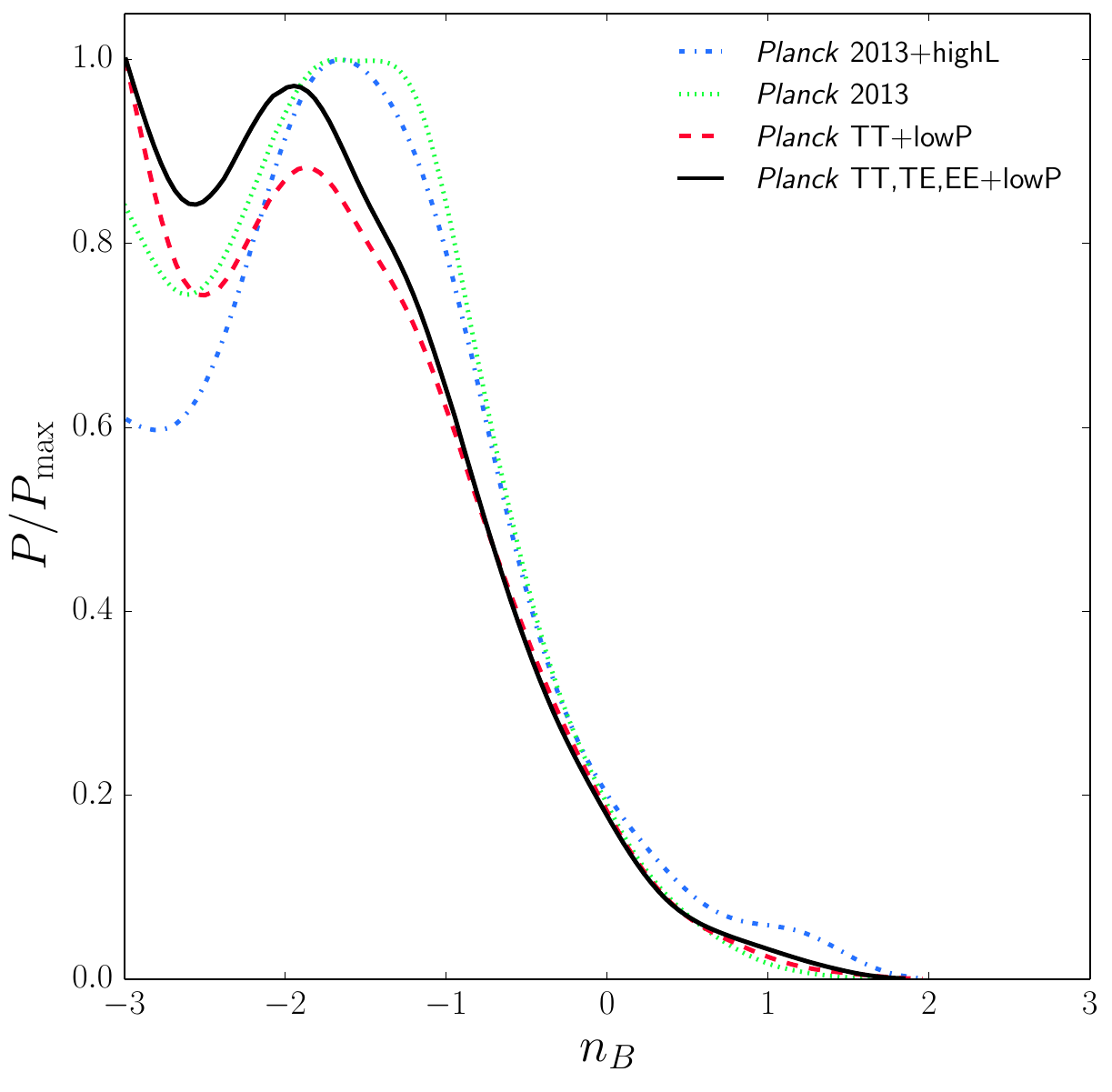}
\caption{Comparison of the constraints on the smoothed PMF amplitude (\emph{top}) and the spectral index (\emph{bottom}) from the 2015 temperature and temperature plus polarization data with the 2013 results for magnetically-induced compensated initial conditions only.}
\label{fig:Constraint1}
\end{figure}

We perform an analysis with the {\Planck} 2015 baseline likelihood. In Table~\ref{tab:tab_data} we report the derived constraints. The constraint on the PMF amplitude is $B_{1\,\mathrm{Mpc}} <4.4\,\mathrm{nG}$ at the 95\,\% confidence limit (CL) for the case that includes temperature and polarization data both at low and high multipoles. The same constraint results when including the polarization only at low $\ell$. 
As in previous analyses, PMFs with positive spectral indices are constrained to lower amplitudes than PMFs with negative spectral indices. 
In Fig.~\ref{fig:Constraint1} we present the results of this analysis compared with the {\Planck} 2013 constraints \citep{planck2013-p11}.
The upper limits from current {\Planck} data are slightly higher than those obtained in 2013  \citep{planck2013-p11}, where the constraint from {\Planck} data alone was
$B_{1\,\mathrm{Mpc}} <4.1\,\mathrm{nG}$. The weaker constraint can be explained by several changes of the 2015 data and likelihood with respect to 2013. The change in the calibration (see \citealt{planck2014-a01}), the different likelihood implementation, and the different models for the foreground residuals are all factors that contribute to the changed upper limit. In fact, all these factors, including also the slightly higher spectral index
$n_{\mathrm s}$ with respect to 2013, are mimicking a slightly larger signal in the temperature anisotropies, which is compatible with larger values of the PMF amplitude.

\begin{table}[tmb]               % table* is a two-column table.  Drop the * for one column.
\begingroup
\newdimen\tblskip \tblskip=5pt
\caption{Mean parameter values and bounds of the central 68\,\% CL from {\Planck} $TT$,$TE$,$EE$ (left column) 
and {\Planck} $TT$ (right column). When consistent with zero, the upper bound of the 
95\,\% CL is reported. 
Note that $H_0$ is a derived parameter. The posterior of the spectral index $n_B$ is strongly prior-dependent since $B_{1\,\mathrm{Mpc}}$ is consistent with zero.}                          % Caption goes here.
\label{tab:tab_data} 
\tabskip=0pt                           % Label goes here.
\nointerlineskip
\vskip -3mm
\footnotesize
\setbox\tablebox=\vbox{
   \newdimen\digitwidth 
   \setbox0=\hbox{\rm 0} 
   \digitwidth=\wd0 
   \catcode`*=\active 
   \def*{\kern\digitwidth}
   \newdimen\signwidth 
   \setbox0=\hbox{+} 
   \signwidth=\wd0 
   \catcode`!=\active 
   \def!{\kern\signwidth}
\halign{ \hbox to 0.9in{$#$\leaderfil}\tabskip=0.3em&
         \hfil$#$\hfil\tabskip=1em&
         \hfil$#$\hfil\tabskip 0pt\cr
\noalign{\doubleline}
\omit\hfil Parameter\hfil&\omit\hfil {\Planck} $TT$,$TE$,$EE$ + lowP\hfil&\omit\hfil {\Planck} $TT$ + lowP\hfil\cr   % Table headings go here.
\noalign{\vskip 3pt\hrule\vskip 5pt}
\omega_{\text b}& 0.0222\pm0.0002& 0.0222\pm0.0002\cr
\omega_{\text c}&  0.1198\pm0.0015& 0.1197\pm0.0022\cr
\theta& 1.0408\pm0.0003& 1.0408\pm0.0005\cr
\tau_\mathrm{reion}& 0.078\pm0.017& 0.075\pm0.019\cr
\log[A_{\mathrm{s}} 10^{-9}]& 3.09\pm0.03& 3.08\pm0.04\cr
n_{\mathrm{s}}& 0.963\pm0.005& 0.964\pm0.007\cr
\noalign{\vskip 1pt}
H_0& 67.77^{+0.68}_{-0.67}**& 67.82^{+0.98}_{-1.00}**\cr
\noalign{\vskip 2pt}
B_{1\,\mathrm{Mpc}}/\mathrm{nG}& <!4.4**& <!4.4*\cr
n_B& <-0.008& <-0.31\cr 
%\log(\tau_\mathrm{rat}) \,& 7.97 ^{+4.03}_{-3.97} & 7.94^{+4.06}_{-3.94} \, \cr
\noalign{\vskip 5pt\hrule\vskip 3pt}}}
\endPlancktable                    % ends one-column \halign
%\endPlancktablewide                 % ends two-column \halign
\endgroup
\end{table}

\begin{table}[tmb]                 % table* is a two-column table.  Drop the * for one column.
\begingroup
\newdimen\tblskip \tblskip=5pt
\caption{Upper bounds of the central 95\,\% CL for the PMF amplitude. C stands for compensated mode, C+P for compensated plus passive modes, $\tau_\mathrm{reion}$ prior indicates the case where instead of the low-$\ell$ polarization likelihood, as a cross-check, we used a Gaussian prior on the optical depth, $\tau_\mathrm{reion}=0.07\pm0.02$.}                          % Caption goes here.
\label{tab:tab_target}                            % Label goes here.
\nointerlineskip
\vskip -3mm
\footnotesize
\setbox\tablebox=\vbox{
   \newdimen\digitwidth 
   \setbox0=\hbox{\rm 0} 
   \digitwidth=\wd0 
   \catcode`*=\active 
   \def*{\kern\digitwidth}
   \newdimen\signwidth 
   \setbox0=\hbox{+} 
   \signwidth=\wd0 
   \catcode`!=\active 
   \def!{\kern\signwidth}
\halign{ \hbox to 1.5in{#\leaderfil}\tabskip=2em&
         \hfil$#$\hfil\tabskip 0pt\cr
\noalign{\doubleline}
\omit\hfil&\omit\hfil $B_{1\,\mathrm{Mpc}}/\mathrm{nG}$\hfil\cr
\noalign{\vskip 3pt\hrule\vskip 5pt}
$TT$,$TE$,$EE$+lowP: C& <4.4\cr
$TT$+lowP: C& < 4.4\cr
$TT$,$TE$,$EE$+lowP: C+P& <4.5\cr
$TT$+lowP: C+P& <4.5\cr
$TT$ + $\tau_\mathrm{reion}$ prior: C+P& <4.4\cr
\noalign{\vskip 5pt\hrule\vskip 3pt}}}
\endPlancktable                    % ends one-column \halign
%\endPlancktablewide                 % ends two-column \halign
\endgroup
\end{table}                        % table* is a two-column table.  Drop the * for one column.

We now include the polarization data in the analysis. Although the impact of PMFs on $TE$ and $EE$ polarization is less important than on temperature anisotropies, we show the results for the case of {\Planck} data, which include also high-$\ell$ $TE$ and $EE$ polarization, in Fig.~\ref{fig:Constraint1}. 
Although the shape of the posterior changes slightly, there is no net improvement on the 95\,\% CL upper bound on $B_{1\,\mathrm{Mpc}}$ with the addition of the high-$\ell$  $TE$ and  $EE$ polarization.

\subsubsection{Constraints with passive tensor contributions}

As described in the previous subsection, in addition to regular magnetically-induced compensated modes, the presence of PMFs prior to neutrino decoupling generates passive modes.
%causes an adiabatic-like mode whose amplitude depends on the PMFs and their generation time; the passive mode.
In Fig.~\ref{fig:CMBAPSPASS} we show that the passive tensor modes may give the dominant magnetic contribution to the CMB power spectra for
a nearly scale-invariant PMF power spectrum. Their inclusion in the analysis may therefore be relevant for the constraints on PMFs.
We include the passive tensor contribution in the MCMC code with the addition of the parameter $\tau_{\mathrm{rat}}$, with the settings described above. 
We perform MCMC analyses with the {\Planck} 2015 likelihood, combining the low-$\ell$ temperature and polarization data either with high-$\ell$ temperature data or with high-$\ell$ temperature and polarization data, i.e., $TT$+lowP and $TT$,$TE$,$EE$+lowP.

\begin{figure}
\includegraphics[width=88mm]{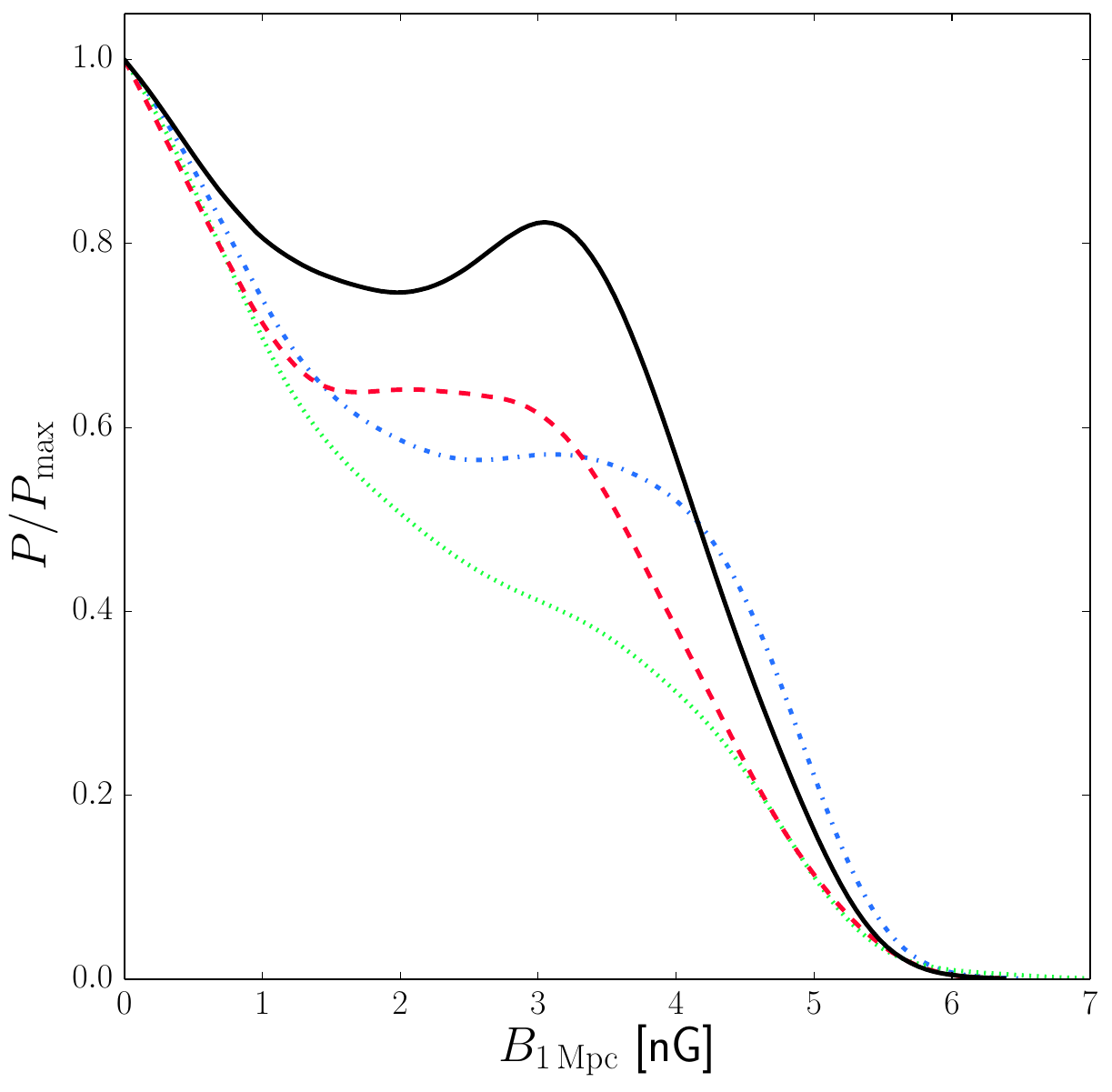}\\
\includegraphics[width=88mm]{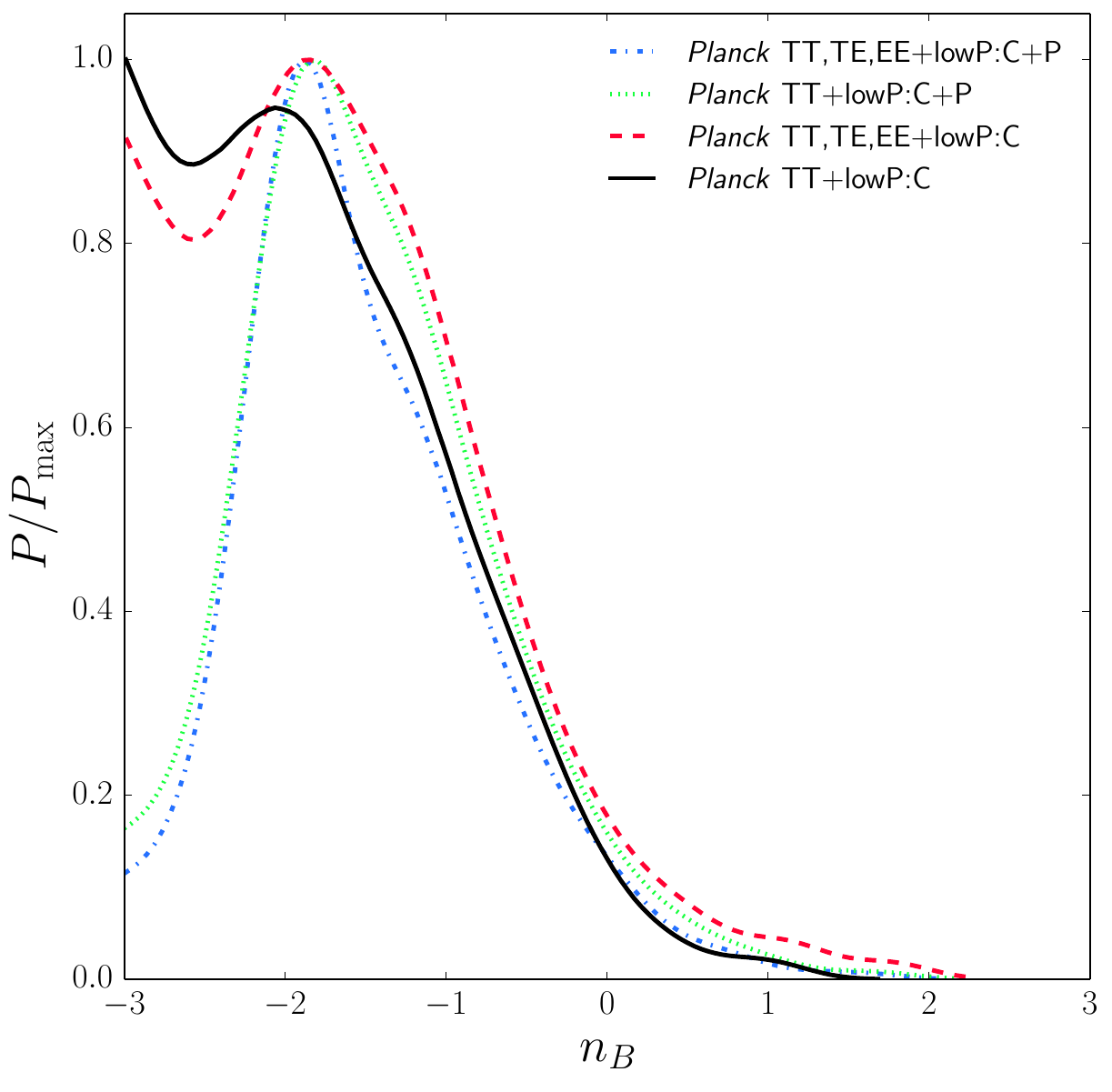}
\caption{Constraints on the smoothed PMF amplitude (\emph{top}) and spectral index (\emph{bottom}) from {\Planck} temperature data with and without the passive tensor contribution. Constraints including both compensated and passive modes are indicated with C+P in the legend, constraints using only compensated modes are marked with C.}
\label{fig:ConstraintPass}
\end{figure}

\begin{figure}
\includegraphics[width=88mm]{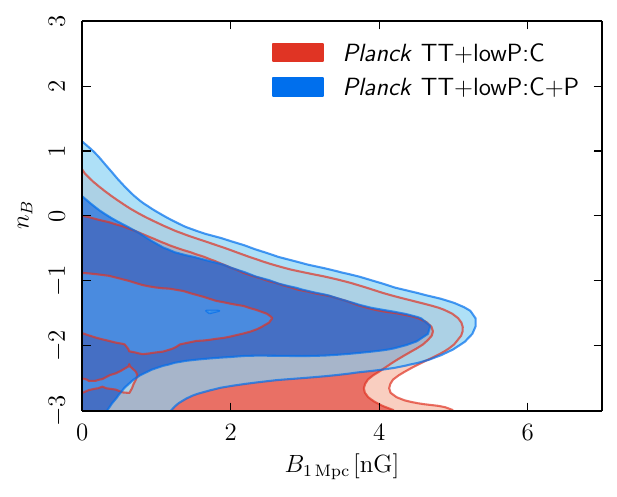}
\caption{PMF amplitude versus the spectral index for the baseline {\Planck} 2015 case. C+P denotes the case where both compensated and passive modes are considered, whereas C indicates the case with only compensated modes.The two contours represent the 68\,\% and 95\,\% confidence levels.}
\label{fig:degenerate}
\end{figure}

Figure~\ref{fig:ConstraintPass} and Table~\ref{tab:tab_target} present the results, compared to the results of the case that includes only compensated contributions.
The  95\,\% CL constraint on the PMF amplitude is $B_{1\,\mathrm{Mpc}}<4.5$\,nG, 
which implies that the addition of the passive tensor contribution does not improve
the constraint on the amplitude of the PMFs. 
This result is expected on the basis of the shape of the angular power spectra due to passive tensor modes and their strong dependence on the
PMF spectral index. This mode is basically a primary tensor mode with and amplitude that depends on the PMFs. Its spectrum flattens on large angular scales
and then decays on intermediate ones for red PMF spectra, whereas it acquires a steeper shape for blue PMF spectra, but with a much lower amplitude
than for compensated vector modes. Therefore, passive tensor modes only contribute significantly for
nearly scale invariant indices. In Fig.~\ref{fig:degenerate} we present the two-dimensional plot for the PMF amplitude and the spectral index. It shows
the strong degeneracy between the two parameters, meaning that the same magnetically-induced power spectrum can be realized with different pairs
of amplitude and spectral index. 
The effect of passive tensor modes on the CMB temperature angular power spectrum is dominant over the primary CMB anisotropies only for a very limited 
range of spectral indices, near the scale-invariant case. Therefore, the degeneracy between amplitude and spectral index reduces the influence of the passive 
tensor mode contribution on the constraints on the amplitude and we do not see any improvement in adding the passive tensor mode.
While the constraint on the amplitude is almost unchanged, the spectral index in the MCMC analysis is sensitive to the contribution of the passive mode. 
In the lower panel of Fig.~\ref{fig:ConstraintPass} we show the different shapes of the posterior distributions for
the PMF spectral index for different data combinations. The inclusion of the passive tensor mode influences the low spectral index part of the posterior, while the compensated modes influence the high spectral index part.
In addition to the analyses that consider the combination of passive and compensated modes, we perform an analysis including only the contribution of the passive tensor mode. 
We obtain  $B_{1\,\mathrm{Mpc}}<6.5$\,nG at 95\,\% CL using \Planck\ $TT$+lowP. This result shows that the contribution of the passive term alone, when considering the spectral index and the generation epoch as free parameters, does not have the constraining power of the combination of passive and compensated modes. 

\begin{figure}
\includegraphics[scale=1]{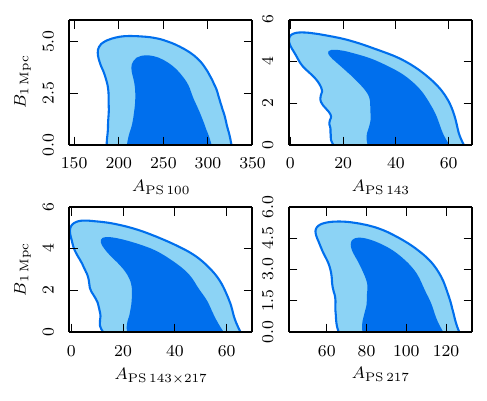}
\caption{Two-dimensional posterior distributions of the PMF amplitude versus the parameter describing the Poissonian term of unresolved point sources for the three
frequencies considered in the likelihood. The two contours represent the 68\,\% and 95\,\% confidence levels.}
\label{fig:Degeneracy}
\end{figure}

\subsubsection{Impact of astrophysical residuals}
\label{sec:foregroundresiduals}

We adopt the {\Planck} likelihood treatment of astrophysical contaminants as described in the {\Planck} likelihood paper. Considering the complexity of this model
and the number of nuisance parameters involved, we investigate whether foreground residuals have an impact on the constraints on PMFs.
Specifically, we investigate possible degeneracies with the foreground parameters by considering the two-dimensional distributions
of the magnetic parameters and foreground parameters. The relevant cases are shown in Fig.~\ref{fig:Degeneracy}. There, we present the two-dimensional distributions of the PMF
amplitude and the Poissonian amplitudes for the three frequencies considered in the {\Planck} high-$\ell$ likelihood: 100\,GHz, 143\,GHz, 217\,GHz, and the 143$\times$217\,GHz cross-spectrum.
We note that especially for the 143\,GHz and the 143$\times$217\,GHz analyses, a weak degeneracy between the two parameters is seen. This result may indicate an impact of the astrophysical residual
modelling on the PMF constraints. To investigate this issue, we perform an analysis by fixing the four parameters for the Poissonian amplitudes to their
best-fit values from the standard Lambda Cold Dark Matter model. This analysis yields a limit of $B_{1\,\mathrm{Mpc}}<3.0$\,nG, which is smaller than the constraint obtained in the case where the parameters
associated to astrophysical residuals are free to vary. This result has no statistical significance, but demonstrates that there is an impact of the astrophysical residuals on the PMF constraints when the data considered 
require a complex model for the residuals. The shape of the dominant PMF contributions to the angular power spectrum on small angular scales is responsible for this degeneracy. In fact,
the steep slope of the vector mode may be degenerate with astrophysical residual contributions, as shown by \cite{2013PhLB..726...45P}.
For comparison we show an analogous plot for the 2013 analysis in Fig.~\ref{fig:deg2013}, which considers the degeneracy for the foreground parameters of the \Planck\ 2013 likelihood. We note how, in contrast to the 2015 analysis, there is only a small degeneracy with the Poissonian amplitude at 143\,GHz. This result shows the importance of the foreground residual modelling for the PMF constraints.

In contrast to the Poissonian terms, we do not observe any degeneracy with the other foreground components, as shown in Fig.~\ref{fig:degall}, including the clustering component of the foreground residuals, which is the other dominant contribution 
on small angular scales at the frequencies considered in this analysis.
The fact that we do not observe a degeneracy in this case is due to the difference in the spectral shape of the PMF contribution and the clustering term. Although 
both are relevant on small angular scales the slightly different shapes break the degeneracy.

\subsubsection{Constraints for specific PMF models}

\Planck\ 2015 results confirm what has been observed in previous analyses, namely that CMB data allow negative PMF spectral indices with larger field amplitudes than positive indices.
The spectral index of the PMFs is the main discriminating factor among possible generation mechanisms.

Some cases are of particular
interest due to their connection with specific classes of generation mechanisms. In particular, PMFs generated during phase transitions or via second order perturbative effects, vector perturbations, etc., are characterized by positive spectral indices, equal or greater than 2. 
To investigate the maximal amplitude allowed by {\Planck} data for fields of this type, we perform two dedicated analyses,
the first with fixed index $n_B=2$, and the second only restricted to positive spectral indices for PMFs.
We include both compensated and passive modes in the analysis, giving $B_{1\,\mathrm{Mpc}}<0.011$\,nG at 95\,\% CL  ($B_{1\,\mathrm{Mpc}}<0.012$\,nG at 95\,\% CL, when considering \Planck\ $TT$,$TE$,$EE$+lowP) for the $n_B=2$  case and $B_{1\,\mathrm{Mpc}}<0.55$\,nG at 95\,\% CL for $n_B>0$.

We consider a third case of interest, the almost scale-invariant fields with $n_B=-2.9$. This specific case is connected to PMF generation from inflation and is studied
to test the strength of the passive tensor modes in constraining the amplitude of the PMFs.
Moreover, we want to compare the 
results obtained from the {\Planck} power spectra with those coming from the non-Gaussianity analysis, presented in the next section, 
which is performed for this spectral index as well.
We obtain $B_{1\,\mathrm{Mpc}}<2.0$\,nG at the 95\,\% CL.  Note that this nearly scale-invariant case is dominated by the tensor passive mode. 
In fact, when we consider only the tensor passive contributions, excluding the compensated ones, 
we obtain the same result as in the passive-compensated combined case, $B_{1\,\mathrm{Mpc}}<2.0$\,nG at the 95\,\% CL.

Together with the passive tensor mode there is also a scalar passive mode, as shown by \cite{2010PhRvD..81d3517S}. When we include this scalar passive contribution in our analysis we obtain again
  $B_{1\,\mathrm{Mpc}}<2.0$\,nG at the 95\,\% CL. We can therefore conclude that the passive scalar contribution is subdominant with respect to the tensor one.
As will become clear in the next section, this result shows that the constraining power of
 the angular power spectrum is comparable to the one of the non-Gaussianity.

In addition to these specific types of PMFs we have performed some analyses with fixed spectral index, choosing 
a grid of values covering the full range we sample. With respect to the case where the spectral index is a free variable,
 the cases with fixed spectral index are expected to give stronger constraints on the 
PMF amplitude, with a trend in agreement with the general case, because one of the two parameters describing the PMF is fixed. 
In Table~\ref{tab:specific} we present the results of these analyses. 
We note how, as expected, the trend of the results with fixed spectral index is in agreement with 
the one of the two-dimensional plot of Fig.~\ref{fig:degenerate} obtained with the generic sampling of the index. 
We note also how the constraint is weakest for $n_B=-1.5$ as expected from the impact on the angular power spectrum. 
\begin{table}[tmb]                 % table* is a two-column table.  Drop the * for one column.
\begingroup
\newdimen\tblskip \tblskip=5pt
\caption{95\,\% CL  upper bounds of the PMF amplitude for fixed spectral index with compensated plus passive tensor modes.}                          % Caption goes here.
\label{tab:specific}                            % Label goes here.
\nointerlineskip
\vskip -3mm
\footnotesize
\setbox\tablebox=\vbox{
   \newdimen\digitwidth 
   \setbox0=\hbox{\rm 0} 
   \digitwidth=\wd0 
   \catcode`*=\active 
   \def*{\kern\digitwidth}
   \newdimen\signwidth 
   \setbox0=\hbox{+} 
   \signwidth=\wd0 
   \catcode`!=\active 
   \def!{\kern\signwidth}
\halign{\hbox to 0.8in{$#$\leaderfil}\tabskip=1em&
        \hfil$#$\hfil&
        \hfil$#$\hfil&
        \hfil$#$\hfil&
        \hfil$#$\hfil&
        \hfil$#$\hfil&
        \hfil$#$\hfil&
        \hfil$#$\hfil&
        \hfil$#$\hfil\tabskip 0pt\cr
%\halign{ \hbox to 1.5in{#\leaderfil}\tabskip=2em&
%         \hfil$#$\hfil\tabskip 0pt\cr
\noalign{\doubleline}
n_B&2&1&0&-1&-1.5&-2&-2.5&-2.9\cr
\noalign{\vskip 3pt\hrule\vskip 5pt}
B_{1\,\mathrm{Mpc}}/\mathrm{nG}&0.011&0.1&0.5&3.2&4.8&4.5&2.4&2.0\cr
\noalign{\vskip 5pt\hrule\vskip 3pt}}}
\endPlancktable                    % ends one-column \halign
%\endPlancktablewide                 % ends two-column \halign
\endgroup
\end{table}                        % table* is a two-column table.  Drop the * for one column.

\subsubsection{Constraints from the BICEP2/\textit{Keck}-\Planck\ joint analysis}

\begin{figure}
\includegraphics[width=88mm]{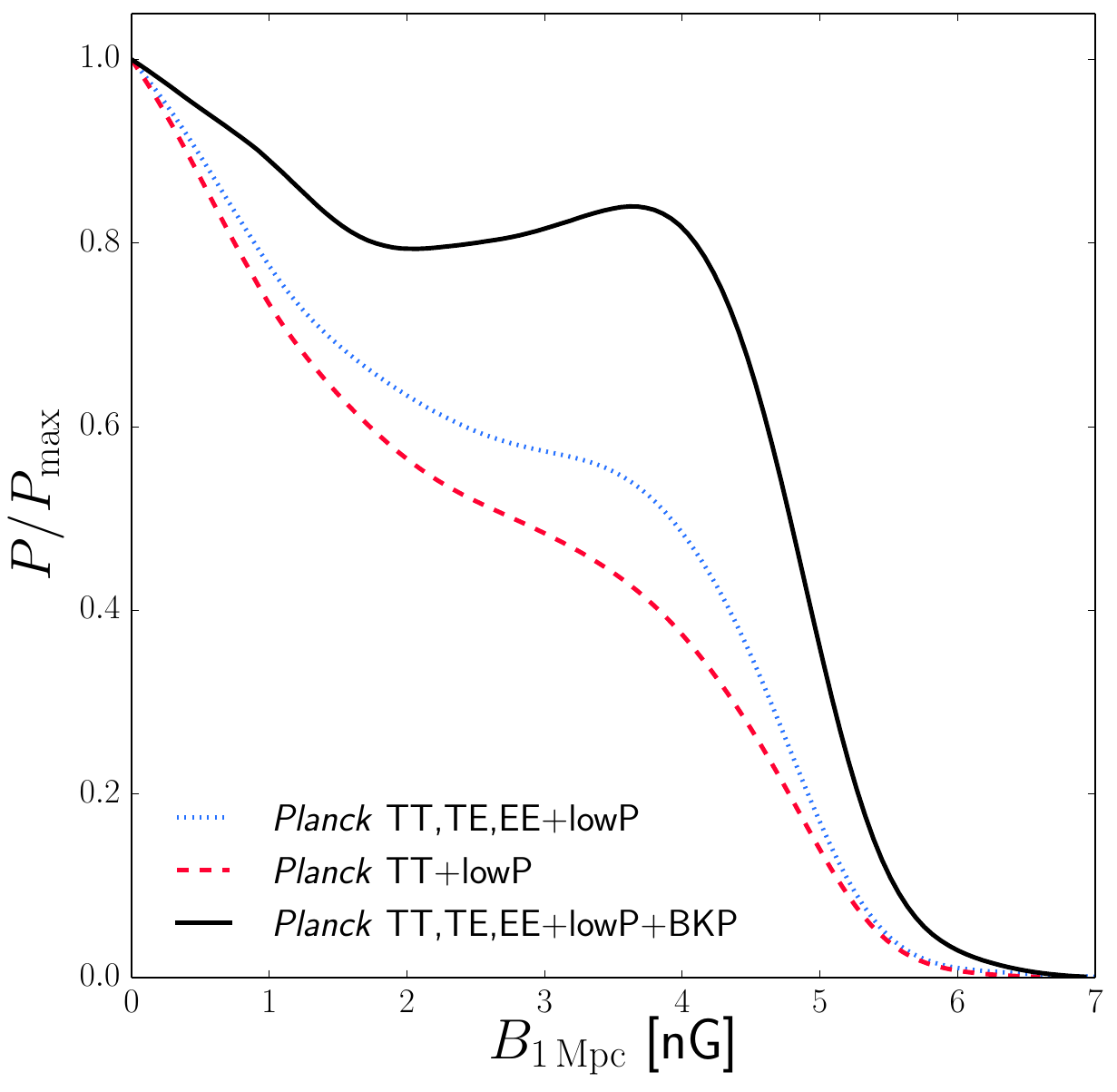}\\
\includegraphics[width=88mm]{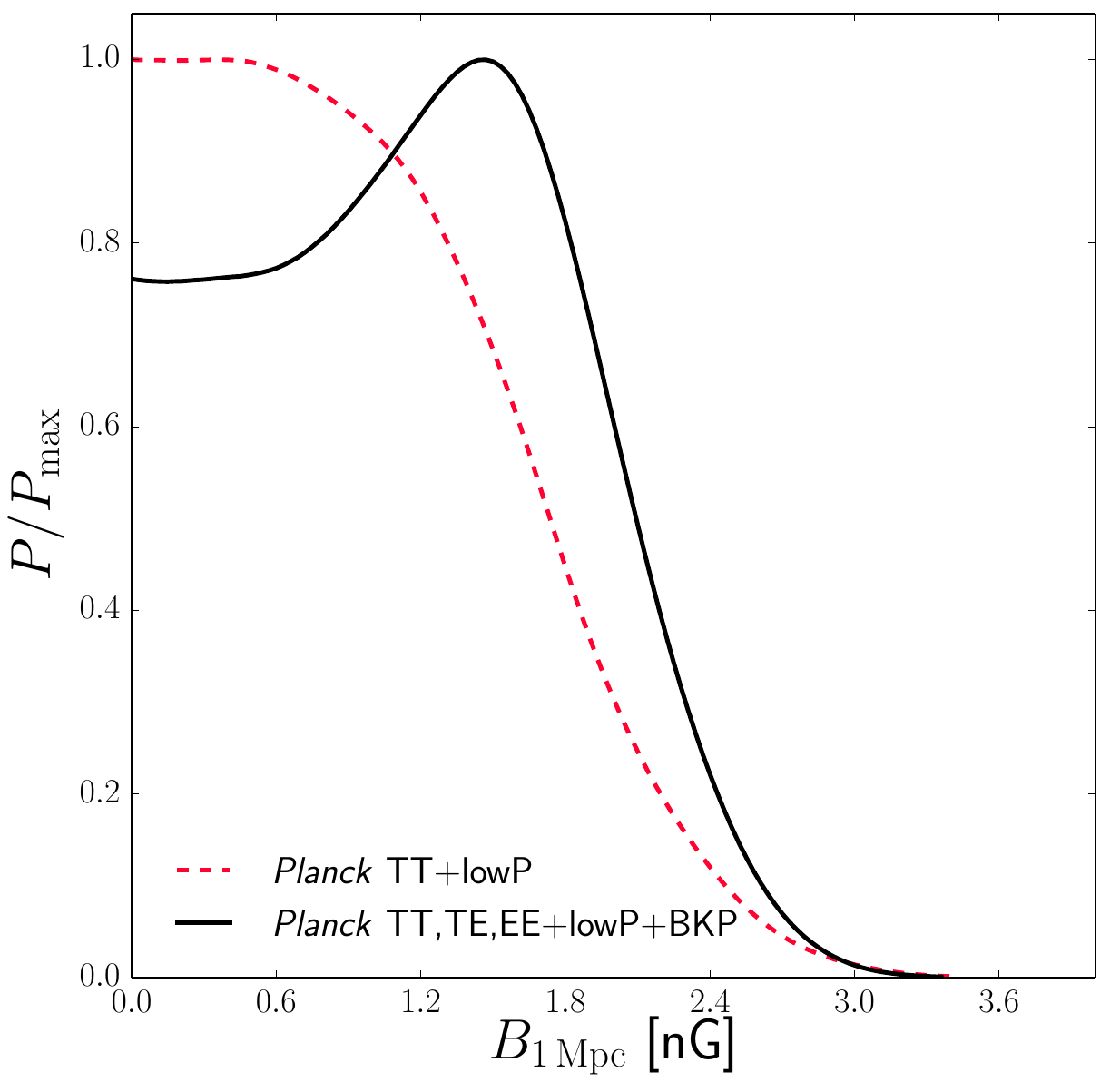}
\caption{Probability distributions for the PMF amplitude including the  BICEP2/\textit{Keck}-\Planck\ cross-correlation, compared with the one based only on \Planck\ data. \emph{Top}: the case in which the spectral index is free to vary, \emph{bottom}: the case with $n_B = -2.9$.}
\label{fig:BKP}
\end{figure}

We perform an analysis using the recent BICEP2/\textit{Keck}-\Planck\ cross-correlation \citep[indicated as BKP;][]{pb2015} in addition to the \Planck\ 2015 data. 
The BKP likelihood is obtained from the $BB$ and $EE$ bandpowers for all cross-spectra between 
the BICEP2/\textit{Keck} maps and the \Planck\ maps at all frequencies \citep{pb2015}.
We consider both compensated and passive contributions and study two cases, one in which we leave the spectral index free to vary and another in which we fix it to $n_B=-2.9$. In the latter case, there is a contribution to the $B$-mode 
polarization on large angular scales from the passive tensor mode.
Figure~\ref{fig:BKP} shows the comparison of the results of these two analyses with the results obtained from \Planck\ data alone. The constraints are $B_{1\,\mathrm{Mpc}}<4.7$\,nG for the case with free spectral index and $B_{1\,\mathrm{Mpc}}<2.2$\,nG  at the 95\,\% CL for the case with $n_B=-2.9$. These are slightly higher upper bounds for the amplitude of PMFs, but they are fully compatible with the results derived from \Planck\ data alone. We note that the posterior distribution for the nearly scale-invariant case changes with the addition of the BKP data but it does not show any significant deviation from the posterior based only on \Planck\ data.

\subsubsection{Constraints with maximally helical contributions}

\begin{figure}
\includegraphics[width=88mm]{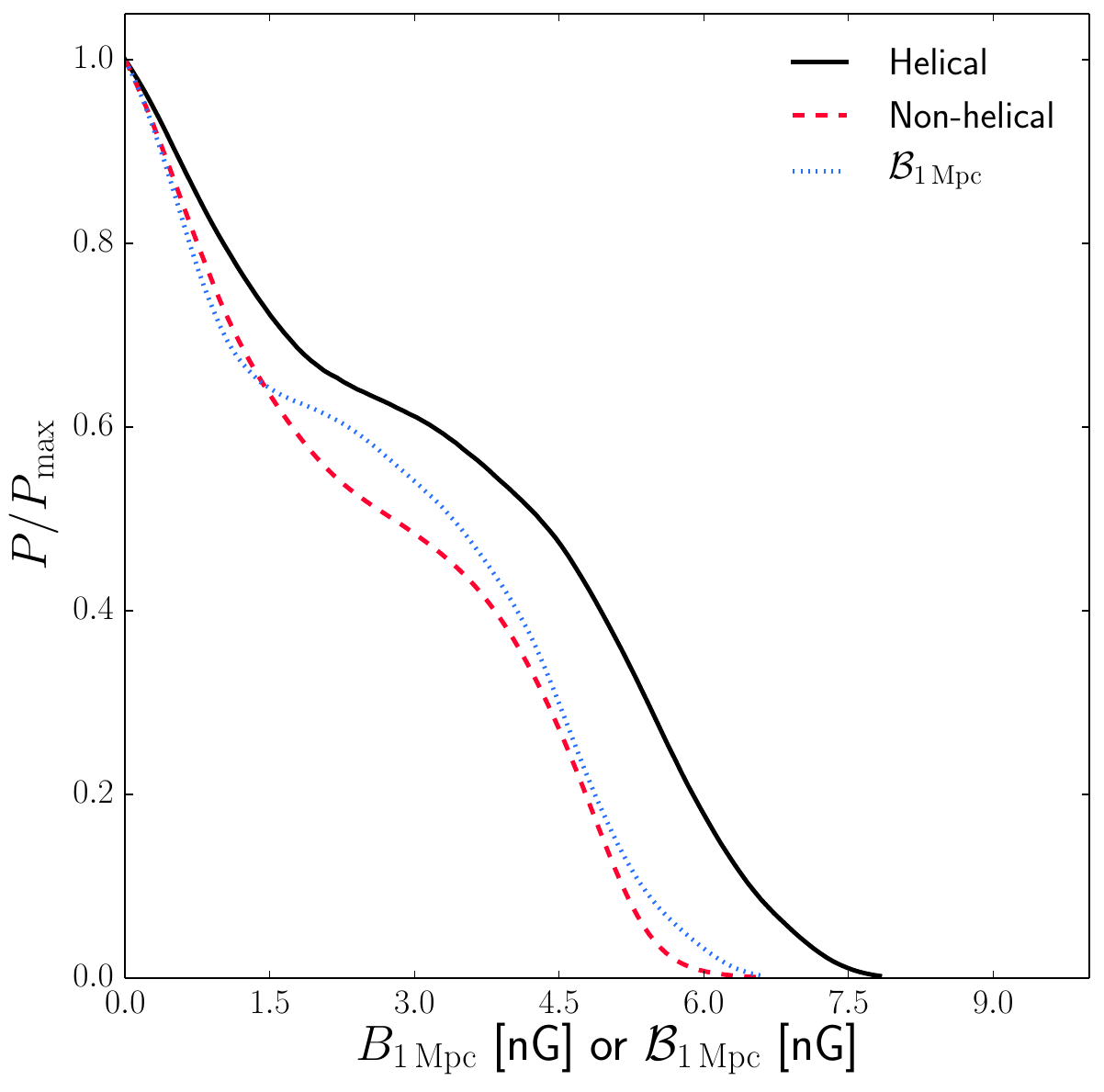}
\caption{PMF amplitude constraint for the helical case (solid black) compared with the non-helical case (dashed red). The dotted blue line shows the constraint on the amplitude of the helical component as an alternative 
interpretation of the constraints on the amplitude of PMFs with a helical component.}
\label{fig:helical}
\end{figure}

We perform an MCMC analysis including the maximally helical contribution.
We restrict our analysis to the case of temperature and polarization with only even cross-correlations. The odd cross-correlations 
$TB$ and $EB$ are present only in the lowP likelihood, therefore only for very low multipoles where the signal from helical PMFs
is negligible. Thus we do not include odd cross-correlators in our analysis.

We perform an analysis using the {\Planck}$TT$+lowP likelihood. The constraint on the PMF amplitude in the maximally helical case is $B_{1\,\mathrm{Mpc}}<5.6$\,nG at the 95\,\% CL. 
A comparison with the corresponding results for the non-helical case is shown in Fig.~\ref{fig:helical}. 
The analysis with the {\Planck} $TT$,$TE$,$EE$+lowP likelihood gives $B_{1\,\mathrm{Mpc}}<5.8$\,nG at the 95\,\% CL. As in the non-helical case, the inclusion of high-$\ell$ polarization does not improve the constraints.
Figure~\ref{fig:CMBAPSH} shows that magnetic fields with a maximally helical component produce smaller CMB fluctuations in temperature and polarization 
than non-helical fields of the same strength. As a result of this, the amplitude of maximally helical magnetic fields are less constrained than
non-helical fields for this \Planck\ 2015 data release.
When considering helical PMFs, we have two components that contribute to the magnetically-induced perturbations, as shown in Eq.~\eqref{HPSpectrum}, 
a symmetric and an antisymmetric part, represented by $P_B$ and $P_H$, respectively. 
These power spectra can be  associated with two amplitudes of the field, $B_{1\,\mathrm{Mpc}}$ associated with the symmetric part and ${\mathcal{B}_{1\,\mathrm{Mpc}}}$ associated with the antisymmetric part (see Eq.~\ref{Hgaussian}).
In the maximally helical case the two amplitudes are not independent from each other, they are related through the conditions $A_H=A_B$ and $n_B=n_H$. Therefore we constrain a single 
amplitude, which can be expressed either through $B_{1\,\mathrm{Mpc}}$ or ${\mathcal{B}_{1\,\mathrm{Mpc}}}$. The constraint $B_{1\,\mathrm{Mpc}}<5.6$\,nG can thus be converted into 
the constraint ${\mathcal{B}_{1\,\mathrm{Mpc}}}< 4.6$\,nG at the 95\,\% CL.
Figure~\ref{fig:helical} shows the posterior distribution for the amplitude expressed as ${\mathcal{B}_{1\,\mathrm{Mpc}}}$ in blue.

\subsubsection{Constraints from the impact of PMFs on the CMB anisotropies via their impact on the thermal history of the Universe}

Primordial magnetic fields are damped on scales smaller than the photon diffusion and free-streaming scale. This leads to heating of ordinary matter (electrons and baryons), 
which affects both the thermal and ionization history of the Universe \citep{1998PhRvD..58h3502S, 2000PhRvL..85..700J, Sethi2005, Schleicher2008b, 2014JCAP...01..009K,Chluba2014PMF}, 
leading to a Compton-$y$ distortion of the CMB and changes in the CMB power spectra through modifications of the Thomson visibility function around decoupling.

Two heating mechanisms have been discussed in the literature, one due to decaying magnetic turbulence at very small scales and the other due to ambipolar 
diffusion \citep[e.g.,][]{Sethi2005}. In this paper, we follow the approach described by \citet{Chluba2014PMF} to incorporate these heating mechanisms.\footnote{As explained by \citet{Chluba2014PMF}, the effect of PMFs on the ionization history was previously overestimated using an approach similar to the one of RECFAST. The reason is that in RECFAST the photoionization rates are evaluated assuming temperature $T\equiv T_{\mathrm e}$, although from a physical point of view the radiation temperature should be used. This reduces the effect on the ionization history by up to one order of magnitude and a consistent treatment is implemented both in {\sc Recfast++} and {\sc CosmoRec} \citep{Chluba2010b}.}

We perform an analysis considering the combination of the heating terms with the gravitational contribution of PMFs. Considering ambipolar diffusion, decaying magnetic turbulence, and 
gravitational effects we obtain an upper limit of $B_{1\,\mathrm{Mpc}}<0.90$\,nG at 95\,\% CL for nearly scale-invariant PMFs with $n_B=-2.9$. We obtain the same result, namely $B_{1\,\mathrm{Mpc}}<0.90$\,nG at 95\,\% CL, when dropping the gravitational effect and considering only the impact of PMFs on the primary CMB anisotropies through their heating effect. These results show that the dominant contribution is given by the heating terms. 
We have also performed analyses with the two terms of ambipolar diffusion and decaying magnetic turbulence considered separately. The results show that the two terms are roughly at the same level in constraining PMFs, with a slightly stronger contribution from the decaying magnetic turbulence term  \citep[see also][]{Chluba2014PMF}.

Together with the  {\Planck} $TT$+ lowP likelihood combination, we have performed an analysis including high-$\ell$ polarization.
In particular, we have considered the case of  {\Planck} $TT$, $TE$, $EE$+ lowP. The result is: $B_{1\,\mathrm{Mpc}}<0.86$ nG at the 95\,\% CL. Due to the nature of the effect of PMFs on the thermal history of the Universe and its impact on the CMB angular power spectra, the polarization data on small angular scales tighten the constraints of this analysis.

\section{Magnetically-induced non-Gaussianities}
\label{sec:non-gaussianities}

The CMB anisotropies induced by PMFs are non-Gaussian.
This is because  magnetic forcing (as described by the magnetic energy momentum tensor) is quadratic in the magnetic fields and therefore the resulting fluctuations are non-Gaussian even for Gaussian fields\footnote{This peculiarity is in common with topological
defects \citep[see][]{Figueroa:2010zx}} \citep{2005PhRvD..72f3002B}. 
There are already published theoretical studies of the passive-mode bispectra \citep{Trivedi:2010gi, Shiraishi:2011dh, Shiraishi:2012rm, Shiraishi:2013vha}, as well as studies of the compensated-mode 
bispectra \citep{Seshadri:2009sy, Caprini:2009vk, Cai:2010uw, Shiraishi:2010yk, Kahniashvili:2010us} 
and of trispectra \citep{Trivedi:2011vt, Trivedi:2013wqa}. 
This illustrates that it is possible to use CMB non-Gaussianities to constrain the PMF amplitude for different generation mechanisms. 
Several non-Gaussianity constraints have previously been used for this purpose \citep{Caprini:2009vk,Seshadri:2009sy,Trivedi:2010gi,Shiraishi:2012rm,Trivedi:2011vt}.
The non-Gaussianity constraints on PMFs are complementary to those derived from the angular power spectra.
In this section we present three different methods for constraining PMFs using non-Gaussianity measurements, all involving the first of the higher-order stastical moments, the bispectrum.
The methods can be applied to either the passive or the compensated modes.

\subsection{Magnetically-induced passive-tensor bispectrum}

The goal of this subsection is to derive an observational limit on the PMF strength from the passive bispectrum. 
The dominant contribution to the passive bispectrum is the large-scale tensor mode, while the scalar mode contributes 
subdominantly to the small scales. 
According to \citet{Shiraishi:2012rm} and \citet{Shiraishi:2013vha}, the signal-to-noise ratio  (integrated over $\ell$) is expected to be almost saturated beyond $\ell = 500$ in estimates based on 
the temperature bispectrum. Including higher multipoles would therefore not bring significant improvements. 
Thus, we take into account the tensor-mode contribution for $\ell \leq 500$ in the following. Here we concentrate on the almost scale-invariant case, $n_B = -2.9$. In this case, the passive-tensor bispectrum is amplified 
in the squeezed-limit configuration with $\ell_1 \ll \ell_2 \approx \ell_3$, as a consequence of the local-type structure of the non-Gaussian 
gravitational waves induced by the PMF, given by 
\begin{eqnarray}
h_{ij}(\vec{k}) \approx - 1.8 
\frac{ \ln(\tau_\nu / \tau_B )}{4\pi \rho_{\gamma,0}} 
{\mathbf \mathcal{M}_{ijkl}}(\hat{\vec{k}})
\int \frac{\dthree{p}}{(2 \pi)^3} B_k({\vec{p}}) \, B_l(\vec{k} - \vec{p})\,. 
\end{eqnarray}
Here $\tau_\nu$ and $\tau_B$ are defined in the same way as in the previous section, while $\rho_{\gamma,0}$ is the present photon energy density. The projection tensor $\mathcal{M}_{ijkl}(\hat{\vec{k}})$ is given by the products of the spin-$\pm$ transverse-traceless tensors as
$\mathcal{M}_{ijkl}(\hat{\vec{k}}) \equiv \sum_{s = \pm} e_{ij}^{(s)}(\hat{\vec{k}}) \, e_{kl}^{(s)*}(\hat{\vec{k}})$, normalized as $\mathcal{M}_{ijij} = 4$. This projection induces a tangled angular dependence on $\vec{k}_i$ in the primordial gravitational wave bispectrum and the resultant CMB bispectrum is given by a non-factorizable combination of $\ell$-modes
\citep{Shiraishi:2011dh, Shiraishi:2012rm, Shiraishi2012}. The resultant CMB temperature and $E$-mode bispectra are almost uncorrelated with the usual scalar-mode bispectra because of their different CMB transfer functions.
To derive constraints, we introduce an amplitude parameter proportional to the amplitude of the magnetically-induced bispectrum, 
\begin{equation}
A_\mathrm{\mathrm{bis}}^{\mathrm{MAG}} = \left(\frac{B_{1\,\mathrm{Mpc}}}{3\,\mathrm{nG}}\right)^6 
\left[ \frac {\ln(\tau_\nu / \tau_B) }{\ln(10^{17})} \right]^3\,, \label{eq:Abis}
\end{equation}
where $B_{1\,\mathrm{Mpc}}$ and $\tau_B$ are treated as free parameters. 
The normalization factors in the last equation are chosen to be comparable to current upper bounds on $B_{1\,\mathrm{Mpc}}$ 
for a PMF created at the GUT epoch, i.e., $\tau_\nu / \tau_B = 10^{17}$. It can be seen that the magnetically-induced bispectrum, which is proportional to $(B_{1\,\mathrm{Mpc}})^6$, has a logarithmic dependence on $\tau_B$. 
An analysis of the constraints from WMAP data is presented by \cite{2014PhRvD..90j3002S},
yielding $A_\mathrm{bis}^{\mathrm{Data}} = -1.5 \pm 1.4$ (68\,\% CL).

In order to constrain the non-factorizable magnetically-induced bispectrum, we use an optimal estimator derived within the so-called separable modal methodology 
(see \citealt{Fergusson:2009nv}, \citealt{Fergusson:2010dm}, \citealt{Shiraishi:2014roa}, and \citealt{2015JCAP...01..007S} for auto-bispectra and \citealt{2014PhRvD..90d3533F}, and \citealt{Liguori:2014} for cross-bispectra), 
where the theoretical bispectrum templates are decomposed in finite subsets of the separable eigenbasis. 
Thus, the bispectrum estimator remains factorizable like in the usual KSW approach \citep{2005ApJ...634...14K}.
Our tensor bispectrum template can be reconstructed well in this modal decomposition with about 400 eigenvectors 
composed of polynomials and a few special functions modelling the CMB temperature transfer functions.

From the foreground-cleaned \texttt{SMICA} temperature map, we obtain observational constraints on the amplitude of the passive-tensor bispectrum.
The observational data and the (Gaussian) simulation maps used in the computation of the linear term and the error bars are inpainted in the same manner as for the {\Planck} 
tensor non-Gaussianity analysis \citep{planck2013-p09a,planck2014-a19},
after including experimental aspects (beam, mask, and anisotropic noise). Our final result is
$A_\mathrm{bis}^{\mathrm{Data}} = -1.6 \pm 1.3$ ($T$ only)
at 68\,\% CL, giving no evidence for a signal at the $2\sigma$ level. 
This {\Planck} temperature constraint is in good agreement with the WMAP one \citep{2014PhRvD..90j3002S}. 
Analogous results have been derived for the combination of $T$- and $E-$modes and the $E$-only case, but since these results are still preliminary, we use the $T$-only mode for the present analysis.

Assuming that the bispectrum is generated by PMFs, its amplitude is given by Eq.~\eqref{eq:Abis}. The amplitude of the bispectrum
depends on the amplitude of the fields to the sixth power and on the logarithm of the ratio $\tau_\nu / \tau_B$, which is greater than unity.
Therefore we have $A_\mathrm{\mathrm{bis}}^{\mathrm{MAG}} \geq 0$.
The result obtained in this analysis 
therefore leads to an upper bound on the strength 
of GUT generated PMFs ($\ln(\tau_\nu / \tau_B)=10^{17}$) of $B_{1\,\mathrm{Mpc}} < 2.8 \,\mathrm{nG}$ (95\,\% CL).

\subsection{Magnetically-induced anisotropic passive scalar bispectrum}

Anisotropic stress from magnetic fields leads to curvature perturbations on super-horizon scales according to \citep{2010PhRvD..81d3517S}
\begin{eqnarray}
\zeta_{\vec{k}}&\approx& 0.9 \ln \left(\frac{\tau_\nu}{\tau_B}\right)\frac{1}{4\pi \rho_{\gamma,0}}\nonumber\\
&&\times\sum_{ij}\left(\hat k_i \hat k_j -\frac{1}{3} \delta_{ij} \right) \int \frac{\dthree{k'}}{(2\pi)^3} B_i(\vec{k'}) \, B_j(\vec{k}-\vec{k'})\,.\nonumber
\end{eqnarray} 
\cite{Shiraishi2012} showed that the three-point correlation of the curvature perturbation $\zeta_{\vec{k}}$ sourced by magnetic fields is
\begin{eqnarray}
\lefteqn{\langle \zeta(\vec{k}_1) \, \zeta(\vec{k}_2) \, \zeta(\vec{k}_3)\rangle \propto}\nonumber\\
&&P_B(k_*) \, P_B(k_1) \, P_B(k_2)\left(\frac{1}{3}\mu^2_{12}+\mu^2_{23}+\mu^2_{31}-\frac{2}{3}-\mu_{12}\mu_{23}\mu_{31}\right)\nonumber\\
&&-P_B(k_*) \, P_{H}(k_1) \, P_{H}(k_2)\left(\mu_{23}\mu_{31}-\frac{1}{3}\mu_{12}\right)\nonumber \nonumber\\
&&\textnormal{+ cyclic permutations},
\end{eqnarray}
where $\mu_{ab}= \vec {\hat k}_a\cdot \vec {\hat k}_b$ and $k_*$ denotes the pivot wavenumber.
We can investigate this primordial non-Gaussianity by estimating the expansion coefficient $c_L$ \citep{Shiraishi2013},
\begin{align}
\langle \zeta(\vec{k}_1) \, \zeta(\vec{k}_2) \, \zeta(\vec{k}_3)\rangle &=(2\pi)^3\delta^{(3)}(\vec{k}_1 + \vec{k}_2 + \vec{k}_3)\nonumber\\
& \sum_{L} c_L \left(P_L(\hat {\vec k}_1\cdot \hat {\vec k}_2)\, P_\zeta(k_1) \, P_\zeta(k_2)+2\,\mathrm{perm.}\right)\nonumber\,,
\end{align}
where $c_0$ is related to the local-form $f_{{\mathrm{NL}}}$ as $c_0=6/5 f^{\mathrm{local}}_{{\mathrm{NL}}}$ and 
$P_L$ is the $L$th order Legendre polynomial.

If the magnetic field is generated at the GUT scale with a nearly scale-invariant spectrum, the Legendre coefficients are related to the field amplitude via
\begin{eqnarray}
c_0&\approx& -2\times 10^{-4} \left(\frac{B_{1\,\mathrm{Mpc}}}{\mathrm{nG}}\right)^6,\label{c0pmf}\\
c_1&\approx& -0.9 \left(\frac{B_{1\,\mathrm{Mpc}}}{\mathrm{nG}}\right)^2 \left(\frac{\mathcal B_{1\,\mathrm{Mpc}}}{\mathrm{nG}}\right)^4,\label{c1pmf}\\
c_2&\approx& -2.8\times 10^{-3} \left(\frac{B_{1\,\mathrm{Mpc}}}{\mathrm{nG}}\right)^6,\label{c2pmf}
\end{eqnarray}
where $B_{1\,\mathrm{Mpc}}$ and $\mathcal B_{1\,\mathrm{Mpc}}$ are the amplitudes of the non-helical and helical magnetic field components (smoothed on a scale of 1\,Mpc), respectively.
Estimating $c_2$ allows us to constrain $B_{1\,\mathrm{Mpc}}$, but estimating $c_1$ does not 
lead to a useful constraint due to 
its dependence on the helical component of the PMF, which is not considered in this analysis (cf.\ Eq.~\ref{c1pmf}).
By the central limit theorem, the estimated value of $c_2$ follows a Gaussian distribution. Therefore, the log-likelihood is given by
\begin{eqnarray}
\label{eq:Gausslikelihood}
\ln{P(\hat c_2(d)|B_{1\,\mathrm{Mpc}})} \approx -\frac{x^2}{2 \sigma^2} - \ln\sigma\,,
\end{eqnarray}
where
\begin{eqnarray}
 x=\left(\hat c_2(d) + 2.8\times 10^{-3} \left(\frac{B_{1\,\mathrm{Mpc}}}{\mathrm{nG}}\right)^6 \right),\nonumber
\end{eqnarray}
with $\hat c_2(d)$ being the estimated value from the data $d$, and
$\sigma^2$ corresponding to the variance of its estimation, which includes cosmic variance and noise variance. In Eq.~\eqref{eq:Gausslikelihood}, we have dropped an irrelevant constant term.
We estimate $c_2$ from \texttt{SMICA}, \texttt{NILC}, \texttt{SEVEM}, and \texttt{Commander} foreground-cleaned maps.
The variance is estimated from the realistic {\Planck} simulations for each foreground-cleaning method \citep{planck2014-a11,planck2014-a12}.
To determine $B_{1\,\mathrm{Mpc}}$ and its confidence region, we use the \texttt{CosmoMC} package \citep{cosmomc} as a generic sampler and obtain the posterior probability of $B_{1\,\mathrm{Mpc}}$, given the likelihood.
This analysis yields upper bounds on the amplitude of the non-helical magnetic field component, $B_{1\,\mathrm{Mpc}}$, which are presented in Table~\ref{tab:cL_constraint}. 
Using the \Planck~data constraint on the local-form $f_{{\mathrm{NL}}}$ \citep{planck2014-a19} and $c_0=6/5\,f^{\mathrm{local}}_{{\mathrm{NL}}}$, we impose an additional constraint of $B_{1\,\mathrm{Mpc}} < 5.5\,\mathrm{nG}$ at 95\,\% CL, which is weaker than the $c_2$ constraints.

\begin{table}[tmb]               % table* is a two-column table.  Drop the * for one column.
\begingroup
\newdimen\tblskip \tblskip=5pt
\caption{\Planck\ constraints on the amplitude of the non-helical magnetic field component, $B_{1\,\mathrm{Mpc}}$ [nG], 
from the \texttt{SMICA}, \texttt{NILC}, \texttt{SEVEM}, and \texttt{Commander} foreground-cleaned maps at 95\,\% CL.}
\label{tab:cL_constraint}                            % Label goes here.
\nointerlineskip
\vskip -3mm
\footnotesize
\setbox\tablebox=\vbox{
   \newdimen\digitwidth
   \setbox0=\hbox{\rm 0}
   \digitwidth=\wd0
   \catcode`*=\active
   \def*{\kern\digitwidth}
   \newdimen\signwidth
   \setbox0=\hbox{+}
   \signwidth=\wd0
   \catcode`!=\active
   \def!{\kern\signwidth}
\halign{\hbox to 0.8in{#\leaderfil}\tabskip 1em&
\hfil$#$\hfil&
\hfil$#$\hfil&
\hfil$#$\hfil&
\hfil$#$\hfil\tabskip 0pt\cr
\noalign{\doubleline\vskip 2pt}
\omit \hfil& \texttt{SMICA} & \texttt{NILC} & \texttt{SEVEM} & \texttt{Commander}\cr
\noalign{\vskip 4pt\hrule\vskip 6pt}
$B_{1\,\mathrm{Mpc}}/\mathrm{nG}$&<4.5& <4.9& <5.0& <5.0\cr
\noalign{\vskip 3pt\hrule\vskip 4pt}}}
\endPlancktable                    % ends one-column \halign
%\endPlancktablewide                 % ends two-column \halign
%\tablenote a Footnote a.\par
%\tablenote b Footnote b.\par
\endgroup
\end{table}                        % table* is a two-column table.  Drop the * for one column.
%----------------------------------

The constraint from the passive-scalar bispectrum with \texttt{SMICA} maps is $B_{1\,\mathrm{Mpc}} < 4.5\,\mathrm{nG}$ (95\,\% CL). 
Thus, the addition of the polarized bispectrum leads to an improved constraint
with respect to the previous limit, $B_{1\,\mathrm{Mpc}} < 5.2\,\mathrm{nG}$, from the {\Planck} 2013 analysis \citep{Shiraishi2013,planck2013-p09a}.

\subsection{Magnetically-induced compensated-scalar bispectrum}

Now we derive the magnetically-induced scalar bispectrum on large and intermediate angular scales using a semi-analytical method.
We compute an effective $f_{\mathrm{NL}}$ based on the comparison between the bispectrum and the power spectrum and derive the constraints on the amplitude of the PMF
using {\Planck} measurements.

We derive the magnetically-induced scalar bispectrum on large angular scales for 
compensated initial conditions, basing our analysis on the treatment presented by \cite{Caprini:2009vk}. 
For simplicity we redefine the parameter describing the amplitude of the PMF in this section.
Instead of using the smoothed amplitude we directly use the root mean square value of the field.
This quantity is finite thanks to the sharp cut-off inserted in the PMF power spectrum to model the
small-scale suppression of the field.
The  mean square of the field is then defined as 
\begin{equation}
\langle{B^2({k})}\rangle=\frac{A_B}{2\pi^2}\frac{k_{\mathrm D}^{n_{B}+3}}{n_{B}+3}\,.
\label{mean-squared}
\end{equation}
The magnetically-induced bispectrum on large angular scales depends on the temperature anisotropy on large angular scales and therefore on the Sachs Wolfe signal induced by the PMFs \citep{Caprini:2009vk,2010JCAP...05..022B}.
We use the expression derived by \citet{2009MNRAS.396..523P} and \citet{Caprini:2009vk},
\begin{equation}
\frac{\Theta_\ell^{\mathrm (0)}(\tau_0,{\vec k})}{2\ell+1}\approx\alpha \, \Omega_{B}({\vec k}) \, j_\ell({\vec k}(\tau_0-\tau_{\mathrm{dec}}))\,,
\label{thetaSW}
\end{equation}
where $\Theta_\ell^{\mathrm (0)}$ is the temperature anisotropy, $\Omega_B({\vec k})=\langle B^2 ({\vec k})\rangle /\rho_{\mathrm{rel}}$, and $\tau_0$ and $\tau_{\mathrm{dec}}$ are the conformal time at present and at decoupling, respectively. For simplicity we have used an approximated expression for the initial conditions instead of the exact one, which involves also the Lorentz force \citep{caprini2015}. We therefore introduce a correction factor,
 $\alpha=0.5$, which numerically includes all the contributions that we do not consider in the expression. The above relation holds for
the compensated mode initial conditions.
The CMB bispectrum on large scales can then be written as
\begin{eqnarray}
\langle 
a_{\ell_1m_1} \, a_{\ell_2m_2} \, a_{\ell_3m_3}\rangle&=&
\frac{(4\pi)^3(-i)^{\ell_1+\ell_2+\ell_3}}{(2\ell_1+1)(2\ell_2+1)(2\ell_3+1)}
\times\nonumber\\ &&\int\frac{\dthree{k}\,d^3q\,d^3p}{(2\pi)^9}\,
\nonumber\\&&Y^*_{\ell_1 m_1}(\hat{\vec{k}}) \, Y^*_{\ell_2  m_2}(\hat{\vec{q}}) \, Y^*_{\ell_3 m_3}(\hat{\vec{p}})
\nonumber\\ &&\times\langle
\Theta^{(0)}_{\ell_1}(\tau_0,\vec k) \, \Theta^{(0)}_{\ell_2}(\tau_0,{\vec q}) \, \Theta^{(0)}_{\ell_3}(\tau_0,{\vec p})\rangle\,.
\label{aaa}
\end{eqnarray}
Substituting Eq.~\eqref{thetaSW}, we note that the bispectrum depends on the 3-point correlation function of magnetic energy density,
\begin{eqnarray}
&&\langle\rho_{B}(\vec k) \, \rho_{B}({\vec q}) \, \rho_{B}({\vec p}) 
\rangle=\frac{1}{(64 \pi)^3}\int\frac{\dthree{\tilde{k}}\,d^3\tilde{q}\,d^3\tilde{p}}{(2\pi)^9}\times\nonumber\\
&&~~~~~\langle 
B_i(\tilde{\vec k}) \, B_i({\vec k}-\tilde{\vec k}) \, B_j(\tilde{\vec q}) \, B_j({\vec q}-\tilde{\vec q}) \, B_l(\tilde{\vec p}) \, B_l({\vec p}-\tilde{\vec p})\rangle\,.\nonumber
\label{rhoBrhoBrhoB}
\end{eqnarray}

The characteristic feature of the magnetically-induced bispectrum, generated by compensated modes, is that, contrary to what often happens for inflationary non-Gaussianities, it is not possible 
to identify an a priori dominant geometric configuration. It is therefore necessary to analyse the bispectrum independently of the
geometric configuration. \cite{Caprini:2009vk} derive an approximate expression for the three-point correlation
function of magnetic energy density, which is independent of the geometric configuration and is tested against the analytic results for the flattened case.
Using this expression, we derive the magnetically-induced bispectrum and specify a geometric configuration only after integrating the magnetic energy density bispectrum in $k$-space.
We show the results for the case corresponding to a local $f_{\mathrm{NL}}$.
The magnetically-induced bispectrum for the nearly scale-invariant case, $n_{B} =-2.9$, is given by
\begin{eqnarray}
f^{\mathrm{eff}}_{\mathrm{NL}}&\approx& \frac{3\pi^9\,\alpha^3}{36\,{\cal A}^2}\frac{n_{B}(n_{B}+3)^2}{2n_{B}+3} 
\frac{\langle{B^2}\rangle^3}{\rho_{\mathrm{rel}}^3} \nonumber\\
&=& 1851\,\frac{n_{B}(n_{B}+3)^2}{2n_{B}+3}\, \left(\frac{\langle{B^2}\rangle}{(10^{-9}\,
  \mathrm{G})^2}\right)^3\,.
\end{eqnarray}
For $n_{B}>-1$, we find
\begin{eqnarray}
f^{\mathrm{eff}}_{\mathrm{NL}}&\approx& \frac{\pi^9\,\alpha^3}{288\,{\cal 
A}^2}\frac{(n_{B}+3)^3}{n_{B}+1}\frac{\langle{B^2}\rangle^3}{\rho_{\mathrm{rel}}^3}
\left(\frac{\ell_{\mathrm{max}}}{\ell_{\mathrm D}}\right)^4\, \frac{1}{\log(\ell_{\mathrm{max}}/\ell_{\mathrm{min}})}\nonumber\\
&=& 5.9\times 10^{-3}\,\frac{(n_{B}+3)^3}{n_{B}+1}\,
\left(\frac{\langle{B^2}\rangle}{(10^{-9}\,\mathrm{G})^2}\right)^3.
\end{eqnarray}
For the specific case $n_{B}=-2$ it is
\begin{eqnarray}
f^{\mathrm{eff}}_{\mathrm{NL}}&\approx& \frac{5\pi^{10}\,\alpha^3}{288 \,{\cal A}^2}\frac{\langle{B^2}\rangle^3}{\rho_{\mathrm{rel}}^3} 
\left(\frac{\ell_{\mathrm{max}}}{\ell_{\mathrm D}}\right)^3\, \frac{\log(\ell_{\mathrm D}/\ell_{\mathrm{
max}})}{\log(\ell_{\mathrm{max}}/\ell_{\mathrm{min}})} \nonumber\\
&=& 5.19\,
\left(\frac{\langle{B^2}\rangle}{(10^{-9}\,\mathrm{G})^2}\right)^3\,,
\end{eqnarray}
where we assume $\cal A=18.98\times 10^{-9}$ as the amplitude of the primordial gravitational potential power spectrum ($\ell^2 C_\ell ={\cal A}/\pi$). 
In all numerical estimates we have taken $\ell_{\mathrm D}=k_{\mathrm D}\ \tau_0= 3000$, $\ell_{\mathrm{max}}= 750$, and $\ell_{\mathrm{min}}= 10$.
The {\Planck} limit on local $f_\mathrm{NL}$ is given in \cite{planck2014-a19}. We use the result of the \texttt{SMICA} KSW $T$+$E$ ISW-lensing-subtracted analysis, namely $f_{\mathrm{NL}}<5.8$ at 68\,\% CL. This limit on $f_\mathrm{NL}$ translates into constraints on the PMF amplitude of  
$\sqrt{\langle B^2\rangle}< 1.8\,$nG for $n_{B} =-2.9 $, $\sqrt{\langle B^2\rangle}< 1.0$\,nG  for 
$n_{B}= -2 $, and $\sqrt{\langle B^2\rangle}< 1.7 $\,nG for $n_{B}= 2$. The constraints on the smoothed amplitude of the field are $B_{1\,\mathrm{Mpc}}< 3$\,nG for 
$n_{B}=-2.9 $,  $B_{1\,\mathrm{Mpc}}< 0.07$\,nG  for $n_{B}= -2 $, and  $B_{1\,\mathrm{Mpc}}< 0.04$\,nG  for $n_{B}=2$. These results show how the constraints are competitive with, in addition to being complementary to, 
the ones given by the CMB angular power spectrum.

\section{Faraday rotation}
\label{sec:Faraday}
\subsection{Constraints on PMFs from the Faraday rotation power spectrum}
The presence of a PMF at the last scattering surface induces a rotation of the
polarization plane of the CMB photons \citep{kosowsky96}. This effect is known as Faraday rotation
(hereafter FR).
The Faraday depth $\Phi$ is proportional to the integral along the line of sight of the magnetic field
component along this direction, $B_\parallel$, and the thermal electron density, $n_\mathrm{e}$, i.e.,
\begin{equation}
\Phi = K \int{ n_\mathrm{e}(x,\hat{\vec{n}}) \, B_\parallel(x,\hat{\vec{n}}) \, dx}.
\end{equation}
The constant is $K = 0.81$ rad\,m$^{-2}$\,pc$^{-1}$\,cm$^{3}$\,$\mu$G$^{-1}$ $=2.6 \times 10^{-26}$\,rad\,nG$^{-1}$. The unit of the Faraday depth $\Phi$ is rad\,m$^{-2}$. In this equation, $\hat{\vec{n}}$ is the unit vector in the line-of-sight direction.

Here we analyse the impact of FR on the polarized CMB power
spectra. We assume that the magnetic field is generated at some
pre-decoupling epoch. We do not consider the generation mechanism itself, but only investigate
the observable effects caused at recombination. Several works have derived the
modification of the Boltzmann equation for the Stokes parameters in the presence of a
homogeneous PMF \citep[see, e.g.,][]{scoccola04} and for a
stochastic distribution \citep[see, e.g.,][]{kosowsky05}.
Here we explore the case of a stochastic distribution. 

As discussed above, a PMF may induce scalar, vector, and tensor perturbations. At
recombination, FR mixes the signatures of different perturbations.
Some previous attempts to constrain FR signatures in the WMAP power spectrum
are presented by
\cite{kahniashvili09} and \cite{pogosian11}. Both assume that $E$-modes are
converted into $B$-modes via FR. They obtain magnetic field strength limits of
$B_{1\,\mathrm{Mpc}} \la 100$\,nG  and suggest an almost scale-invariant spectrum, i.e., $n_\mathrm{B} \approx -2.9$
for a power-law distribution. 

As in previous works, we assume a PMF distribution described by a power law
as in Eq.~\eqref{gaussian}.
Note that any
helical part of the field does not contribute to the Faraday
rotation \citep{campanelli04}.
The generation of
magnetically-induced $B$-modes through FR of $E$-modes is described by \citep{kosowsky05}
\begin{eqnarray}
C_{\ell}'^{BB} &=& N_{\ell}^{2} \sum_{\ell_{1}\ell_{2}} \frac{(2\ell_{1} +
1)(2\ell_{2} + 1)}{4 \pi (2\ell + 1)}\times\nonumber\\
&& N_{\ell_{2}}^{2} \, K(\ell,\ell_{1},\ell_{2})^{2} \,
C_{\ell_{2}}^{EE} \, C_{\ell_{1}}^{\alpha} \, \left(C_{\ell_{1}0\ell_{2}0}^{\ell0}\right)^{2}
\label{eq:cl_bb}
\end{eqnarray}
and the rotation of primordial $B$-modes into magnetically-induced $E$-modes by
\begin{eqnarray}
C_{\ell}'^{EE} &=& N_{\ell}^{2} \sum_{\ell_{1}\ell_{2}} \frac{(2\ell_{1} +
1)(2\ell_{2} + 1)}{4 \pi (2\ell + 1)}\times\nonumber\\
&&N_{\ell_{2}}^{2} \, K(\ell,\ell_{1},\ell_{2})^{2} \,
C_{\ell_{2}}^{BB} \, C_{\ell_{1}}^{\alpha} \, \left(C_{\ell_{1}0\ell_{2}0}^{\ell0}\right)^{2},
\label{eq:cl_ee}
\end{eqnarray}
where $N_{\ell}= (2(\ell-2)! / (\ell+2)!)^{1/2}$ is a normalization factor,
$K(\ell,\ell_{1},\ell_{2}) = -1/2 \, (L^2 + L_1^{2} + L_2^2 - 2L_1L_2 - 2L_1L +2L_1
- 2L_2 - 2L)$ with $L\equiv \ell(\ell+1)$, $L_1\equiv \ell_1(\ell_1+1)$,
$L_2\equiv \ell_2(\ell_2+1)$, and $C_{\ell_{1}0\ell_{2}0}^{\ell0}$ is a 
Clebsch-Gordan coefficient.
The power spectrum of the rotation angle is related to the
one of the Faraday depth through
\begin{equation}
\label{eq:cl_phi}
C_{\ell}^{\alpha} = \nu_{0}^{-4} C_{\ell}^{\Phi},
\end{equation}
where $\nu_{0}$ is the observed frequency, and
\begin{equation}
\label{eq:cl_rm}
C_{\ell}^{\Phi} \approx \frac{9\ell(\ell+1)}{(4\pi)^{3} e^{2}} \frac{B_{1\,\mathrm{Mpc}}^{2}}{\Gamma( n_{B}+3 / 2)} \left( \frac{\lambda}{\tau_{0}} \right)^{n_{B}+3} \int_{0}^{x_\mathrm{D}}{dx \, x^{n_{B}} \, j_{\ell}^{2}(x)}.
\end{equation}
Here, $x_\mathrm{D} = \tau \, k_\mathrm{D}$, where $\tau$ is the conformal time and $k_\mathrm{D}$ is given by Eq.~\eqref{kd_def}.

In Eqs.~\eqref{eq:cl_bb} and \eqref{eq:cl_ee}, $C_{\ell}^{EE}$ and $C_{\ell}^{BB}$ are the
primordial power spectra, whereas $C_{\ell}'^{EE}$ and $C_{\ell}'^{BB}$
are the ones including the effect of Faraday rotation, i.e., the
observed ones.
We use the observed $E$-mode spectrum $C_{\ell}'^{EE}$ at 70\,GHz as a
proxy for the primordial one, $C_{\ell}^{EE}$, in Eq.~\eqref{eq:cl_bb} to calculate
predicted $B$-mode spectra to be compared with the observed one.

\begin{figure}
\includegraphics[width=88mm]{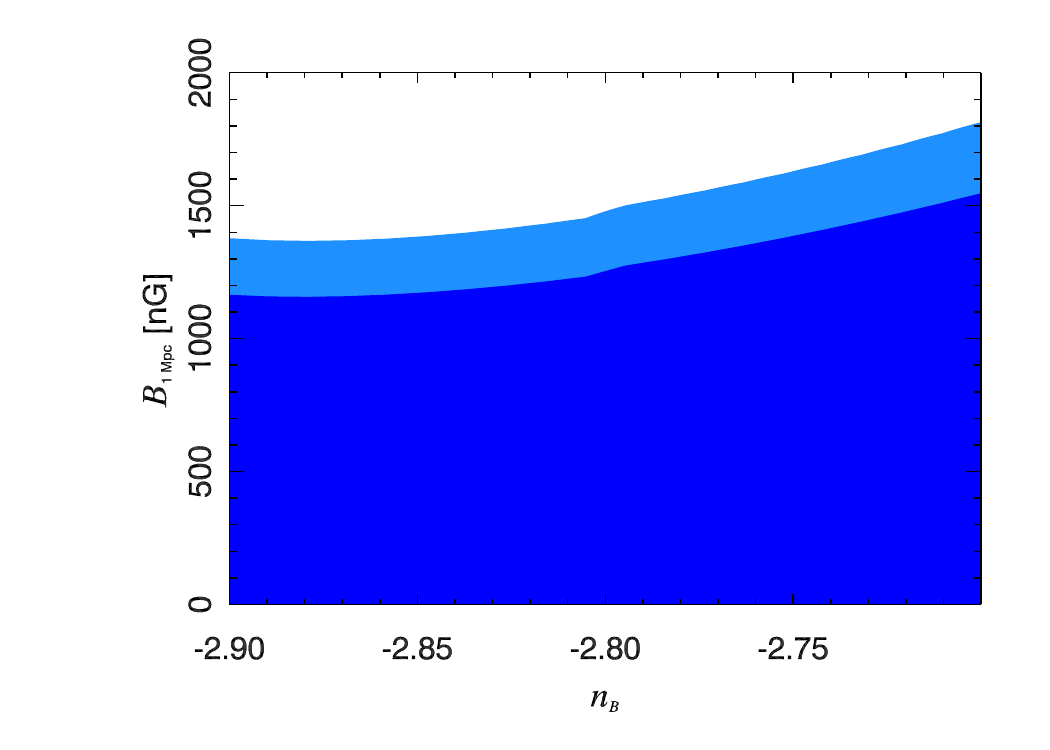}
\caption{Probability contours of PMF strength vs.\ spectral index of the PMF power spectrum as constrained by the 70\,GHz observations.}
\label{fig:contour}
\end{figure}

The 70\,GHz observations give $B_{1\,\mathrm{Mpc}} <$ (1040, 1380)\,nG (68\,\%, 95\,\% CL). The reduced $\chi^2$ is 1.35. The magnetic field spectral index for the stochastic distribution remains unconstrained. 
In Fig.~\ref{fig:contour}, we show the probability contours derived from the 70\,GHz data for the magnetic field strength $B_{1\,\mathrm{Mpc}}$ and the spectral index of the PMF power spectrum, $n_{B}$. 
The upper bounds obtained from the FR analysis are very high compared to the other methods.
They are also slightly weaker than those previously obtained from FR analyses with WMAP-5 \citep{kahniashvili09}, WMAP-7 \citep{pogosian11}, and WMAP-9 \citep{ruiz-granados14} data. 

%%%%%%%%%%%%%%%%%%%%%%%%%%%%%%%%%%%%%%%%%%%%%%%%%%%%%%%%%%%%%%%%%%%%%%%%%%%%%%%%%%%%%%%%%%

\subsection{Robustness of the results in the presence of foregrounds}

The magnetic fields of the Milky Way contribute to the net Faraday rotation.
Although the precise geometry of these magnetic fields remains
uncertain (see, e.g., \citealt{ruiz-granados10} and \citealt{jansson12}), the Galactic Faraday depth could be a foreground for the primordial Faraday depth, at least on large scales.

To quantify the impact of the Galactic FR on the detection of primordial magnetic fields, we use Galactic observations of polarized synchrotron emission at 1.4\,GHz and 23\,GHz and the synthesized all-sky Faraday rotation map derived from extragalactic radio source emission provided by \cite{oppermann14}. In addition, we use simulations of the Galactic Faraday rotation obtained by using an axisymmetric Galactic magnetic field model for the halo field described by \cite{ruiz-granados10}. Maps of Stokes $Q$ and $U$ are provided by \cite{wolleben06} at 1.4\,GHz and by \cite{bennett13} at 23\,GHz. Both frequencies are dominated by polarized synchrotron emission from within the Milky Way and are used to obtain the Galactic Faraday depth \citep[see][for details]{ruiz-granados14b}. For computing the power spectrum of the FR coming from simulations and observations at 1.4 and 23\,GHz, we use the polarization proccessing mask provided by WMAP-9\footnote{\url{http://lambda.gsfc.nasa.gov/product/map/dr5/m\_products.cfm}}.

\begin{figure}
\includegraphics[width=88mm]{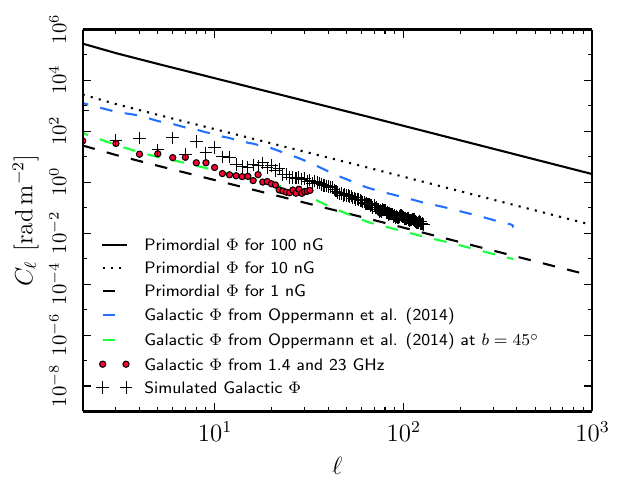}
\caption{{Power spectra of the primordial Faraday depth $\Phi$ for magnetic field strengths of 100 (solid line), 10 (dotted line), and 1\,nG (dashed line), the Galactic Faraday depth from the all-sky Faraday map of \cite{oppermann14} (blue dashed line), the Galactic Faraday depth from \cite{oppermann14} at 45\deg\ latitude (green dashed line), the Galactic Faraday depth derived from Galactic emission at 1.4 and 23\,GHz (red circles), and the Galactic Faraday depth for a Galactic magnetic field model (black crosses).}}
\label{fig:impact}
\end{figure}

In Fig.~\ref{fig:impact}, we show the power spectra of Galactic $\Phi$ derived from polarized measurements at 1.4\,GHz and 23\,GHz, simulations, and for the all-sky Faraday rotation map provided by \cite{oppermann14}\footnote{\url{http://www.mpa-garching.mpg.de/ift/faraday}}.
The fluctuations in the Galactic Faraday sky are not isotropic. Therefore, their statistics are not completely described by a power spectrum. \cite{oppermann14} model the Galactic Faraday depth as the product of an isotropic Gaussian random field and a latitude-dependent function. In Fig.~\ref{fig:impact}, we show two angular power spectra derived from the results of \cite{oppermann14}. For the first one, we generate Gaussian realizations from their angular power spectrum, multiply them with their latitude profile, and pass them through the \texttt{anafast} routine of \texttt{HEALPix}. Averaging the result over 1000 realizations gives the blue dashed line in Fig.~\ref{fig:impact}. For comparison, we also show the angular power spectrum of \cite{oppermann14} multiplied with the square of their profile function at a latitude of $|b|=45\deg$, which gives the strength of the foreground Faraday rotation at a typical latitude used for CMB analysis.
We plot also the power spectrum of the primordial Faraday rotation for PMFs of strength 1, 10, and 100\,nG, respectively, and a spectral index of $n_B = -2.9$.

Our main conclusion, as indicated by Fig.~\ref{fig:impact}, is that observations of the Galactic Faraday depth show that primordial magnetic field strengths lower than 10\,nG would require a detailed knowledge of the Galactic FR, at least at multipoles lower than $\ell \approx 50$, where the foreground rotation would be dominant.
The exact detection limit for PMFs achievable through a study of Faraday rotation depends on Galactic latitude and on the extent to which the correction for foreground rotation is possible.

The upper limit found from the FR analysis is $B_{1\,\mathrm{Mpc}} < 1380$\,nG and no restriction can be obtained for the spectral index $n_B$. 
In any case, our FR constraint on PMFs is well above the expected contamination level by Galactic FR, and therefore this contamination is currently not an issue. This is consistent with the prediction by \cite{de2014} that the Galactic Faraday depth would not be measurable with {\Planck}.

\section{Constraints on PMFs from Alfv\'en waves}
\label{alfven}

Here we investigate the signature statistical anisotropy induced by Alfv\'en waves, which delivers yet another constraint on PMFs.
In \cite{planck2013-p09a} we constrained the Alfv\'en waves in the early Universe, where some arbitrary origin (including stochastic PMFs) was assumed for primordial vector perturbations. 
Given no evidence for Alfv\'en waves from that analysis, we now consider stochastic PMFs 
as the source of primordial vector perturbations and constrain an average background magnetic field and the energy density of stochastic PMFs\footnote{For this analysis we follow 
the approximation of considering the stochastic background of PMFs as split into an average background field, which emulates the average effect of the stochastic fields,
and the stochastic fields at the perturbative level.}. 
PMFs may produce  Alfv\'en waves in the early Universe, which leave observable imprints on the CMB via the Doppler and
integrated Sachs-Wolfe effects.  
\cite{DKY} show that Alfv\'en waves in the early Universe generate a
fractional CMB anisotropy
\begin{eqnarray}
\frac{\Delta T}{T_0}(\hat {\vec{n}},\vec{k}) \approx \vec{n}\cdot
\vec{\Omega}(\vec{k},\tau_{\mathrm{last}})  
= \vec{n}\cdot \vec{\Omega}_0(\vec{k})\, \varv_A \, k\,\tau_{\mathrm{last}}\,\hat {\vec{n}}_0\cdot \vec{k}\,, 
\end{eqnarray}
where $\vec{k}$ denotes a Fourier mode vector, $\hat {\vec{n}}$ a sky direction, $\hat {\vec{n}}_0$ the unit vector in the direction of the homogeneous background magnetic field $\bar {\vec{B}}$, and
$T_0$ is taken again as $2.7255\,\mathrm{K}$ \citep{Fixsen2009}. Here $\vec{\Omega}(\vec k,\tau_{\mathrm{last}})$ and $\vec{\Omega}_0(\vec{k})$  denote the gauge invariant
linear combination of vector perturbations at last scattering and at an initial time, respectively.
In this analysis, we assume a non-helical stochastic PMF, $\vec{B}$, to be the sole source of initial vector fluctuations, $\vec{\Omega}_0=\varv_A/\bar B\,\vec{B}$ \citep{DKY,2008PhRvD..78f3012K}. 
The Alfv\'en wave velocity, $\varv_A$, is given by \citep{DKY}
\begin{eqnarray} 
\varv_A =\frac{\bar B}{2\sqrt{\pi(\rho_{r}+p_{r})}}\approx 2.2\times10^{5}\,\mathrm{m}\,\mathrm{s}^{-1}\,\frac{\bar B}{1\,\mathrm{nG}}\,,\label{v_A}
\end{eqnarray}
where  $\rho_{r}$ and $p_{r}$ are the co-moving density and pressure of the photons. 

\citet{2008PhRvD..78f3012K} show that Alfv\'en waves in the early Universe produce
correlations between harmonic modes separated by $\Delta \ell=0, \pm
2$, and $\Delta m=0, \pm 1, \pm 2$. We give the explicit form of the correlations in Appendix \ref{anisotropy_alfven}.
Investigating these imprints, we impose a constraint on the Alfv\'en waves in the early Universe.  
In the weak Alfv\'en wave limit, the CMB data log-likelihood $\mathcal{L}$ can be expanded as
\begin{eqnarray}
\mathcal{L}&\approx&\left.\mathcal{L}\right|_{h=0}+\left.\frac{\partial \mathcal{L}}{\partial h}\right|_{h=0}h +\left. \frac{1}{2}\frac{\partial^2 \mathcal{L}}{\partial h^2}\right|_{h=0}h^2\nonumber+\mathcal O(h^3)\,,\label{like}
\end{eqnarray}
where $h=B_{1\,\mathrm{Mpc}}^2 \, \varv^4_A/\bar B^2$.
The first term on the right hand side is simply equal to the likelihood of the standard cosmological model and 
the first and second derivatives of the likelihood are obtained by
\begin{eqnarray}
\frac{\partial \mathcal{L}}{\partial h}&=&\mathcal H -\langle \mathcal H \rangle,\;\;\;\;\;\;\;\frac{\partial^2 \mathcal{L}}{\partial h^2}=-\langle \mathcal H^2\rangle +\langle \mathcal H \rangle  \langle \mathcal H \rangle\,,\label{L_der}
\end{eqnarray}
where
$\langle\ldots \rangle$ denotes the ensemble average of signal and noise.
The Hessian $\mathcal H$ is given by
\begin{eqnarray}
\mathcal H=\frac{1}{2}\left [\tens{C}^{-1} {\vec{a}} \right]^{\dagger} \frac{\partial \tens{C}}{\partial h} \left [\tens{C}^{-1} {\vec{a}} \right]\,,\label{H}
\end{eqnarray}
where
$\vec{a}$ is the vector consisting of the spherical harmonic
coefficients, $a_{\ell m}$, of the CMB anisotropy data, and $\tens C$ is their
covariance matrix.

In our analysis, we consider the foreground-cleaned {\tt SMICA} map, where we apply
the  common mask \citep{planck2014-a12}.
We assume the fiducial \Planck\ cosmological model and use realistic \Planck\ simulations to estimate the ensemble average values for
signal and noise, as required in Eq.~\eqref{L_der}. The quantity $\tens{C}^{-1} {\vec{a}}$, required in Eq.~\eqref{H}, is determined via the messenger
field method \citep{elsner2013}.
Some of the parameters ($n_B, \theta_B,\phi_B$) influence the signature correlation nonlinearly.
Due to these nonlinear parameters, we use the \texttt{CosmoMC} package \citep{cosmomc} as a generic sampler for the log-likelihood in Eq.~\eqref{like} and obtain
the posterior probability for the Alfv\'en wave parameters $\{B_{1\,\mathrm{Mpc}}^2/\bar B^2\,\varv^4_A, n_B, \theta_B, \phi_B \}$.
As discussed previously, we assume the initial vector fluctuations to be entirely sourced by a non-helical stochastic PMF, $\vec{B}$.
In Table~\ref{tab:alfven}, we show upper bounds on this combination of parameters at 68\,\% and 95\,\% CL, after
marginalizing over the spectral index $n_B$ and the direction $\theta_B,\phi_B$. 

Other theoretical models with correlations across multipoles with $\Delta \ell=\pm 1, \pm 2$  are investigated in \citet{planck2014-a18} and \citet{planck2014-a24}.  
The Planck data show no evidence in favour of these models.

%--------------------------------
\begin{table}[tmb] % table* does a two-column table
\begingroup % this + \endgroup at the end keep table things local
\newdimen\tblskip \tblskip=5pt
\caption{
\Planck\ constraints on the Alfv\'en wave amplitude $B_{1\,\mathrm{Mpc}}^2/\bar B^2\,\,\varv^4_A$.
The Alfv\'en wave velocity $\varv_A$ is normalized to the speed of light.}%@
\label{tab:alfven}
%\nointerlineskip
\vskip -3mm
\footnotesize % good font size for a table, but can be changed
\setbox\tablebox=\vbox{ %
\newdimen\digitwidth % see \S\,16.12 for the purpose of the next 10 lines
\setbox0=\hbox{\rm 0}
\digitwidth=\wd0
\catcode`*=\active
\def*{\kern\digitwidth}
\newdimen\signwidth
\setbox0=\hbox{+}
\signwidth=\wd0
\catcode`!=\active
\def!{\kern\signwidth}
\halign{\hbox to 1.0in{#\leaderfil}\tabskip=1.5em& \hfil$#$\hfil& \hfil$#$\hfil\tabskip 0pt\cr
% template goes here. See examples.
\noalign{\doubleline}
%  \noalign{\vskip -1pt}
% \omit&\multispan2 \hfil Confidence level \hfil\cr
% \noalign{\vskip -6pt}
% \omit&\multispan2\hrulefill\cr
\omit \hfil{Confidence Level}\hfil& 68\,\%& 95\,\%\cr
% heading goes here. See examples.
\noalign{\vskip 3pt\hrule\vskip 3pt}
$B_{1\,\mathrm{Mpc}}^2/\bar B^2\,\varv^4_A$& <3.4\times 10^{-7}& <1.7\times 10^{-5}\cr 
% % table lines go here. See examples.
\noalign{\vskip 3pt\hrule\vskip 3pt}
}}
\endPlancktable % this command is defined in Planck.tex.
\endgroup
\end{table}

\section{Conclusions}
\label{sec:conclusions}

\subsection{Methodology}

In this paper, we have presented constraints on a stochastic background of primordial magnetic fields using {\Planck} data. PMFs may have left different types of imprints on the CMB. This is why the CMB can be regarded as one of the best laboratories for investigating and constraining PMFs. The richness of {\Planck} data, which provide several different probes based on the statistics of CMB anisotropies, allows us to constrain PMFs using different methods, deriving constraints that are complementary to each other.
Aiming for a broad perspective on PMFs, we have taken advantage of these different possibilites offered by {\Planck}.
In particular, we have exploited the impact of PMFs on the CMB anisotropy angular power spectra in temperature and polarization, both through the magnetically-induced modes and their effect on the ionization history of the Universe. In addition we have considered the Faraday rotation induced by them on the CMB polarization. Beyond the two-point statistics probed by the angular power spectrum, we have investigated higher-order statistical moments of the CMB imprinted by PMFs. In particular, we have analysed the CMB bispectrum for the presence of magnetically-induced passive and compensated modes. Finally, we have considered the correlations between harmonic modes induced by PMFs.

\subsection{Constraints for non-helical fields from the angular power spectra}

The constraints based on the CMB angular power spectra have been derived using the {\Planck} likelihood. The general analysis, which considers only the contributions from compensated modes, provides the constraint $B_{1\,\mathrm{Mpc}}<4.4$\,nG at 95\,\% CL, with positive spectral indices constrained to lower amplitudes of the fields, which we have shown to be robust under the inclusion of high-$\ell$ polarization data. In fact, the impact of PMFs on the $EE$ and $TE$ polarization spectra is negligible compared to the dominant contribution given by the magnetically-induced vector modes on the $TT$ spectra on small angular scales. 
The inclusion of the passive tensor contribution does not improve the constraints on the amplitude but affects the posterior distribution for the PMF spectral index. The results confirm that CMB data constrain positive spectral indices to smaller amplitudes. The inclusion of the contribution of passive modes, thanks to their sensitivity to the slope of the PMF spectrum, strongly disfavours nearly scale-invariant magnetic spectra for amplitudes greater than very few nanogauss (see Fig.~\ref{fig:degenerate}).
This is due to the large contribution that magnetically-induced passive tensor modes with nearly scale-invariant spectra give to the CMB anisotropies on large angular scales.

We have also performed an analysis using both the \Planck\ 2015 data and the BICEP2/\textit{Keck}-\Planck\ cross-correlation. The results are fully compatible with the analysis based on \Planck\ data alone, with only a slightly higher upper limit on the PMF amplitude.

Our likelihood analysis for PMF properties is sensitive to CMB foreground residuals in the data, since these contribute to small angular scales in the CMB spectra as well. 
In our analysis we have used the foreground residual treatment provided by the {\Planck} likelihood. 
Since some of the foreground models assume an angular power spectrum with a shape similar to the one given by magnetically-induced perturbations, we have investigated the issue of 
possible degeneracies between PMFs and foreground residuals. In particular, we have noticed that there is a degeneracy with the Poissonian terms modelling unresolved point 
sources at 100, 143, 143$\times$217 and 217\,GHz. Fortunately, other foreground contributions do not show any degeneracy thanks to their different spectral shapes.
We have further tested how severely foreground residuals might affect the obtained constraints on PMF amplitudes. Since PMFs that are consistent with the data do not significantly 
affect other cosmological parameters, we have fixed the amplitudes of foreground residuals to the values obtained under the {\Planck} $\Lambda$CDM cosmology. 
The limit in this case, $B_{1\,\mathrm{Mpc}}<3.0$\,nG at 95\,\% CL, is slightly tighter than the limit obtained without fixing the foreground residual amplitudes, $B_{1\,\mathrm{Mpc}}<4.4$\,nG at 95\,\% CL. 
This test has no statistical significance in the constraints on PMFs but illustrates the impact foreground residuals can have on constraints on the PMF amplitude.

Our new constraints are compatible with previous constraints from other experiments and from the previous {\Planck} release \citep{planck2013-p11}. 
The slightly higher upper limits with respect to the 2013 {\Planck} release are due to changes in the 2015 data. 
In particular, the changed calibration and  slightly different slope of the power spectrum of cosmological perturbations allow for stronger PMFs, with possible contributions from the different foreground residual treatment.

\subsection{Constraints on maximally helical PMFs}

We also constrain maximally helical PMFs. We restrict our analysis to the maximally helical case because of the absence of $TB$ and $EB$ information in the {\Planck} 2015 high-$\ell$ likelihood. Maximal helicity decreases the amplitude of the magnetically generated CMB fluctuations and, as a consequence, we obtain $B_{1\,\mathrm{Mpc}}<5.6$\,nG at 95\,\% CL in this case.
% prevents us from constraining the degree of helicity.} Maximal helicity decreases the amplitude of the magnetically generated CMB fluctuations and, as a consequence, we obtain $B_{1\,\mathrm{Mpc}}<5.6$\,nG at 95\,\% CL in  this case. 

\subsection{Selected scenarios}

We have further investigated two specific PMF models of interest: causally generated fields with a spectral index of $n_{B}=2$ and fields with an almost scale-invariant power spectrum with $n_{B}=-2.9$. The constraints for these extreme cases are $B_{1\,\mathrm{Mpc}}^{n_{B}=2}<0.011$\,nG and $B_{1\,\mathrm{Mpc}}^{n_{B}=-2.9}<2.1$\,nG at 95\,\% CL, respectively.

The impact of PMFs on the ionization history of the Universe directly affects the CMB temperature and polarization power spectra. 
In  particular, we have also considered the main dissipative effects operating during and after recombination, namely ambipolar diffusion 
and energy cascading in MHD turbulence in the prediction for the CMB spectra in temperature and polarization. 
These modify the primary CMB power spectra in addition to the gravitational contributions of the magnetic modes. 
For the nearly scale-invariant case we have obtained the constraint $B_{1\,\mathrm{Mpc}}^{n_{B}=-2.9}<0.9$\,nG at 95\,\% CL. This limit is tighter than when neglecting the effect on the ionization history. However, uncertainties related to the modelling of the heating mechanism \citep[see discussion by][]{Chluba2014PMF} suggests that further investigation of this promising avenue is needed.

\subsection{Non-Gaussianity-based constraints}

For the non-Gaussianity analyses we have focused on the passive modes with a nearly scale-invariant power spectrum, $n_B=-2.9$, and the compensated scalar modes.
These are the dominant contributions on large angular scales, where the non-Gaussianity analyses are performed.   

In our first CMB non-Gaussianity analysis, we have considered passive tensor modes for PMFs with nearly scale-invariant spectra. 
These contribute predominantly to the CMB fluctuations on large angular scales. For this case, we have calculated the resulting CMB bispectrum and compared it with the observational limit. 
We have used a bimodal decomposition to estimate the amplitude of the {\Planck} bispectrum in the squeezed configuration, in which the observational limit on the amplitude 
of the bispectrum can be translated into a constraint on the amplitude of PMFs. 
Using the temperature bispectrum we obtain $B_{1\,\mathrm{Mpc}}^{n_{B}=-2.9}<2.8$\,nG for fields that were generated at the GUT phase transition.

For our second non-Gaussianity analysis we have used a different approach to the magnetically-induced bispectrum. We have considered the passive contributions by tensor and scalar modes 
for nearly scale-invariant fields, but instead of using the bispectrum amplitude we have used the local type of non-Gaussianity induced by PMFs, considering only the non-helical part. 
The local bispectrum from multipoles with $\Sigma_{n=1}^3 \ell_n=$ even (where $\ell_n$ stands for the multipoles $\ell_1\,,\ell_2\,,\ell_3$ normally used to express the bispectrum, see for example Eq.~\ref{aaa})
has been decomposed with coefficients determined by the amplitude of the PMF. A likelihood estimation allows us to constrain the PMF amplitude with CMB foreground-cleaned maps. We have applied this method to all four component separation methods used by {\Planck}. The {\tt {SMICA}} map, which is expected to contain the least foreground residuals \citep{planck2014-a11}, gives the constraint $B_{1\,\mathrm{Mpc}}^{n_{B}=-2.9}<4.5$\,nG, improving previous constraints derived from the scalar bispectrum for WMAP data.

Our third non-Gaussianity analysis focuses on the compensated scalar modes. In this case, we have used an analytic estimate of the bispectrum on large angular scales. We have used an improved estimate of the source term \citep{caprini2015} with respect to previous results \citep{Caprini:2009vk}. This analytic estimate can be compared with the observed local $f_{\mathrm{NL}}$ from {\Planck}, giving $B_{1\,\mathrm{Mpc}}^{n_{B}=-2.9}<3.0$\,nG.

The results from the different non-Gaussianity analyses (although coming from different methods) are all consistent and at the level of those derived with the likelihood analysis using only the CMB angular power spectra.

\subsection{Constraints from Faraday rotation}

We have further considered the effects of Faraday rotation on the primary CMB polarization anisotropies. In this context, we have used the $EE$- and $BB$-polarization power spectra. We have derived the constraints on the PMF amplitude using a $\chi^2$ analysis based on the LFI 70\,GHz low-$\ell$ ($\ell<30$) polarization power spectra.
The resulting constraint is $B_{1\,\mathrm{Mpc}}<1380$\,nG. 
The upper limits from Faraday rotation are larger than those derived from magnetically-induced perturbations, thermal effects, and non-Gaussianity. On one hand, the Faraday rotation signal rapidly vanishes with increasing frequency (see Eq.~\ref{eq:cl_phi}) and thus is strong only for lower frequencies. On the other hand, since the $BB$ spectrum is the results of the rotation of the $EE$ spectrum, it has a stronger contribution on smaller angular scales. Our analysis includes only the low multipoles of the 70 GHz data, where the signal is lower due to its spectral shape.
The combination of data sets and type of signal results in less stringent constraints. However, even with the restricted subset of {\Planck} data available, the constraints are only slightly weaker than derived in previous analyses performed with WMAP \citep{kahniashvili09, pogosian11,ruiz-granados14}.

To estimate the impact of Galactic Faraday rotation on the results, we have analysed synthetic Galaxtic Faraday maps as well as radio synchrotron data at 1.4 GHz and 23 GHz. We have accounted for the fact that the signal is not isotropic on the sky but depends on the latitude of the observations. We have derived an estimate of the power spectrum for the Faraday depth, shown in Fig.~\ref{fig:impact}, and compared it with the predictions for different values of the PMF amplitude. Our results show that the threshold for which the Galactic contamination may become relevant is around 10 nG. This amplitude is much below our current constraints, which can therefore be considered clean from Galactic contamination.

\subsection{Constraints on Alfv\'en waves}

To complete the round of different types of analyses involving different probes, we have investigated the correlation induced between different modes in harmonic space by Alfv\'en waves produced by the presence of PMFs \citep{2008PhRvD..78f3012K}. This correlation has not been used to constrain the PMF amplitude directly, but the Alfv\'en wave parameter, which is a combination of the stochastic background amplitude, the Alfv\'en velocity, and the assumed mean background field.
We have again used the {\tt {SMICA}} foreground-cleaned map to derive the upper limit $B_{1\,\mathrm{Mpc}}^2\,\varv^4_A/\bar B^2<1.7\times 10^{-5}$. From this constraint we deduce that the data do not show any evidence of Alfv\'en waves, a conclusion that was also reached in previous analyses \citep{planck2013-p09a} carried out with more generic assumptions on the origin of the Alfv\'en waves. The absence of Alfv\'en waves is also compatible with the results from other models for the harmonic space correlations that are not related to the PMFs \citep{planck2014-a18,planck2014-a24}.

\subsection{Concluding summary}

The results presented show that CMB anisotropies are one of the best probes for investigating the nature of PMFs. 
The \Planck\ 2015 data offer the possibility to use the PMF's signatures either in the angular power spectra in temperature and polarization or in higher-order statistics, where both measurements can be tackled by different methodologies. All the independent constraints we obtain are consistent with each other.

The {\Planck} 2015 data constrain the PMF amplitude at the nanogauss level. Different signatures are sensitive to different contributions and may be optimal for specific types of PMF. In particular, the analysis that uses the gravitational impact of PMFs on the CMB angular power spectra is dominated by the compensated vector contribution on small angular scales and therefore is able to constrain PMFs without any assumptions on their generation mechanism. 
On the contrary, two of the three analyses of non-Gaussianities are dominated by passive tensor modes, which can provide significant constraints only for nearly scale-invariant PMFs.

The future of both classes of methods, the angular power spectra and the non-Gaussianities, is bright, but three avenues are particularly promising. 
The helicity of PMFs will be constrained by $TB$ and $EB$ cross-correlations, which will be included in the next {\Planck} release.
The study of the PMF's impact on the ionization history is expected to further improve with future \Planck\ polarization data.
Non-Gaussianities are a distinctive signature of PMFs and further studies may provide more and more refined predictions of the magnetically-induced passive and compensated CMB bispectra and trispectra, which will improve the future {\Planck} analyses.

\begin{acknowledgements}
The Planck Collaboration acknowledges the support of: ESA; CNES and CNRS/INSU-IN2P3-INP (France); ASI, CNR, and INAF (Italy); NASA and DoE (USA); STFC and UKSA (UK); CSIC, MINECO, JA, and RES (Spain); Tekes, AoF, and CSC (Finland); DLR and MPG (Germany); CSA (Canada); DTU Space (Denmark); SER/SSO (Switzerland); RCN (Norway); SFI (Ireland); FCT/MCTES (Portugal); ERC and PRACE (EU). A description of the Planck Collaboration and a list of its members, indicating which technical or scientific activities they have been involved in, can be found at \url{http://www.cosmos.esa.int/web/planck/planck-collaboration}. This research used resources of the National Energy Research Scientific Computing Center, 
a DOE Office of Science User Facility supported by the Office of Science of the U.S. Department 
of Energy under Contract No. DE-AC02-05CH11231.
Some of the results in this paper have been derived using the \texttt{HEALPix} package.
\end{acknowledgements}

\bibliographystyle{aat}

\bibliography{Bib,Planck_bib}
%PMF,PMFOne,PMF_NG,Bispectrum,PMF_Faraday,JKim_refs}

\begin{appendix}

\section{Impact of foregrounds on PMF constraints from the angular power spectra}

\begin{figure}
%\centering
\includegraphics{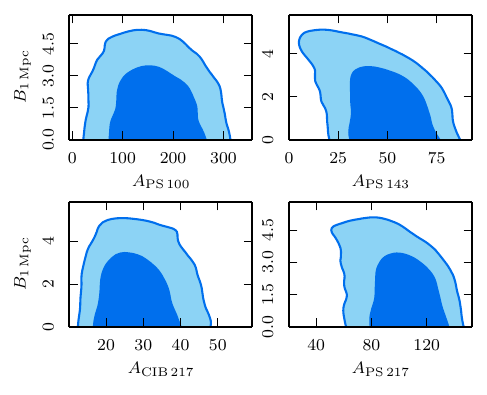}
\caption{Probability contours for the PMF amplitude and the foreground parameters for the \Planck 2013 likelihood.}
\label{fig:deg2013}
\end{figure}

\begin{figure}
%\centering
\includegraphics{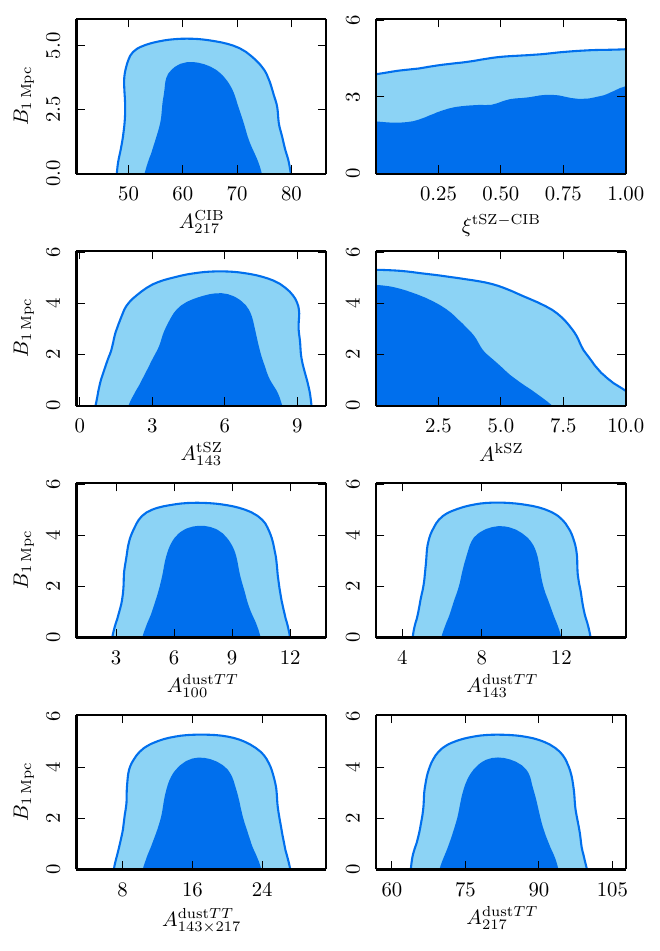}
\caption{Probability contours for the PMF amplitude and the foreground parameters.}
\label{fig:degall}
\end{figure}

In Fig.~\ref{fig:deg2013} we present the two-dimensional proability distributions of PMF amplitude and foregroun d parameters for the \Planck 2013 likelihood, which show only a mild degeneracy with the Poissonian amplitude for the 143 GHz.
In Fig.~\ref{fig:degall}, we plot the two-dimensional probability distributions of the PMF amplitude and the foreground parameters (for their description see \citealt{planck2014-a13}), except for the Poissonian terms, which have been discussed in Sect.~\ref{sec:foregroundresiduals}. These plots do not show any degeneracy. 
\onecolumn
\twocolumn

\section{Statistical anisotropy induced by Alfv\'en waves}
\label{anisotropy_alfven}

It has been shown that the presence of Alfv\'en waves in
the early Universe leads to specific correlations of the CMB in harmonic
space \citep{2008PhRvD..78f3012K}.  The signature correlations induced by Alfv\'en waves are as follows:
\begin{eqnarray} 
\langle a_{\ell m}\,a^*_{\ell m}\rangle &=&C_\ell+\frac{\ell(\ell+1)}{(2\ell-1)(2\ell+3)}\Biggl\{(\ell^2+\ell-3)\cos^2\theta_B\nonumber\\
&&+\ell(\ell+1)- (3\cos^2\theta_B-1)m^2\left[1-\frac{3}{\ell(\ell+1)}\right]\Biggr\} I^{\ell,\ell}_d\,;\nonumber
\end{eqnarray}
\begin{eqnarray} 
\langle a_{\ell,m}\,a^*_{\ell,m\pm 1}\rangle &=& -\sin2\theta_B \exp[\pm i \phi_B]\,\frac{\ell^2+\ell-3}{(2\ell-1)(2\ell+3)} \left(m\pm\frac{1}{2}\right)\nonumber\\
&&\times\sqrt{(\ell\mp m)(\ell\pm m+1)}\,I^{\ell,\ell}_d\,;\nonumber
\end{eqnarray}
\begin{eqnarray} 
\langle a_{\ell,m}\,a^*_{\ell,m\pm 2}\rangle &=& -\frac{1}{2}\sin^2\theta_B \exp[\pm i 2\phi_B]\,\frac{\ell^2+\ell-3}{(2\ell-1)(2\ell+3)}\nonumber\\
&&\times\sqrt{(\ell\mp m)(\ell\mp m-1)(\ell\pm m+1)(\ell\pm m+2)}\,I^{\ell,\ell}_d\,;\nonumber
\end{eqnarray}
\begin{eqnarray} 
\langle a_{\ell,m}\,a^*_{\ell+2,m}\rangle&=&-(3\cos^2\theta_B-1)\frac{(\ell+3)\ell}{2(2\ell+3)\sqrt{(2\ell+1)(2\ell+5)}}\nonumber\\
&&\times\sqrt{((\ell+1)^2-m^2)((\ell+2)^2-m^2)}\,I^{\ell,\ell+2}_d\,;\nonumber
\end{eqnarray}
\begin{eqnarray} 
\langle a_{\ell,m}\,a^*_{\ell-2,m}\rangle&=&-(3\cos^2\theta_B-1)\frac{(\ell+1)(\ell-2)}{2(2\ell-1)\sqrt{(2\ell-3)(2\ell+1)}}\nonumber\\
&&\times\sqrt{((\ell-1)^2-m^2)(\ell^2-m^2)}\,I^{\ell,\ell-2}_d\,;\nonumber
\end{eqnarray}
\begin{eqnarray} 
\langle a_{\ell,m}\,a^*_{\ell+2,m\pm 1}\rangle&=&\sin2\theta_B\exp[\pm i\phi_B]\frac{(\ell+3)l}{2(2\ell+3)\sqrt{(2\ell+1)(2\ell+5)}}\nonumber\\
&&\times\sqrt{((\ell+1)^2-m^2)(\ell\pm m+2)(\ell\pm m+3)}\, I^{\ell,\ell+2}_d\,;\nonumber
\end{eqnarray}

\begin{eqnarray} 
\lefteqn{\langle a_{\ell,m}\,a^*_{\ell-2,m\pm 1}\rangle=}\nonumber\\
&&-\sin2\theta_B\exp[\pm i\phi_B]\frac{(\ell+1)(\ell-2)}{2(2\ell-1)\sqrt{(2\ell-3)(2\ell+1)}}\nonumber\\
&&\times\sqrt{(\ell^2-m^2)(\ell\mp m-1)(\ell\mp m-2)}\, I^{\ell,\ell-2}_d\,;\nonumber
\end{eqnarray}
\begin{eqnarray} 
\lefteqn{\langle a_{\ell,m}\,a^*_{\ell+2,m\pm 2}\rangle=}\nonumber\\
&&-\frac{1}{2}\sin^2\theta_B\exp[\pm i 2\phi_B]\frac{(\ell+3)\ell}{2(2\ell+3)\sqrt{(2\ell+1)(2\ell+5)}}\nonumber\\
&&\times \sqrt{((\ell\pm m+1)(\ell\pm m+2)(\ell\pm m+3)(\ell\pm m+4)} I^{\ell,\ell+2}_d\,;\nonumber
\end{eqnarray}
\begin{eqnarray} 
\lefteqn{\langle a_{\ell,m}\,a^*_{\ell-2,m\pm 2}\rangle=}\nonumber\\
&&-\frac{1}{2}\sin^2\theta_B\exp[\pm i 2\phi_B]\frac{(\ell+1)(\ell-2)}{2(2\ell-1)\sqrt{(2\ell-3)(2\ell+1)}}\nonumber\\
&&\times\sqrt{((\ell\mp m-3)(\ell\mp m-2)(\ell\mp m-1)(\ell\mp m)} I^{\ell,\ell-2}_d\,.\nonumber
\end{eqnarray}
Here $C_\ell$ is the power spectrum in the absence of Alfv\'en waves, 
$\theta_B$ and $\phi_B$ are the spherical angles of the direction of the background magnetic field $\bar{\vec{B}}$, and
$I^{\ell \ell'}_d$ is given by
\begin{eqnarray} 
I^{\ell \ell'}_d&=& 2T^2_0\left(\frac{\tau_{\mathrm{last}}}{\tau_0}\right)^2\, (2\pi)^{n_B+7}\frac{B^2_\lambda}{\Gamma(n_B/2+3/2)} \frac{\varv^4_A}{\bar B^2}\label{Iellell}\\
&&\times\int \,d\ln k\,\left(\frac{k}{k_\lambda}\right)^{n_B+3}\exp\left(-2\frac{k^2}{k^2_\mathrm{D}}\right) j_\ell(k\tau_0)\,j_{\ell'}(k\tau_0)\,,\nonumber
\end{eqnarray}
where $k_\mathrm{D}$ is the co-moving wave number of the dissipation scale due to photon viscosity
and given approximately by $10/(c\tau_{\mathrm{last}})$ \citep{DKY}. 
The dissipative damping effect becomes significant at multipoles $\ell\gtrsim 500$ \citep{DKY}.  
We note that the damping scale considered in this context is different from the one considered in Sect.~\ref{sec:APS}.
The damping effects considered in each case are related to two different aspects and are specific to the two topics treated; 
in the study of the impact of PMFs on CMB anisotropies,
the damping scale considered is due to the dissipation of the PMFs themselves as investigated by \cite{1998PhRvD..58h3502S},
in this section instead, we consider a damping scale derived from the damping of the vector perturbations generated
by the PMFs, not the PMFs themselves. The latter damping scale is derived by \cite{DKY}. 
In Eq.~\eqref{Iellell}, $B^2_{\lambda}$ denotes the total power of the non-helical PMF smoothed at the spatial scale of $\lambda$.
There are general cases for dipole and quadrupole coupling, where cosmological parameters are assumed to vary with position \citep{Moss2011}.
If we allow a position-dependent parameter to be a vector and treat the PMF as a position-dependent parameter, 
the correlation induced by Alfv\'en waves may be incorporated into the framework of this approach.
\end{appendix}
\end{document}